\documentclass[fleqn,usenatbib]{mnras}

\usepackage{newtxtext,newtxmath}

\usepackage[T1]{fontenc}

\hypersetup{
    colorlinks=true,
    linkcolor=red,
    filecolor=magenta,
    urlcolor=cyan,
}

\newcommand{\msun}{\mbox{${\rm M}_\odot$}}
\newcommand{\mstar}{\mbox{${M}_{\rm star}$}}
\newcommand{\cmjj}{\mbox{${\rm cm^{-3}}$}}
\newcommand{\lya}{\mbox{${\rm Ly}\alpha$}}
\newcommand{\lyb}{\mbox{${\rm Ly}\beta$}}
\newcommand{\kms}{\,{\rm{km\,s}^{-1}}}
\newcommand{\cc}{{\rm cm^{-3}}}

\defcitealias{Chen:2020aa}{Paper I}

\usepackage{soul}
\usepackage{graphicx}
\usepackage{amsmath}
\usepackage{subfigure}
\usepackage{multirow}
\usepackage{threeparttable}
\usepackage{array}
\usepackage{natbib}
\usepackage{xcolor}
\usepackage{subfigure}
\usepackage{hyperref}
\usepackage[version=4]{mhchem}
\usepackage{bm}
\usepackage{longtable}
\usepackage[normalem]{ulem}

\title[CUBS V: Thermodynamics of the Cool CGM at $z\lesssim 1$]{The Cosmic Ultraviolet Baryon Survey (CUBS) V: On the Thermodynamic Properties of the Cool Circumgalactic Medium at $\bm{z\lesssim 1}$}

\author[Z. Qu et al.]{
Zhijie Qu$^{1}$\thanks{E-mail: quzhijie@uchicago.edu},
Hsiao-Wen Chen$^{1}$,
Gwen C.\ Rudie$^{2}$,
Fakhri S.\ Zahedy$^{2}$,
Sean D. Johnson$^{3}$, 
\newauthor
Erin Boettcher$^{4, 5, 6}$,
Sebastiano Cantalupo$^{7}$,
Mandy C. Chen$^{1}$,
Kathy L. Cooksey$^{8}$,
David DePalma$^{9}$,
\newauthor
Claude-Andr\'e Faucher-Gigu\`ere$^{10}$,
Michael Rauch$^2$,
Joop\ Schaye$^{11}$,
and Robert A. Simcoe$^9$
\\
$^{1}$Department of Astronomy \& Astrophysics, The University of Chicago, 5640 S. Ellis Ave., Chicago, IL 60637, USA\\
$^{2}$The Observatories of the Carnegie Institution for Science, 813 Santa Barbara Street, Pasadena, CA 91101, USA\\
$^{3}$ Department of Astronomy, University of Michigan, Ann Arbor, MI 48109, USA\\
$^4$ Department of Astronomy, University of Maryland, College Park, MD 20742, USA\\
$^5$ X-ray Astrophysics Laboratory, NASA/GSFC, Greenbelt, MD 20771, USA\\
$^6$ Center for Research and Exploration in Space Science and Technology, NASA/GSFC, Greenbelt, MD 20771, USA\\
$^7$ Department of Physics, University of Milan Bicocca, Piazza della Scienza 3, I-20126 Milano, Italy \\
$^8$ Department of Physics and Astronomy, University of Hawai'i at Hilo, Hilo, HI 96720, USA\\
$^9$ MIT-Kavli Institute for Astrophysics and Space Research, 77 Massachusetts Ave., Cambridge, MA 02139, USA\\
$^{10}$ Department of Physics \& Astronomy, Center for Interdisciplinary Exploration and Research in Astrophysics (CIERA), Northwestern University, \\
1800 Sherman Avenue, Evanston, IL 60201, USA \\
$^{11}$ Leiden Observatory, Leiden University, PO Box 9513, NL-2300 RA Leiden, the Netherlands
}

\date{Accepted XXX. Received YYY; in original form ZZZ}

\pubyear{2022}

\begin{document}
\label{firstpage}
\pagerange{\pageref{firstpage}--\pageref{lastpage}}
\maketitle

\begin{abstract}
This paper presents a systematic study of the photoionization and thermodynamic properties of the cool circumgalactic medium (CGM) as traced by rest-frame ultraviolet absorption lines around 26 galaxies at redshift $z\lesssim1$.
The study utilizes both high-quality far-ultraviolet and optical spectra of background QSOs and deep galaxy redshift surveys to characterize the gas density, temperature, and pressure of individual absorbing components and to resolve their internal non-thermal motions.
The derived gas density spans more than three decades, from $\log (n_{\rm H}/\cmjj) \approx -4$ to $-1$, while the temperature of the gas is confined in a narrow range of $\log (T/{\rm K})\approx 4.3\pm 0.3$.
In addition, a weak anti-correlation between gas density and temperature is observed, consistent with the expectation of the gas being in photoionization equilibrium.
Furthermore, decomposing the observed line widths into thermal and non-thermal contributions reveals that more than 30\% of the components at $z\lesssim 1$ exhibit line widths driven by non-thermal motions, in comparison to $<20$\% found at $z\approx 2$-3.
Attributing the observed non-thermal line widths to intra-clump turbulence, we find that massive quenched galaxies on average exhibit higher non-thermal broadening/turbulent energy in their CGM compared to star-forming galaxies at $z\lesssim 1$.
Finally, strong absorption features from multiple ions covering a wide range of ionization energy (e.g., from \ion{Mg}{II} to \ion{O}{IV}) can be present simultaneously in a single absorption system with kinematically aligned component structure, but the inferred pressure in different phases may differ by a factor of $\approx 10$.
\end{abstract}

\begin{keywords}
surveys -- galaxies: haloes -- intergalactic medium -- quasars: absorption lines
\end{keywords}



\section{Introduction}

The circumgalactic medium (CGM), a baryon reservoir surrounding galaxies, represents a critical interface in the galactic baryon cycle that drives galaxy evolution (see \citealt{Tumlinson:2017aa, Donahue2022} for recent reviews).
The CGM is the reservoir that gathers both feedback energy and material ejected from galaxies, as well as accreting gas, which will feed star formation \citep[e.g.,][]{Naab2017}.
Theoretical simulations suggest that at high redshift ($z\gtrsim 2$-3), the hot CGM begins to form as a result of feedback from star formation in the galaxy and accretion shocks from continuous accretion \citep[e.g.,][]{Faucher-Giguere:2011aa, vandeVoort:2011aa, Correa:2018aa}.
Meanwhile, the gas assembly of galaxy transitions from direct intensive accretion from the intergalactic medium (IGM) to continuous cooling accretion from the hot CGM \citep[e.g., ][]{Keres:2005aa, Nelson:2013aa}, which also affects star formation in the galaxy.

Quasar absorption spectroscopy is a powerful tool for characterizing the diffuse CGM and IGM over a broad redshift range.
Previous absorption-line surveys along random sightlines have yielded an accurate accounting of the cosmic evolution of different ions based on the observed redshift dependence of  
their column density distribution functions 
\citep[e.g.,][for a comprehensive list of empirical studies]{Rahmati2016}.
In parallel, galaxy-centric absorption spectroscopy using background QSOs has revealed that the CGM is multiphase and contributes significantly to the total baryonic mass budget \citep[e.g.,][]{Tumlinson:2017aa, Johnson:2017aa, Rudie:2019aa, Zahedy:2020}. 
In addition to these statistical measurements, detailed absorption profile analyses to determine the relative absorption strengths between different ions provide further constraints on the physical properties of the gas such as its density, ionization state, metallicity, and temperature \citep[e.g.,][]{Zahedy:2019aa,Zahedy:2021aa, Cooper:2021aa, Haislmaier:2021,Sameer:2021aa}.
However, studies of these derived (but more fundamental) quantities remain scarce and little is known about how these physical properties evolve with time.

Constraining the thermodynamic properties of the CGM is directly relevant to understanding how the CGM evolves and how it is coupled with the host galaxies. 
It requires knowledge of the gas density, temperature, turbulent velocity, and the presence or absence of non-thermal energy/pressure sources such as cosmic rays and magnetic fields.
Constraints on the gas density provide an estimate of the total gas mass; constraints on the gas temperature provide an estimate of the cooling efficiency that relates to the ability of a galaxy to sustain star formation; and constraints on the turbulent velocity provide an estimate of additional sources of pressure that help maintain the dynamic state of the gas.

Beyond the local universe, direct measurements of a volume-filling hot CGM in X-ray bands are currently unattainable for $\lesssim L^*$ galaxies \citep[e.g., ][]{Bregman:2007aa}.
Instead, the cool, photoionized CGM may serve as a tracer of the ambient hot gas through a simple pressure balance assumption  \citep[e.g.,][]{Voit:2019aa}.
Obtaining a robust characterization of the thermodynamic properties of the cool CGM requires high signal-to-noise (S/N) and high-resolution absorption spectra for resolving the complex absorption profiles of different ions from different phases \citep[e.g.,][]{Rudie:2019aa,Zahedy:2019aa}.
In addition, deep galaxy redshift surveys around the background QSOs are necessary for connecting the observed absorption properties with star formation properties in the host galaxies as well as with the galaxy environment.

The Cosmic Ultraviolet Baryon Survey (CUBS) is designed to track the CGM evolution over a majority of cosmic time (last eight-billion years; $z\approx0$ to $z\approx 1$), with both high-quality QSO absorption spectra and deep galaxy survey data available (\citealt{Chen:2020aa}; hereafter \citetalias{Chen:2020aa}; see also Section \ref{sec:data}). 
In this study, we leverage available galaxy survey data and far-ultraviolet (FUV) and optical quasar absorption spectra from CUBS to investigate how the thermodynamic properties of the CGM have evolved since $z\approx 1$.  We
have compiled a sample of galaxies at $z\lesssim 1$ for which absorption components have been detected in the CGM.
The galaxy sample includes new $z\!\approx\!1$ galaxies identified in the CUBS program (designated as CUBSz1; CUBS VI, in preparation) and galaxies at $0.2\!\lesssim\!z\!\lesssim\!0.6$ from the literature, which also include previously published galaxies from the CUBS program.
For each absorption system, we carry out detailed photoionization modeling and line width analyses (Section \ref{sec:analysis}) to determine the underlying gas density, temperature, and turbulent velocity (Section \ref{sec:scaling}).
Our analysis shows that in comparison to the CGM measurements at $z\approx 2-3$ \citep[e.g.,][]{Rauch1996,Simcoe:2006aa,Kim2016,Rudie:2019aa}, non-thermal motions play a more significant role in setting the thermodynamic state of the cool CGM at $z\lesssim 1$ (Section \ref{sec:T_NT}).  In addition, our analysis also reveals that multiphase gas is prevalent in the CGM at $z\approx 1$ (Section \ref{sec:everywhere}) and that the gas pressure  between different phases can differ by more than a factor of 10 (Section \ref{sec:mpc}).

\section{Data}
\label{sec:data}

The CUBS program covers 15 fields around a UV bright QSO at $z_{\rm QSO}\gtrsim 0.8$ with {\it GALEX} NUV magnitude of $AB({\rm NUV})< 17$.
For each field, high-quality FUV and optical absorption spectra of the QSO, along with deep imaging and spectroscopic galaxy survey data, are available for probing the diffuse CGM and IGM along the QSO sightline at $z\lesssim 1$.
The survey design and data of the CUBS program are described in detail in \citetalias{Chen:2020aa}.
Here we briefly summarize the available data and introduce a new galaxy sample at $z\approx 1$ for investigating the thermodynamic properties of the CGM.
In this study, we also include literature samples with similar data quality and analysis methods.
Totally, there are 42 unique galaxy systems (20 in CUBSz1 and 22 from the literature), among which 26 systems (9 in CUBSz1 and 17 in the literature) have multiple absorption transitions detected for constraining the physical properties of individual components (e.g., density and temperature).

\subsection{QSO absorption spectra and galaxy gurveys}
We obtained medium-resolution FUV QSO absorption spectra using the {\it Hubble Space Telescope} ({\it HST}) Cosmic Origins Spectrograph (COS; \citealt{Green2012}) and the G130M and G160M gratings (PID$=$15163; PI: Chen).
Multiple central wavelengths were adopted to obtain a contiguous spectral coverage from 1100 \r{A} to 1800 \r{A}.
We coadded pipeline-reduced individual exposures using custom software developed by one of us (Johnson; see  \citealt{Chen:2018aa}; \citetalias{Chen:2020aa} for details).  This custom applies additional corrections in the wavelength calibrations of COS spectra.  It takes advantage of all usable transitions to achieve a wavelength calibration accuracy of $\lesssim 5\,\kms$ (\citealt{Johnson:2013aa}; see also \citealt{Wakker2015aa}).  The final coadded spectra have a typical spectral resolution of FWHM $\approx 20 \kms$ and a median S/N of 12-31 per resolution element over the full spectral range.
At $z\!\approx\!1$, HST/COS FUV spectra cover a wide range of ions, including \ion{H}{I}, \ion{He}{I}, \ion{C}{II}, \ion{N}{II} to \ion{N}{IV}, \ion{O}{I} to \ion{O}{V}, \ion{S}{II} to \ion{S}{V}, \ion{Ne}{IV} to \ion{Ne}{VI}, \ion{Ne}{VIII}, and \ion{Mg}{X}.

We obtained high-resolution optical spectra of the QSOs using the Magellan Inamori Kyocera Echelle Spectrograph (MIKE; \citealt{Bernstein2003}), covering a wavelength range from 3300\!~\! \r{A} to 9300\!~\! \r{A} with spectral resolution of FWHM\,$\approx\!8$-10\,$\kms$.
The echelle spectra were processed and combined using custom software \citep{Zahedy:2016aa}, leading to a typical S/N of 22-63.
Complementary to the FUV spectra, the MIKE spectra provide additional coverage for \ion{Mg}{I}, \ion{Mg}{II}, \ion{Fe}{II}, and \ion{C}{IV} at $z\!\approx\!1$.

The CUBS galaxy survey component includes three different elements, each targeting a different combination of survey depth and field of view (see \citetalias{Chen:2020aa} and \citealt{Cooper:2021aa} for a more detailed description).
The deepest element was completed using VLT/MUSE \citep{Bacon2010}, covering the inner $1'\times 1'$ region centered around the QSO (PID$=$0104.A-0147; PI: Chen).
The MUSE observations reached a limiting magnitude of $AB(r)\approx 25$, enabling identifications of faint galaxies down to stellar mass of $\mstar\!\approx\!10^8$-$10^9 \msun$ at projected distance $d\!\lesssim\!250$ kpc from the QSO sightline at $z\!\approx1\!$.
Additional galaxy survey data were obtained using LDSS3 and IMACS \citep{Osip:2008aa, Dressler2011} on the Magellan Telescopes to reach
a limiting magnitude of $AB(r)\!=\!24$ and $22.5$, respectively.  These wide-field survey data enable a detailed investigation of the galaxy environment beyond the MUSE footprint to $\approx 1-5$ Mpc.

\subsection{CUBSz1: a new galaxy sample probing the CGM at $\bm{z \approx 1}$}
\label{sec:cubsz1}

A new sample of galaxies was assembled from the CUBS program for investigating the CGM at $z\approx 1$ (i.e., CUBSz1).
The galaxies were selected based on their close proximity to the QSO sightline with no prior knowledge of whether or not an absorption feature was known. 
The focus on $z\!\approx\!1$ was motivated by the need to bridge the gap in existing CGM studies between $z\!\approx\!2$ and $z\!\lesssim\!0.4$ in order to understand the rapid decline in the cosmic star formation rate (SFR) density between the two cosmic epochs \citep[e.g.,][]{Madau:2014aa}.  

To summarize, we selected galaxies spectroscopically identified at $d<200$ kpc from the QSO sightlines.  The adopted search radius of 200 kpc corresponds to roughly $1.5\,r_{\rm vir}$ (halo radius) for typical $L_*$ galaxies in dark matter haloes of mass $M_h=10^{12}\,\msun$ at $z\!=\!1$.
In addition, we focused on selecting galaxies at $z\!>\!0.883$ to ensure that the observed \ion{He}{I}\,$\lambda\,584$ line strength could be adopted as a proxy for the neutral hydrogen column density $N({\rm HI})$ when hydrogen \lya\ and \lyb\ lines are not covered by the available COS spectra.
The helium abundance is roughly constant from $z\!\approx\!2$-3 to the local universe \citep{Cooke:2018aa}.  Therefore, \ion{He}{I} column densities, $N({\rm HeI})$, provide a good proxy of $N({\rm HI})$ after appropriate ionization corrections are accounted for.  
This exercise yielded 26 galaxies at $d\!<\!200$ kpc and $z\!\approx\!0.89$-1.21 in six CUBS fields.  Considering galaxies with line-of-sight velocity $|\Delta\,v_g|\!\lesssim\!500 \kms$ and projected distance of $\lesssim 1$ Mpc as part of the group led to a total of 20 unique galaxies or galaxy groups at $z\!\approx\!0.89$-1.21 in this new galaxy sample.
The typical number of galaxy members in a group was between 1 and 3.

We searched for associated absorption features within $\pm 500\,\kms$  of each galaxy in the COS and MIKE QSO spectra.
Among the 20 galaxies and galaxy groups, nine systems have detected absorption transitions from a wide range of ionization states, from low ionization state species (such as \ion{Mg}{II}, \ion{S}{II}) to high ionization species (such as \ion{Ne}{VIII}), while the remaining 11 systems exhibit either no detectable absorption signals or no meaningful constraints can be placed due to contaminating features.
In this study, we focus on investigating the thermodynamic properties of the cool CGM based on the observed low-to-intermediate ionization state ions.   Detailed galaxy and absorber properties are presented in a separate paper (CUBS VI, in preparation).
 
\subsection{Literature samples for probing the CGM at $\bm{z<1}$}

To explore how the observed CGM properties evolve with redshift, we also considered previously published galaxy samples in the analysis.
There have been extensive efforts in extracting physical properties of the CGM using UV absorption transitions \citep[e.g.,][]{Stocke:2013,Savage:2014aa, Shull:2014aa, Werk:2014aa, Keeney:2017aa, Lehner:2019aa, Prochaska:2017aa}.
Here, we considered galaxy-CGM systems with spectrally resolved component-by-component absorption properties available, including those published in CUBS III \citep{Zahedy:2021aa}, CUBS IV \citep{Cooper:2021aa}, and the COS-LRG survey targeting passive luminous red galaxies \citep[LRG;][]{Chen:2018aa,Zahedy:2019aa}.
These samples extended the CGM thermodynamic study to lower redshifts, $z\!\approx\!0.2$-0.6.
Altogether, 17 unique galaxies or galaxy groups were collected from the literature 
for the thermodynamic study.
We adopted the measured Doppler line widths $b$ of all ions from previous authors for all these systems.  For gas densities, we applied a 0.2 dex offset to the published values from CUBS IV, because these were based on an updated UVB from \citet[hereafter FG20]{Faucher-Giguere:2020aa}.  Both the COS-LRG and CUBS III samples adopted an ultraviolet background (UVB) from \citet[commonly referred to as HM05]{Haardt:2001aa}, which has been shown to lead to on average 0.2 dex higher gas density than what would be obtained using FG20 \citep[e.g.,][]{Zahedy:2021aa}.  As described below, we chose to carry out the photo-ionization analysis using HM05 to minimize the offsets applied to the literature sample in order to bring all density estimates to a common background for consistency.

\section{Analysis}
Constraining the thermodynamic properties of the diffuse cool CGM at $z\lesssim 1$ requires knowledge of the density, temperature, and turbulent velocity of the gas.
Here, we describe the three stages to constrain these physical properties: (1) a Voigt profile fitting routine to obtain the ionic column densities, line widths, and the associated uncertainties; (2) a photoionization model analysis to determine the gas density based on the relative abundances of different ions; (3) a line width profile analysis to delineate thermal and non-thermal contributions to the observed line widths for constraining turbulent velocities.

\label{sec:analysis}
\subsection{Voigt profile analysis}
Medium-resolution {\it HST} COS FUV spectra and high-resolution MIKE optical spectra enable the decomposition of each absorption feature into multiple components and provide the discriminating power to determine the velocity alignment of individual components between different ionic species \citep[e.g.,][]{Zahedy:2021aa, Cooper:2021aa}.
For each resolved component, a model Voigt profile is fit to the data to determine the velocity centroid ($v$), column density ($N$), and line width (i.e., the Doppler parameter; $b$).
The number of absorption components to be considered in the decomposition and the corresponding velocity centroids of individual components are determined based on a global inspection of the observed \ion{H}{I}, \ion{Mg}{II}, and \ion{O}{IV} lines.
In particular, the \ion{Mg}{II} doublet transitions are relative narrow and are detected in high-resolution and high S/N MIKE spectra of ${\rm FWHM}\!\approx\!8$-10 $\kms$ and $S/N\!\approx\!50$ (in comparison to ${\rm FWHM}\!\approx\!20$ $\kms$ and $S/N\!\approx\!20$ from {\it HST} COS).
These transitions provide the strongest constraint for the velocity centroids of individual components.

However, the \ion{Mg}{II} lines are only detected in relatively high-density (i.e., low-ionization state) gas.
When \ion{Mg}{II} is absent, the component decomposition relies on comparisons between different \ion{H}{I}, \ion{O}{III}, and \ion{O}{IV} lines.
\ion{H}{I} (for galaxies at $z\lesssim 0.94$), \ion{O}{III}, and \ion{O}{IV}
have multiple strong transitions in the FUV window, which allow accurate identifications of real signal in the presence of contaminating features based on the anticipated line ratios and enable a robust decomposition using only COS spectra.

Nine galaxies/galaxy groups in the CUBSz1 sample exhibit associated absorption features in the QSO spectra and these absorption systems are decomposed into 26 kinematically aligned absorption components, each with multiple ionic transitions detected at consistent velocity controids.
For additional ionic transitions, the line centroids are fixed in the Voigt profile models to the velocity components identified in \ion{H}{I}, \ion{Mg}{II}, and \ion{O}{IV}.
The final best-fit parameters, including $N$ and $b$, are determined using a Bayesian framework (e.g., \citealt{Zahedy:2021aa}; implemented using {\texttt{emcee}}; \citealt{Foreman-Mackey:2013aa}), and the posterior distribution of each parameter is recorded for subsequent photoionization and line width analyses.

 \subsection{Photoionization modeling}
 \label{sec:photo}
\begin{figure*}
\begin{center}
\includegraphics[width=0.973\textwidth]{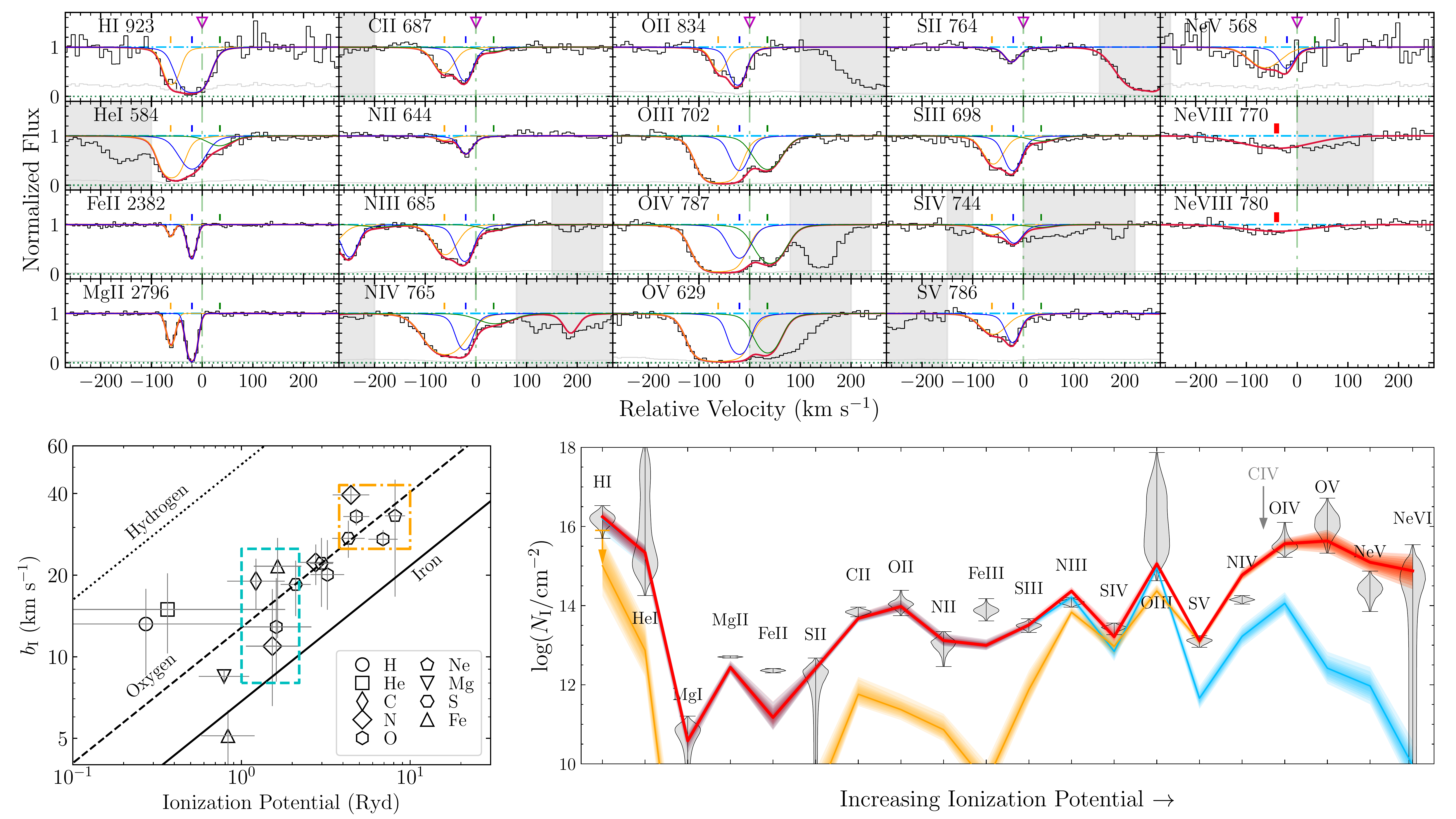}
\end{center}
 \vspace{-0.3cm}
\caption{An example of a multiphase component in the CUBSz1 sample.  This is a pLLS of $\log\,N_{{\rm HI}}/{\rm cm}^{-2}\!=\!16.2\pm 0.1$ at $-62\,\kms$ in the top spectral panels (the orange component), which occurs in the vicinity of a group of galaxies at $z = 0.9373$ (zero relative velocity in the spectral panels) along the sightline toward CUBS J0333$-$4102.
The observed flux and associated 1-$\sigma$ uncertainty are the black and gray histograms. The shaded gray regions are blended regions excluded from fitting.
The red curve is the combined flux from the Voigt-profile components shown by the other colored curves.
The thin color vertical bars indicate the low to intermediate ionization states adopted in the photoionization modelings, while the thick red vertical bars indicate the high ionization state transitions.
The magenta triangle at the zero velocity is the systemic redshift of the absorption system defined by the closet galaxy.
The {\it bottom-left} panel shows the observed line widths $b_I$ of individual transitions and the corresponding ionization potentials of the associated ions.
For each ion, the vertical bar represents the 1-$\sigma$ uncertainty and
the horizontal bar marks the ionization energy range from production to further ionization of each specie.  The symbol is placed at the logarithmic median of this range.
There is a clear trend that higher ionization species display larger $b$ values.
Ions associated with two photoionized phases are highlighted in cyan and orange boxes, bracketing the three intermediate ions (\ion{N}{III}, \ion{S}{IV}, and \ion{O}{III}) with contributions from both.
The dotted, dashed, and solid lines are the expected $b_I$ values, when converting the ionization energy to thermal energy for hydrogen, oxygen, and iron, respectively.
The {\it bottom-right} panel shows comparisons between observed (points with error bars) and predicted (color bands) column densities from our photoionization analysis. 
The data points are sorted by the ionization potentials of individual ions, and the red band shows the combined model prediction with 90\% uncertainties from two different phases. 
The high-density phase model (the cyan line and shaded region) is constrained by \ion{H}{I}, \ion{O}{II}, \ion{S}{II}, and \ion{S}{III}, while the low density phase model (the orange line and shaded region) is constrained by broad \ion{H}{I} (upper limit to $\log\,N_{{\rm HI}, {\rm broad}}/{\rm cm}^{-2}$ with a fixed $b_{{\rm HI}, {\rm broad}}\!=\!35\,\kms$), \ion{S}{V}, \ion{N}{IV}, \ion{O}{IV}, and \ion{O}{V}.  In the absence of the FUV ions, \ion{C}{IV} provides a valuable alternative to constrain the low-density phase.
Here, we assume a relative solar abundance pattern for oxygen and sulfur with $[{\rm S}/{\rm O}]\!=\!0$, and allow
deviations from the solar pattern for other ions.
}
\label{fig:mpc_system}
\end{figure*}

A grid of photoionization models are constructed to infer gas densities and chemical abundances of individual absorption components based on the measured ionic column densities.
We perform a series of calculations using Cloudy \citep[v17;][]{Ferland:2017RMxAA}, assuming photoionization equilibrium (PIE).
The adopted ultraviolet background (UVB) is an updated version of the \citet{Haardt:2001aa} UVB (i.e., HM05 in Cloudy).
A three-dimensional grid of PIE models are calculated by varying \ion{H}{I} column density from $\log (N_{\rm H{\small I}}/{\rm cm^{-2}}) = 13$ to $19$; gas density from $\log (n_{\rm H}/\cmjj) = -6$ to 1; and the metallicity from $\log (Z/Z_\odot) = -4$ to 1, all in steps of $0.25$ dex.
A Markov chain Monte Carlo (MCMC) approach is adopted to search for the best solution of individual absorption components in the PIE model grid (implemented using {\texttt{emcee}}; \citealt{Foreman-Mackey:2013aa}).
The joint likelihood of the photoionization model for each component over a suite of ions is calculated following
\begin{eqnarray}
\ln p(N_{\rm H{\small I}}, n_{\rm H}, Z) = \sum_{I} \ln\,p_{N_{I}}(\hat{N}_{I}),
\end{eqnarray}
where $\hat{N}_{I}$ is the model column density of ion $I$ expected for a particular combination of $N_{\rm H{\small I}}$, $n_{\rm H}$, and $Z$, and $p_{N_{I}}$ is the posterior probability distribution of the column density extracted from the MCMC chain obtained in the Voigt profile analysis. In the lower right panel of Figure \ref{fig:mpc_system}, the violin plot shows the column density posterior distribution for measurements (e.g., \ion{H}{I}, \ion{Mg}{II}, and \ion{O}{IV}), upper limits for non-detection (e.g., \ion{Mg}{I} and \ion{S}{II}), broad allowable range for saturated features (e.g., \ion{He}{I} and \ion{O}{III}).

Of the 26 absorption components identified in CUBSz1, 20 have multiple associated ions available to constrain the gas density.
Wherever possible, we combine oxygen and sulfur ions to determine a best-fit photoionization model, because they are both $\alpha$-elements and share a similar abundance pattern.
In the CGM, a simple photoionization scenario can typically explain the observed relative abundances between different ions from low-ionization species such as \ion{O}{I} to intermediate ionization species such as \ion{O}{IV}, and possibly \ion{O}{V} and \ion{Ne}{V} with HM05 as the incident field \citep[e.g.,][]{Savage:2005aa}.
To explain higher ionization species such as \ion{Ne}{VI} and \ion{Ne}{VIII} under photoionization becomes significantly more challenging, because the preferred gas density becomes unphysically low, $\log(n_{\rm H}/{\cc}) \lesssim -5$, and the inferred cloud size becomes unrealistically large, $\gtrsim 1\!~\rm Mpc$ \citep[e.g.,][]{Savage:2005aa, Meiring:2013aa}.
Therefore, to explain \ion{Ne}{VI} and \ion{Ne}{VIII}, collisional ionization, a much more intensive (harder) radiation field or non-equilibrium ionization will be needed \citep[e.g.,][]{Oppenheimer:2013aa, Hussain:2017aa}.  We focus on the gas that produces low and intermediate-ionization species in this study, and defer the discussion of those extreme ions to a subsequent paper. 

For each absorption component, we start with a single-phase (i.e., single-density) photoionization model to explain the low-to-intermediate ionization state ions (i.e., \ion{H}{I} to \ion{O}{IV}).
In most cases (16 of 20 components in the CUBSz1 sample), the measured ionic column densities can be reproduced by a single-phase model.
However, four absorption components cannot be explained by a single-phase photoionization model, which are referred to as multiphase components.
For these cases, we consider two different densities 
with the high-density phase typically constrained by \ion{O}{II}, \ion{S}{II}, and \ion{S}{III} and the low-density phase  constrained by \ion{S}{V}, \ion{O}{IV}, and sometimes \ion{O}{V}.
The decomposition of these two phases is guided by the observed line widths \citep[e.g.,][]{Cooper:2021aa}.  An example is shown in Figure \ref{fig:mpc_system} for illustration.  

The multiphase component shown in Figure \ref{fig:mpc_system} is a partial Lyman limit system (pLLS) with neutral hydrogen column density $\log\,N_{{\rm HI}}/{\rm cm}^{-2}=16.2\pm 0.1$ at $z=0.9373$ toward QSO J\,0333$-$4102 at $z_{\rm QSO}=1.124$.  This component occurs at $d\!=\!70$-180 kpc and line-of-sight velocity $v\,=\,-62\,\kms$ (relative to the closest galaxy) from a group of at least two star-forming galaxies.
The observed low-ionization lines, including \ion{O}{II}, \ion{S}{II}, \ion{S}{III}, \ion{N}{II}, \ion{C}{II}, and \ion{Fe}{III}, all share a comparable line width with $b\!\approx\!15\,\kms$, while high-ionization transitions, including \ion{S}{V}, \ion{O}{IV}, \ion{O}{V}, \ion{N}{IV}, and \ion{Ne}{V}, display a broader line width of $b\!\approx\!30\,\kms$ ({\it left} panel of Figure \ref{fig:mpc_system}).
At the same time, intermediate ionic transitions (i.e., \ion{N}{III}, \ion{S}{IV}, and \ion{O}{III}) exhibit a line width of $b\!\approx\!20\,\kms$ in between the low- and high-ionization lines.
Our photoionization analysis shows that this component is best described by a two-phase model with a 
high-density phase of $\log (n_{\rm H}/\cc) = -2.21_{-0.10}^{+0.09}$ primarily responsible for the low-ionization lines and a low-density phase of $\log (n_{\rm H}/\cc) = -3.95\pm0.11$ for high-ionization transitions.  Both phases contribute significantly to the intermediate-state ions (lower right panel of Figure \ref{fig:mpc_system}).  Therefore, attributing these intermediate-state ions solely to high-density (or low-density) phase would result in an underestimate (or overestimate) of the gas density.

To summarize, the inferred gas densities from the photoionization model analysis for all components in the CUBSz1 sample, along with published values for the literature samples, are presented in Column (3) of Table \ref{tab:sample}.  Low-ionization phase components are marked by ``l'' in the system ID, while high-ionization phase components are marked as ``h''.

\subsection{Line width analysis}
The observed line widths of individual ionic transitions are set by the underlying gas temperature and non-thermal motions of ions, which in turn provide empirical constraints on the gas pressure.
Specifically, thermal broadening depends on temperature $T$ and atomic mass $m_I$ following $b_{{\rm T}} = (2k_{\rm B}T/m_I)^{1/2}$, while non-thermal broadening is constant for all transitions in a single phase, $b_{{\rm NT}}$, independent of $m_I$.
Then, the expected model line width for one ion is expressed following
\begin{eqnarray}
\hat{b}_{I}^2(T, b_{{\rm NT}})= \frac{2 k_{\rm B}T}{m_I} + b_{{\rm NT}}^2,
\end{eqnarray}
where we implicitly assumed that the non-thermal line profile can be described by a Gaussian function.

The best-fit gas temperature and non-thermal velocity, along with associated uncertainties, are obtained using an MCMC approach implemented by \texttt{emcee}.
For each component in the CUBSz1 sample, the likelihood of observing a suite of ions of line width $b_{I}$ is calculated following
\begin{eqnarray}
\ln\,p(T,b_{{\rm NT}}) = \sum_{I} \ln\,p_{b_{I}}(\hat{b}_{I}),
\end{eqnarray}
where $p_{b_{I}}$ is the posterior distribution of $b_{I}$ for ion $I$.
For the literature samples, the likelihood is calculated by interpreting the published error uncertainties as the 1-$\sigma$ confidence interval of a Gaussian distribution function.  The likelihood of observing a suite of ions of line width $b_{c,I}$ is therefore
\begin{eqnarray}
\ln p(T, b_{{\rm NT}})\propto -\sum_{I} \frac{(\hat{b}_{I}-b_{I})^2}{2\sigma_{b_{I}}^2},
\end{eqnarray}
where $b_{I}$ and the associated uncertainties, $\sigma_{b_{I}}$ are adopted from the literature \citep{Zahedy:2019aa, Zahedy:2021aa, Cooper:2021aa}.

The derived temperature and non-thermal velocity, as well as the estimated 68\% confidence
interval for each quantity, are presented in Columns (7) and (8) of Table \ref{tab:sample}, along with the measured $b$ values for \ion{H}{I} and metal lines in Columns (4) and (5).  In the thermal motion dominated regime,  we infer an 84\% upper limit for the non-thermal $b$ value.
Similarly, in the non-thermal motion dominated regime, we present an 84\% upper limit on the gas temperature.


\section{Results}
\label{sec:results}
The analysis described in Section \ref{sec:analysis} results in measurements of  density, temperature, and internal non-thermal motion for 20 spectrally-resolved absorption components identified near nine galaxies or galaxy groups at $z\approx 1$ in CUBSz1.  Combining CUBSz1 and available literature samples enables a detailed investigation of whether and how these physical quantities are correlated with each other and how they evolve over cosmic time.
In this section, we focus our attention on the thermal properties of the gas, the energy partition between thermal and internal turbulent motions within cool clouds in the CGM, and the pressure balance between different phases.

\subsection{Scaling relations between density, temperature, and non- thermal motion}
\label{sec:scaling}

As summarized in Table~\ref{tab:sample}, the derived CGM density spans more than three decades from $\log\,(n_{\rm H}/\cmjj)\!\approx\!-4$ to $-1$ in all systems included in the joint CUBSz1 and literature samples.
However, different samples show different median values of the gas density.
The CUBSz1 sample at $z\!\approx\!1$ has a median of $\log\,(n_{\rm H}/\cmjj)\!\approx\!-3.3$ with an intrinsic scatter of $0.9$ dex, while
the literature samples at lower redshifts ($0.2\!\lesssim\!z\!\lesssim\!0.6$) exhibit a consistent median value of $\log\,(n_{\rm H}/\cmjj)\!\approx\!-2.4$ with a scatter of $0.8$ dex (i.e., CUBS III, CUBS IV, and COS LRG; \citealt{Zahedy:2019aa, Zahedy:2021aa, Cooper:2021aa}).
The significance of the gas density difference is 3.6\,$\sigma$ (Kolmogorov–Smirnov test; KS $p=2\!\times\!10^{-4}$) between the $z\!\approx\!1.0$ and $0.2\lesssim\!z\!\lesssim 0.6$ samples.

There are two possible reasons for the difference in median gas density.
First, it may be due to a physical difference in the properties of the CGM, with the literature samples at $z\approx0.2-0.6$ driven by high-density systems found in Lyman limit systems and/or in massive haloes hosting elliptical galaxies, whereas CUBSz1 contains absorbers originating primarily in lower-mass star-forming haloes.
Alternatively, differences in the coverage of different ionization species between the two epochs may affect the identification of multiphase gas. In particular, \ion{O}{IV} is a prominent tracer of low-density photoionized gas, but it is not covered by {\it HST} COS at $z \lesssim 0.5$.
In the absence of constraints for \ion{O}{IV}, it is significantly more challenging to detect the low-density phase, resulting in a density distribution skewed to a higher median value at low redshift $z\lesssim 0.5$.

\begin{figure*}
\begin{center}
\includegraphics[width=0.49\textwidth]{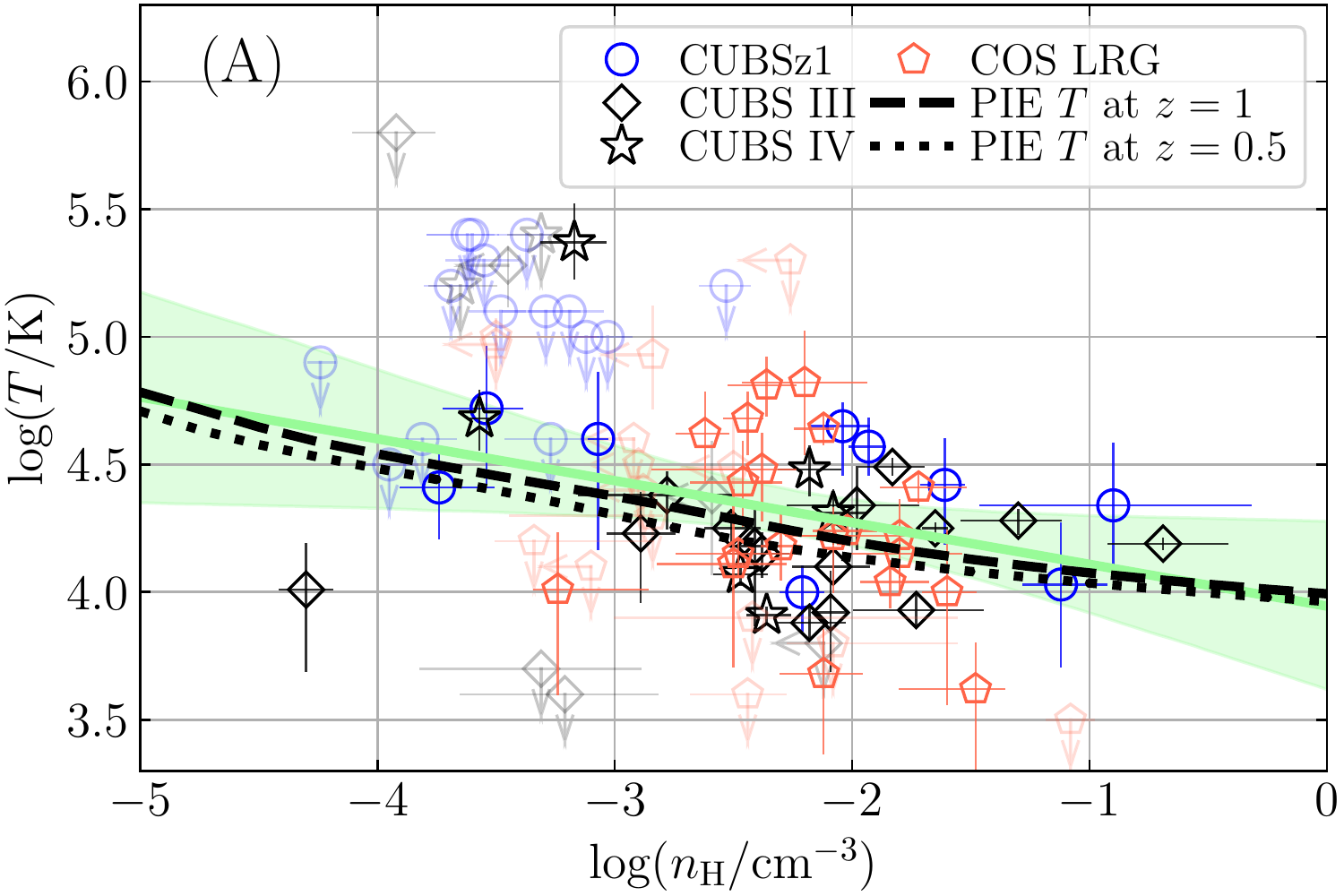}
\includegraphics[width=0.49\textwidth]{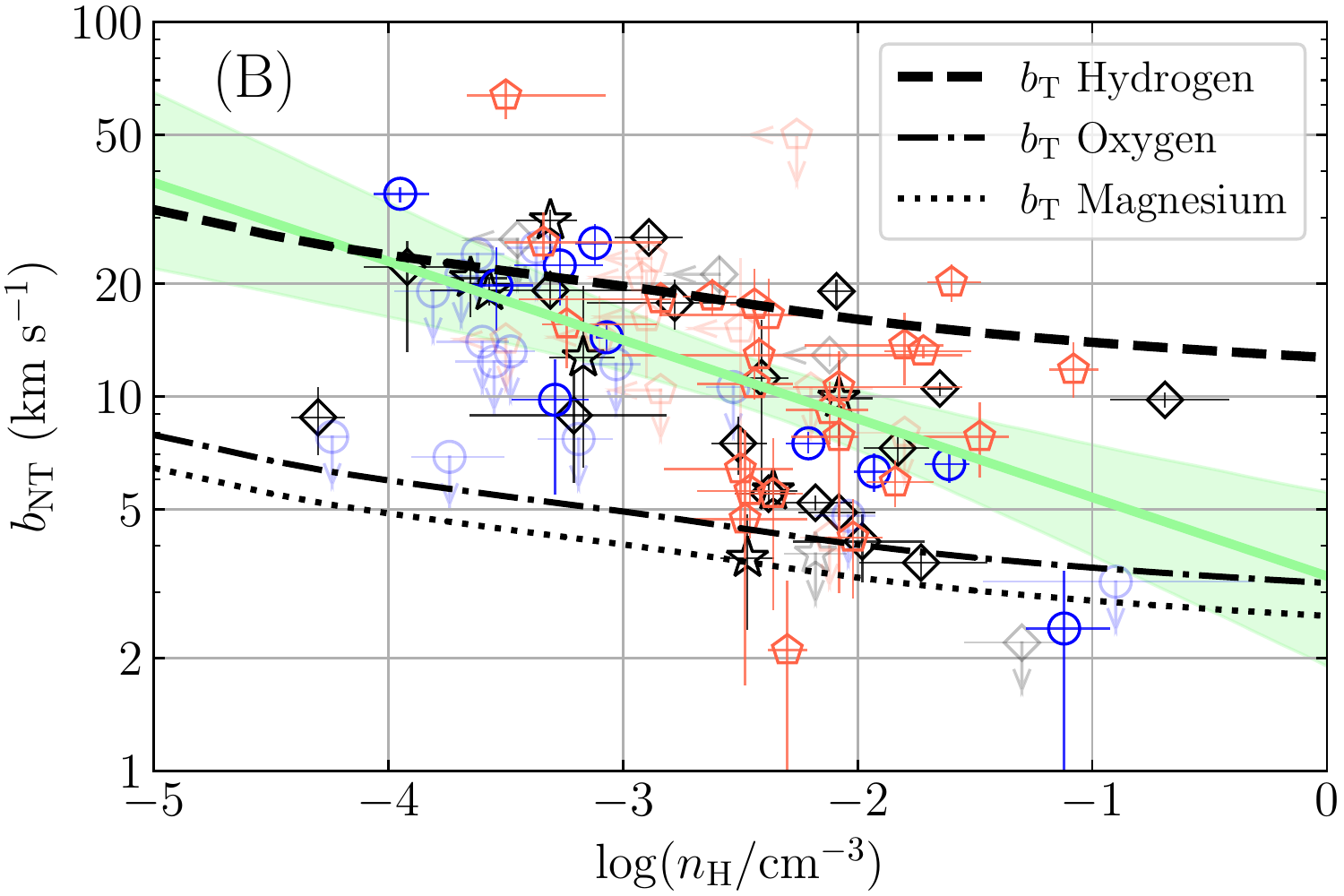}
\vskip 0.2cm
\includegraphics[width=0.49\textwidth]{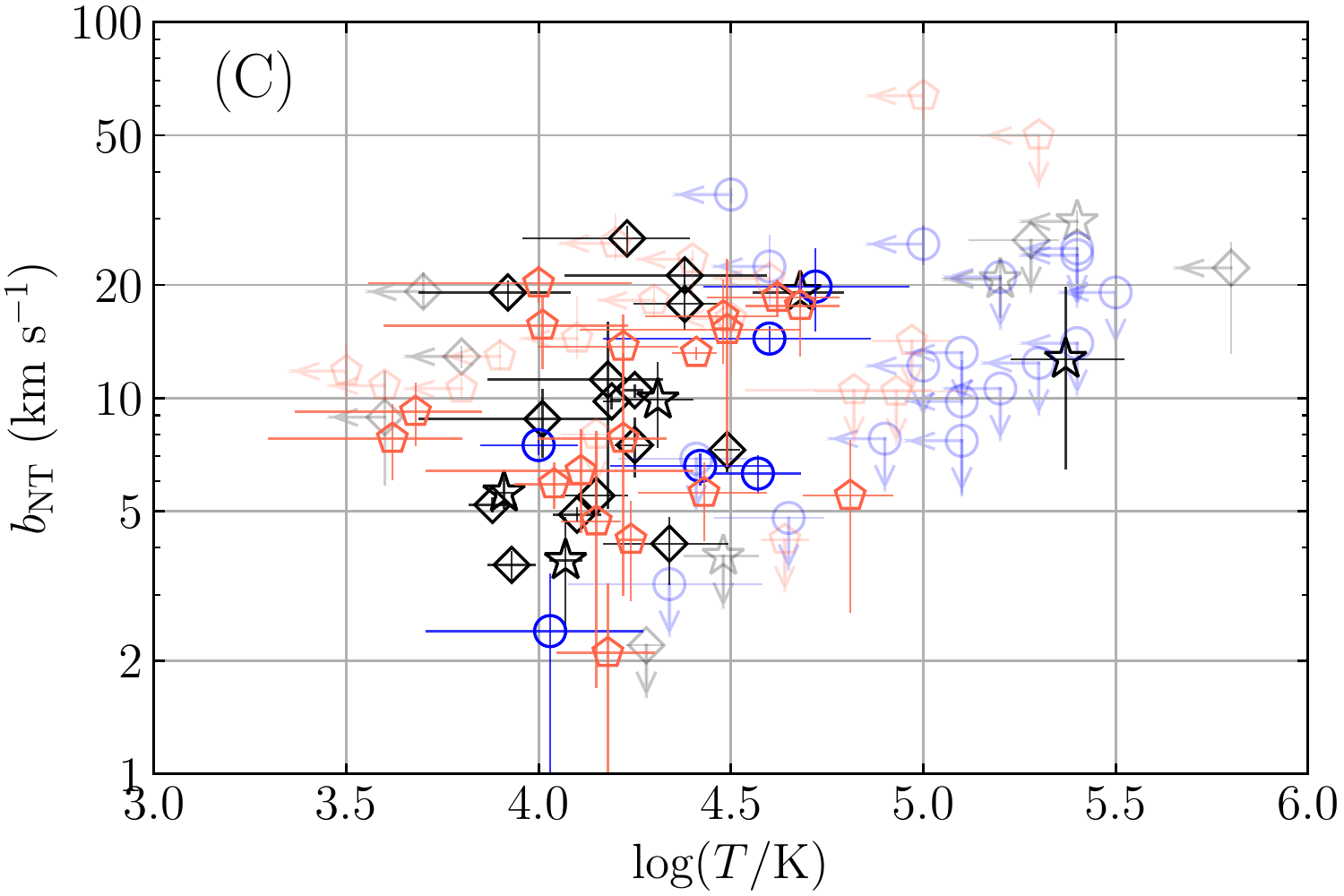}
\includegraphics[width=0.49\textwidth]{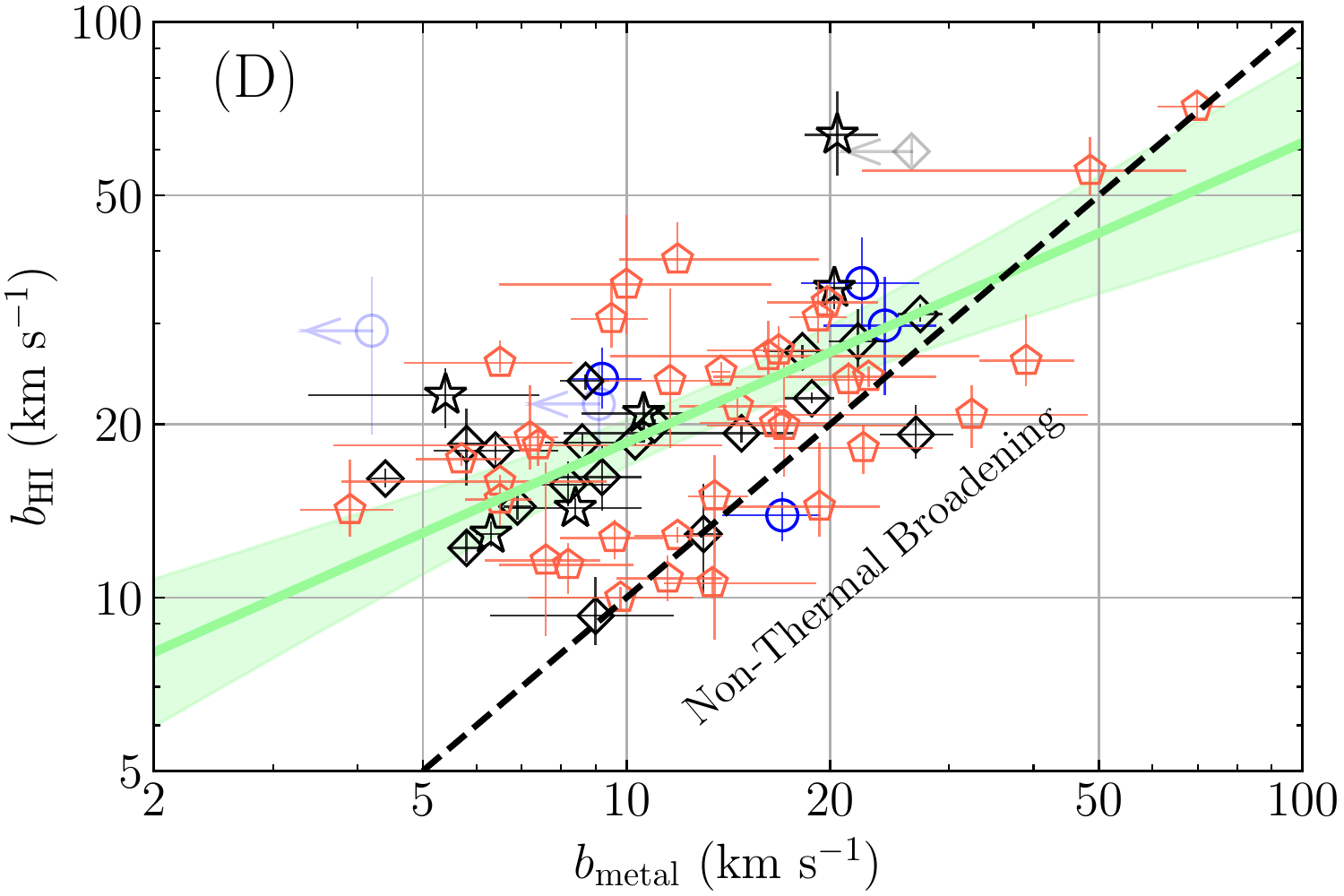}
\end{center}
 \vspace{-0.3cm}
\caption{Summary of the thermal and non-thermal motion properties of the photo-ionized CGM for the joint CUBSz1 and literature samples.  
Panels show: (A) the density $n_{\rm H}$ and temperature $T$ phase diagram; (B) the distributions between $n_{\rm H}$ and non-thermal line width $b_{\rm NT}$; (C) between $T$ and $b_{\rm NT}$; and (D) between \ion{H}{I} line width $b_{\rm HI}$, and the mean line widths of metal transitions, respectively.  In all panels, absorption components with only upper limits available on temperature or density are indicated by downward or leftward arrows.  These are greyed out for clarity.  
We find that the $n_{\rm H}$-$T$ phase diagram is best described by a shallow power-law of $\log\,(T/{\rm K})\!=\!(3.92\pm0.17)\!-\!(0.16\pm 0.07)\,\log\,(n_{\rm H}/\cmjj)$ (green solid curvey with shaded area showing 95\% confidence interval), which agrees well with the expectations of PIE models.  The PIE models are calculated using the HM05 UVB for sub-solar gas metallicity ($< 0.5\,Z_\odot$) at two different redshifts, $z=0.5$ (dotted line) and $z=1.0$ (dashed line).  
In addition, a weak anti-correlation is found between $b_{\rm NT}$ and $n_{\rm H}$,
with a best-fit power-law model of $\log\,(b_{\rm NT}/{\rm km~s^{-1}})\!=\!(0.52\pm0.11)\!-\!(0.21\pm0.04)\,\log\,(n_{\rm H}/\cmjj)$.  
At the same time, no correlation is found between $T$ and $b_{\rm NT}$  with a Pearson correlation coefficient of $r\!=\!0.19$.  Finally, $b_{\rm HI}$ is found to be weakly correlated with $b_{\rm metal}$, with $ \log\,(b_{\rm HI}/{\rm km~s^{-1}})\!=\!(0.75\pm 0.09)\!+\!(0.52\pm0.08)\log\,(b_{\rm metal}/{\rm km~s^{-1}})$, and non-thermal motions dominate the observed line width for broad components.  In the absence of \ion{H}{I} coverage, this weak correlation enables an estimate of $b_{\rm HI}$ (and therefore total gas pressure; see the discussion in Section \ref{sec:mpc} below) based on the observed $b_{\rm metal}$.
}
\label{fig:T_bNT}
\end{figure*}

At the same time, the decomposed gas temperatures from the observed line widths show that the photoionized gas is typically cool even for the low-density phase, with a narrow range of temperature characterized by a median of $\log (T/{\rm K}) \approx 4.3$ and a scatter of 0.3 dex.
Figure \ref{fig:T_bNT} summarizes the thermal and non-thermal motion properties of photo-ionized CGM for the joint CUBSz1 and literature samples.
In Panel (A), we show a weak anti-correlation (generalized Kendall correlation coefficient $\tau=-0.08$ including upper limits of the temperature) between $T$ inferred from the line width analysis and $n_{\rm H}$ returned by the photoionization modeling, which can be constrained by a power-law function, $T\propto n_{\rm H}^A$ with a slope of $A=-0.16 \pm 0.07$.
The best-fit power-law model, $\log\,T/{\rm K} = A\log\,n_{\rm H}/{\rm cm}^{-3} +B$, was determined under a Bayesian framework in order to account for uncertainties in both densities and temperatures.  The likelihood of obtaining a set of measurements ($T_i$, $n_{{\rm H}_i}$)
is calculated following
\begin{eqnarray}
    p(A, B) = &\Pi_i\int\frac{1}{
    (\sigma^2_{x, i}+\sigma^2_{\rm p})^{1/2}
    (\sigma^2_{y, i}+\sigma^2_{\rm p})^{1/2}} \times \notag \\
    & {\rm exp} \left( -\frac{(t-x_i)^2}{2(\sigma^2_{x, i} +\sigma^2_{\rm p})} - 
    \frac{(At+B-y_i)^2}{2(\sigma^2_{y, i}
    +\sigma^2_{\rm p})}\right) {\rm d} t,
\end{eqnarray}
where $x_i=\log\,n_{{\rm H}_i}/{\rm cm}^{-3}$, $y_i=\log\,T_i/{\rm K}$, and $\sigma_{\rm p}$ represents the intrinsic scatter of the sample on the logarithm scale.
Upper limits are incorporated into the likelihood calculation using a one-sided Gaussian probability function.
As a comparison, we calculate the PIE temperatures with the HM05 UVB at $z=0.5$ and $z=1.0$, and a subsolar metallicity ($Z<0.5Z_\odot$) for the cool CGM.  Because the dependence of PIE temperature on the gas metallicity is minimal in the sub-solar regime that is typical for the CGM, the large scatter observed in the data may indicate fluctuations in the local radiation field.
This empirical relationship is consistent with the expectation from the PIE assumption in our photoionization modeling (Section \ref{sec:analysis}).

The line width analysis reveals a typical internal non-thermal width of cool CGM absorbers of $b_{\rm NT} = 12\kms$ with a scatter of $10 \kms$.
As shown in Figure \ref{fig:T_bNT}\,B, $b_{\rm NT}$ is weakly anti-correlated with $n_{\rm H}$ with a power-law slope of $-0.21\pm0.04$ and generalized Kendall correlation coefficient $\tau=-0.28$.
We compare this to the expectation from thermal broadening of different elements at the PIE temperature (dashed, dash-dotted, and dotted curves for hydrogen, oxygen, and magnesium, respectively).
The $b_{\rm NT}$ value greatly exceeds the curves for oxygen and magnesium, which demonstrates that internal non-thermal motions dominate the line widths of metal lines at all densities, but only becomes significant for hydrogen absorption lines at low densities.  In contrast, no correlation is found between $T$ and $b_{\rm NT}$ with a Generalized Kendall correlation coefficient of $\tau=0.04$.
As discussed in Section \ref{sec:T_NT} below, the majority of the cool CGM is subsonic, while a small fraction of the components are subject to supersonic motions (Figure \ref{fig:T_bNT}\,C).

In addition, a modest correlation is observed between $b_{\rm metal}$ and $b_{\rm H{\small I}}$ (Figure \ref{fig:T_bNT}\,D) with a best-fit power-law slope of $0.51\pm0.08$ and an intrinsic scatter of $0.12\pm0.02$ in $\log\,(b_{\rm HI}/\kms)$ (Kendall $\tau =0.30$).
In the absence of $b_{\rm H{\small I}}$, this modest $b_{\rm H{\small I}}$-$b_{\rm metal}$ correlation provides an effective tool for inferring $b_{\rm H{\small I}}$ from $b_{\rm metal}$ with an uncertainty of $30\%$ ($1\sigma$).  
This is particularly useful when the broad \ion{H}{I} features from the low-density phase are obscured by the stronger \ion{H}{I} lines from the high-density phase in multiphase components (Section \ref{sec:mpc}).
For these components, we use $b_{\rm metal}$ to infer $b_{\rm HI, broad}$, which is in turn a tracer of the total gas pressure following $P\propto b_{\rm H {\small I}}^2$ (adopted in Section \ref{sec:mpc}).

Given the large dynamic range in $n_{\rm H}$, we also investigate whether or not including additional scaling with $n_{\rm H}$ would reduce the scatter in the mean $b_{\rm H {\small I}}$-$b_{\rm metal}$ correlation.
Applying a plane fit to $b_{\rm H {\small I}} (n_{\rm H}, b_{\rm metal})$, we find $ \log\,(b_{\rm HI}/{\rm km~s^{-1}}) = (0.73\pm 0.07) + (0.01\pm0.02) \log\,(n_{\rm H}/{\rm cm^{-3}}) + (0.56\pm 0.08)\log\,(b_{\rm metal}/{\rm km~s^{-1}})$, demonstrating that $n_{\rm H}$ plays a negligible role in shaping the $b_{\rm H{\small I}}$-$b_{\rm metal}$ relation.
We therefore conclude that the best-fit $b_{\rm H {\small I}}$-$b_{\rm metal}$ correlation in Figure \ref{fig:T_bNT} is applicable for absorption features originating in a broad range of gas density.  

\subsection{Energy partition between thermal and non-thermal motions in the cool, photoionized CGM}
\label{sec:T_NT}

The exercise presented in Section \ref{sec:scaling} now enables a more detailed look at the energy partition between thermal and internal non-thermal motions in cool photoionized clouds in the CGM.
The thermal and non-thermal energy densities are calculated based on a combination of $n_{\rm H}$, $T$, and $b_{\rm NT}$. 
Observations of hydrogen lines are of particular importance, because of the large mass ratio between hydrogen and metal ions.
Recall that {\it HST}/COS spectra provide a spectral resolving power of FWHM $\approx\!20 \,\kms$, corresponding to a limiting $b$ value of $12 \kms $, while MIKE spectra offer FWHM $\approx\!8 \kms$ with a limiting $b$ value of $\approx 5 \kms $.  At $T\gtrsim 10^4$ K, where the majority of the components are found (see Figure \ref{fig:T_bNT}\,A), the expected thermal line width for hydrogen is $\gtrsim\!12\,\kms$, well resolved by COS.  On the other hand, the anticipated line widths for metal lines are scaled down according to $1/\sqrt{m_I}$. Combining hydrogen and metal lines, when line width constraints from MIKE are not available, therefore provides the largest discriminating power for resolving thermal and non-thermal contributions\footnote{We have tested the robustness of the thermal and non-thermal decomposition using hydrogen and UV metal lines from COS spectra alone.  Using components with \ion{Mg}{II} detected in the MIKE spectra, we experiment with computing $b_{\rm NT}$ and $T$ by including or excluding \ion{Mg}{II} and \ion{Fe}{II} lines.  We find that excluding the line width measurements from MIKE would lead to a slight increase in the estimated $b_{\rm NT}$ (by $\approx 2\,\kms$) and a corresponding decrease in $T$, but the values remain consistent to within the uncertainties. }.
Without \ion{H}{I}, it becomes challenging to distinguish between thermal and non-thermal contributions to the observed line widths.
For this exercise, we only consider those 67 components with detected \ion{H}{I}.

\begin{figure*}
\begin{center}
~
\includegraphics[width=0.485\textwidth]{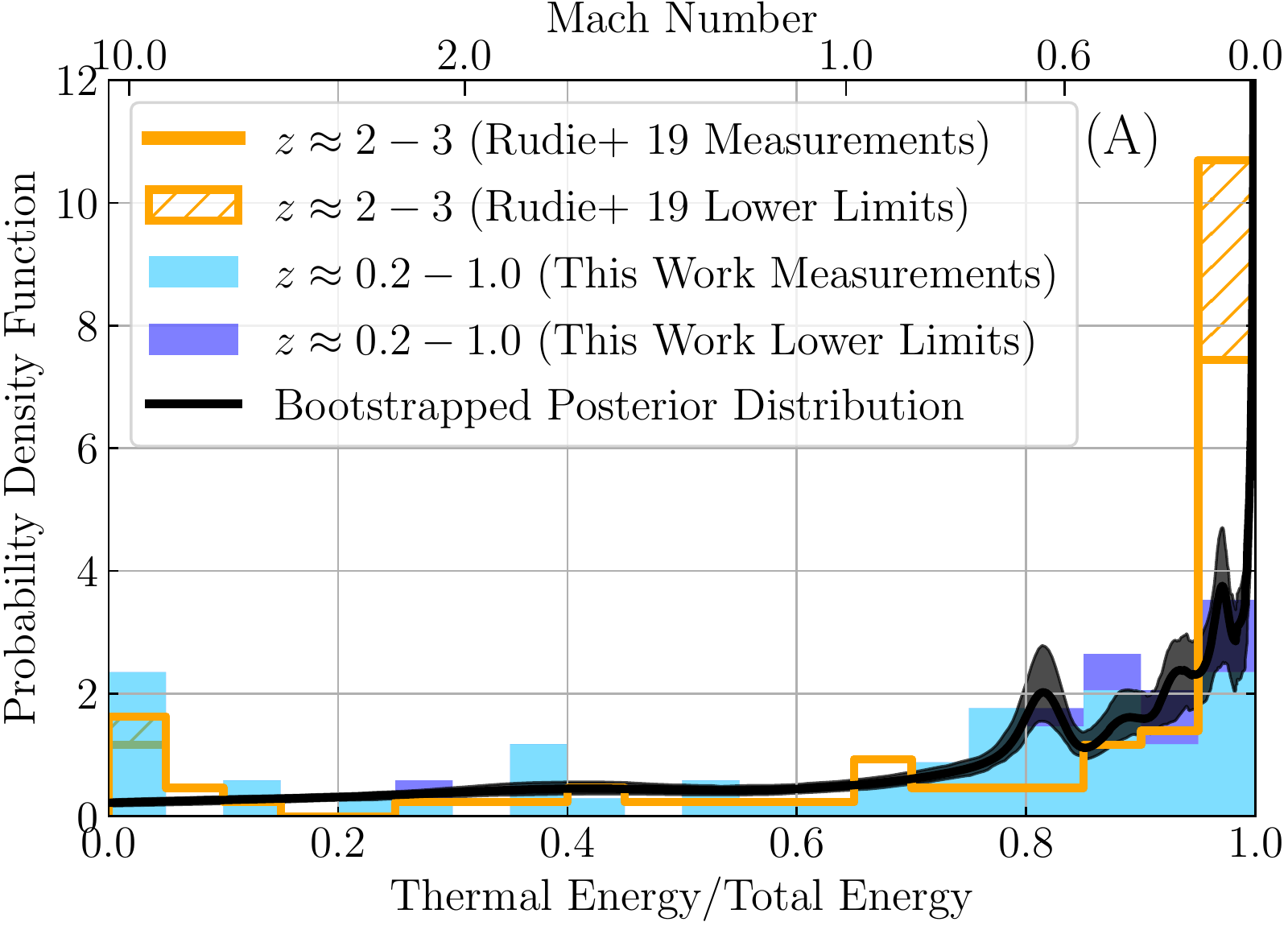}
\includegraphics[width=0.485\textwidth]{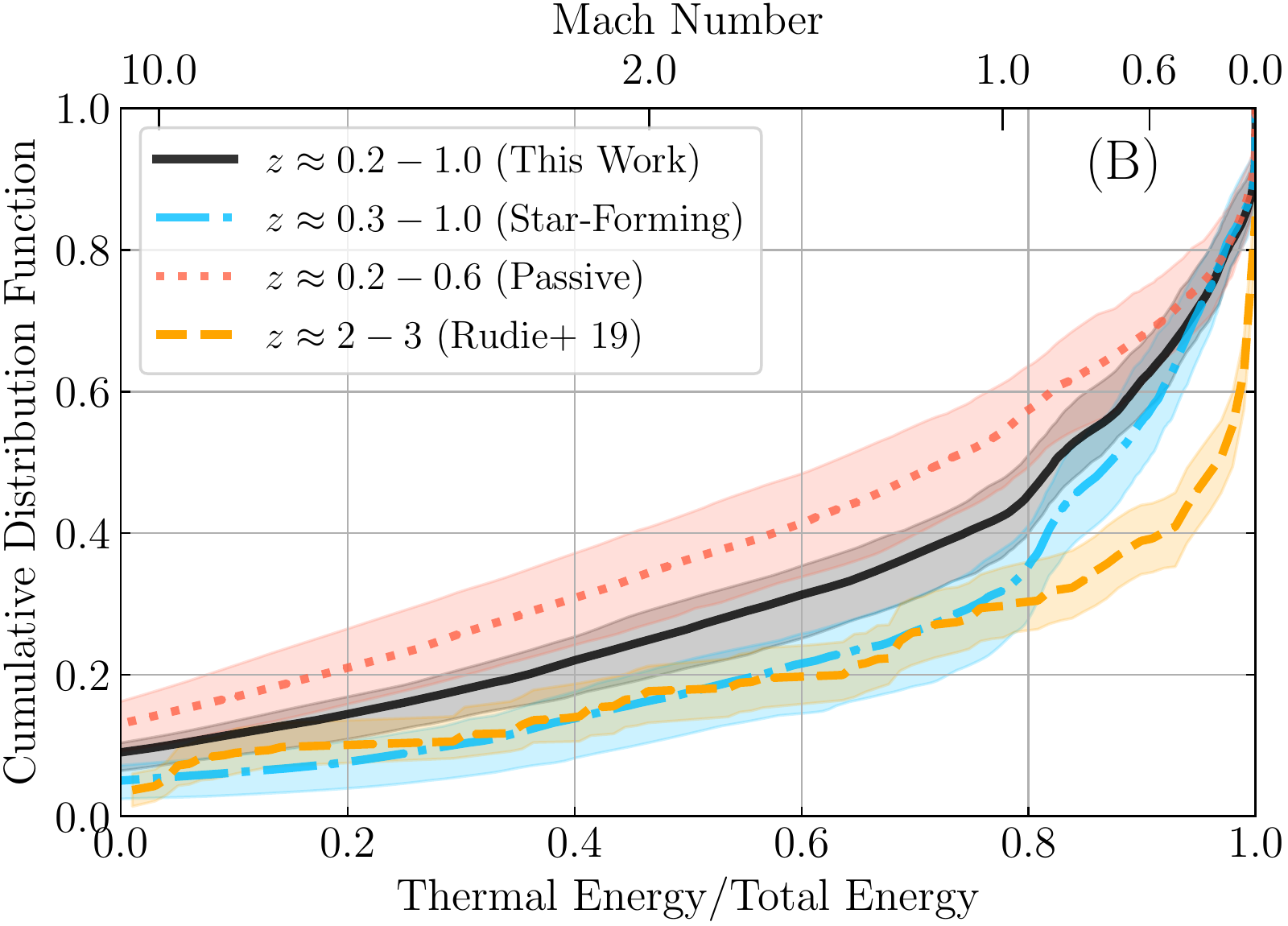}
\vskip 0.2cm
\includegraphics[width=0.49\textwidth]{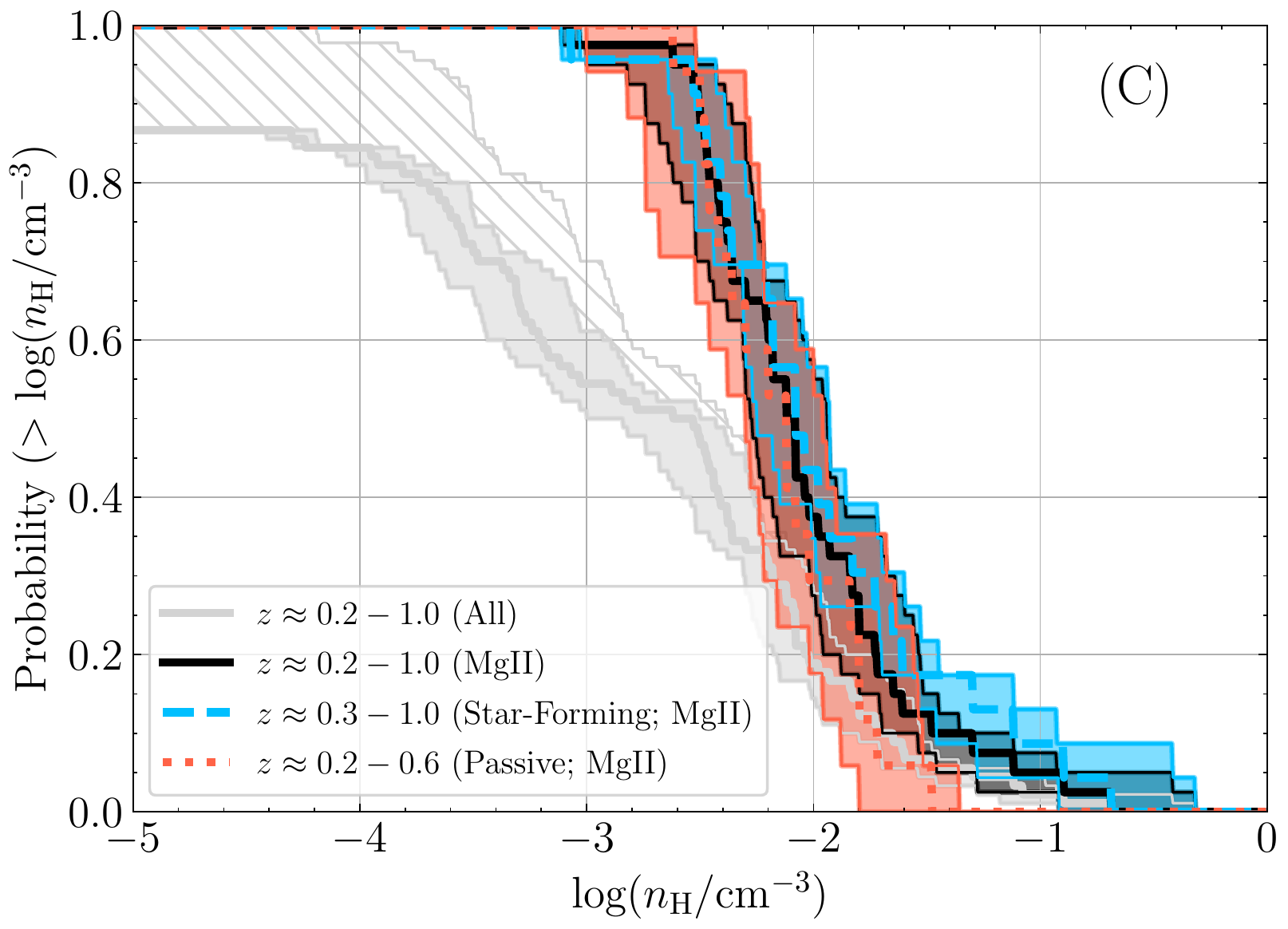}
\includegraphics[width=0.49\textwidth]{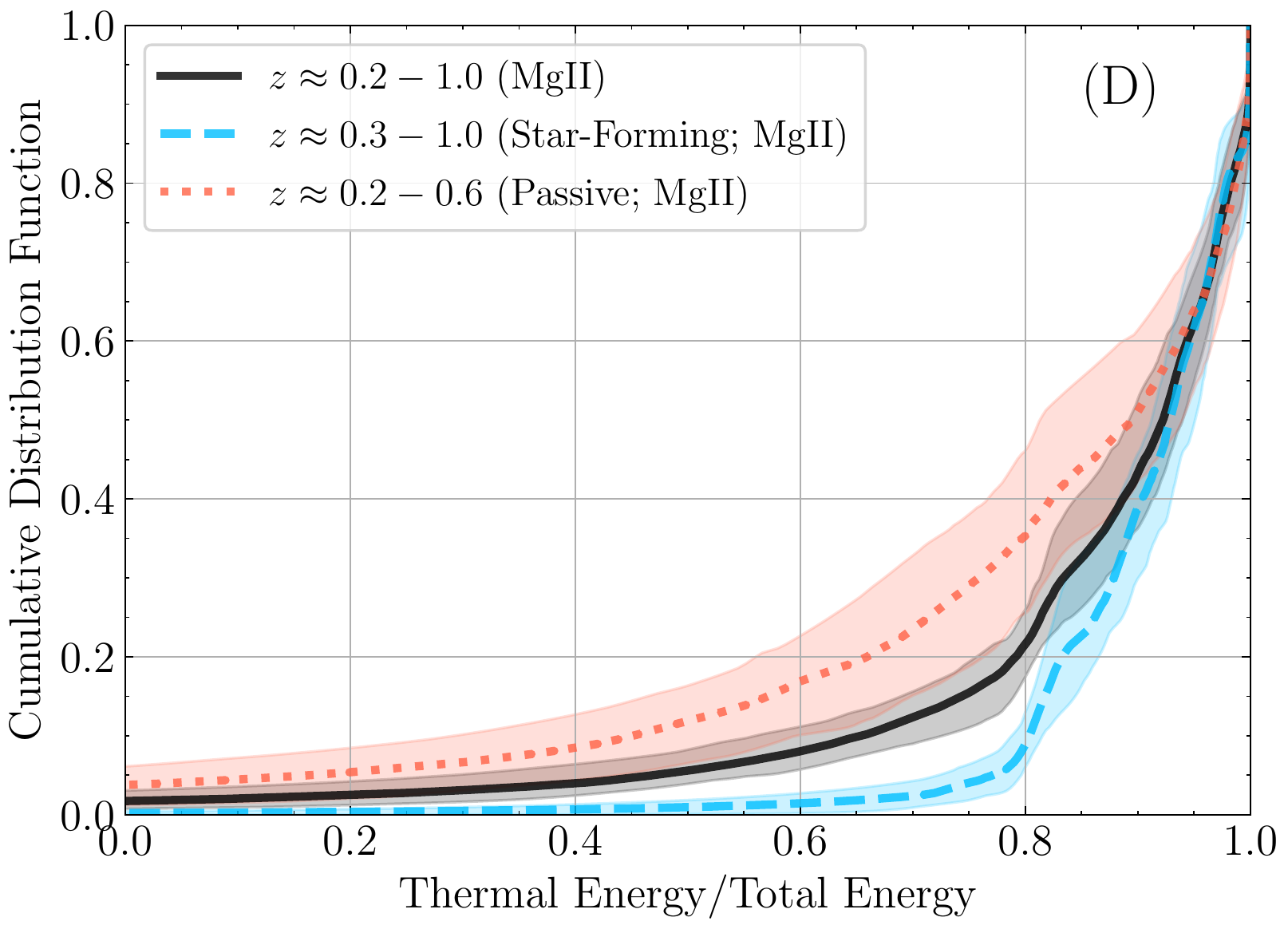}
\end{center}
 \vspace{-0.3cm}
\caption{Comparisons of the thermodynamic state of the cool CGM between different samples.  Panel (A) shows the observed distributions of thermal energy fraction (see Equation \ref{eq:efrac}), which translates directly to the Mach number (Equation \ref{eq:mach}) of the system displayed at the top x-axis of the panel.
Histograms display the distribution function constructed using the best-fit $T$ and $b_{\rm NT}$, while the black shaded region displays the mean posterior distribution of $E_{\rm T}/E_{\rm tot}$ and the 68\% confidence intervals extracted from bootstrap. 
Panel (B) shows the cumulative distributions of the thermal energy ratio with the shaded region representing 1-$\sigma$ uncertainties. 
At low redshift, the cool CGM of passive galaxies (the red dotted line) exhibits more power in turbulence than the CGM around star-forming galaxies (the light blue, dash-dotted line).  A significant fraction of the total gas energy lies in supersonic motions with Mach number ${\cal M}>1$ around passive, evolved galaxies.
There is also a weak (2.9 $\sigma$) difference in the CGM between low- and high-redshift star-forming galaxies at large thermal energy ratios $E_{\rm T}/E_{\rm tot} > 0.8$.
The $z\lesssim1$ sample is further divided into subsamples of different densities to investigate whether the distinction between passive and star-forming haloes arises as a result of the underlying differences in gas density.
Panel (C) displays the cumulative distributions of gas densities for the full $z\lesssim1$ sample in grey (i.e., 67 components with detected \ion{H}{I}).  The hatched region includes upper limits in density estimates.  The black curve shows the distribution of only components with \ion{Mg}{II} detected (35 components with both detected \ion{H}{I} and \ion{Mg}{II}), which pre-selects high density clumps.  The red and cyan bands show these \ion{Mg}{II}-selected  high-density clumps associated with passive (17 components) and star-forming galaxies (18 components), respectively.  The cumulative distributions of $E_{\rm T}/E_{\rm tot}$ for the \ion{Mg}{II}-selected components are shown in Panel (D),  confirming that the distinction between passive and star-forming haloes exists, independently of the underlying density of the cool gas.
}
\label{fig:Tratio}
\end{figure*}

We calculate the thermal and non-thermal energies, $E_{\rm T}$ and $E_{\rm NT}$, respectively, following
\begin{eqnarray}
E_{\rm T} &=& \frac{3}{2}k_{\rm B} T = \frac{3}{4} m_{\rm H} b^2_{\rm HI, T}\notag \\
E_{\rm NT} &=& \frac{3}{2} \mu m_{\rm H} v^2_{\rm NT} = \frac{3}{4} \mu m_{\rm H} b^2_{\rm NT},
\end{eqnarray}
where we have adopted the Boltzmann constant $k_{\rm B}$ and a mean molecular weight of $\mu = 0.6$ appropriate for ionized gas, and applied $b=\sqrt{2}\,\sigma_v$ with $\sigma_v$ representing the 1D velocity dispersion.
Then we calculate the total internal energy by summing $E_{\rm T}$ and $E_{\rm NT}$ (ignoring cosmic-ray and magnetic energy)
\begin{eqnarray}
E_{\rm tot} &=& \frac{3}{4} m_{\rm H} b^2_{\rm HI, T} + \frac{3}{4} \mu m_{\rm H} b^2_{\rm NT} \notag \\
    &=& \frac{3}{4} m_{\rm H} [b^2_{\rm HI} - (1-\mu) b^2_{\rm NT}],
\end{eqnarray}
which leads to a thermal-to-total energy ratio of
\begin{eqnarray}
\label{eq:efrac}
    \frac{E_{\rm T}}{E_{\rm tot}} = \frac{b^2_{\rm HI}-b^2_{\rm NT}}{b^2_{\rm HI} - (1-\mu) b^2_{\rm NT}}.
\end{eqnarray}

The expression for the thermal energy fraction in Equation \ref{eq:efrac} involves one observed quantity $b_{\rm HI}$ and one derived parameter $b_{\rm NT}$, providing the smallest propagated uncertainty.
Alternatively, following a similar derivation in \citet{Rudie:2019aa}, the thermal energy fraction can also be expressed as a function of the Mach number (${\cal M}{\equiv} v_{\rm NT}/c_{\rm s}$ where $c_s$ is the sound speed of the cool gas $c_{\rm s}^2 = \gamma k_{\rm B}T/ \mu m_{\rm H}$) and the polytropic index $\gamma$ following
\begin{eqnarray} \label{eq:mach}
    \frac{E_{\rm T}}{E_{\rm tot}} = \frac{1}{1+\frac{\gamma(\gamma-1)}{2} {\cal M}^2}, 
\end{eqnarray}
where $\gamma=5/3$ is adopted for an adiabatic monatomic gas.

Figure \ref{fig:Tratio} shows the distributions of $E_{\rm T}/E_{\rm tot}$ over the full range from 0 to 1 for all available samples at $z\lesssim 1$ (the light blue histogram) and $z=2-3$ (orange open histogram), respectively.  Including components with only upper limits available, the $E_{\rm T}/E_{\rm tot}$ distributions for low- and high-redshift samples are shown in dark blue and orange hatched histograms, respectively. 
To account for uncertainties, we also show the coadded posterior distribution of $E_{\rm T}/E_{\rm tot}$ in black with the band width indicating the 68\% confidence interval determined from a bootstrap routine.

At high redshift $z=2$\,-\,3, \citet{Rudie:2019aa} investigate the thermal energy ratio of the CGM of star-forming galaxies.
The high-redshift sample shows a median thermal energy ratio of $0.97_{-0.03}^{+0.01}$, or
up to 20\% of the gas (i.e., absorption components) may be driven by non-thermal motion (see also \citealt{Rauch1996}, \citealt{Simcoe:2006aa}, and \citealt{Kim2016}).
This indicates that the thermal energy is the dominant contributor to the internal energy of cool clouds in the high-$z$ CGM.
In contrast, a median of $E_{\rm T}/E_{\rm tot}$ of $0.82_{-0.03}^{+0.04}$ is found for the $z\lesssim 1$ sample, or
$\approx 30$\% of the gas may be dominated by energy of internal non-thermal motions. 
A 2-sample KS test returns a $p$-value of $1.4\times10^{-5}$, indicating a $4.2 \sigma$ difference in the thermal energy ratio between the two epochs.
A major contributor to the difference in the thermal energy ratio between the two samples is the increase in turbulence power from a median value of $b_{\rm NT}\approx 5.8\kms$ in the $z=2$-3 sample to $b_{\rm NT}\approx 12.9\kms$ in the $z\lesssim 1$ sample.
Note however that while the $z=2$-3 and $z\lesssim 1$ samples share a similar median temperature of $\log\,(T/{\rm K}) \approx 4.3$-4.4, the high-redshift sample exhibits a significantly larger scatter in gas temperature of 0.7 dex (in comparison to 0.3 dex for the $z\lesssim1$ sample).
Attributing the large scatter to an intrinsic difference in the temperature distribution would lead to an increase in the mean gas temperature in the high redshift sample (i.e., $10^{\sigma/2}$ times larger than the median assuming a log-normal distribution).
This difference of the intrinsic scatter also contributes to the difference of the thermal energy ratio between the low- and high-redshift samples.

Figure \ref{fig:Tratio} shows that non-thermal broadening is significant in the observed cool CGM at $z\lesssim 1$.  The $z\lesssim 1$ sample is further divided into two sub-samples based on the host galaxy type: star-forming galaxies (CUBS III, CUBS IV, and the CUBSz1 sample; 33 absorption systems), and passive galaxies (COS LRG; 35 absorption systems).  
Figure \ref{fig:Tratio}\,B shows a positive correlation between the significance of non-thermal processes and a lack of star formation in the host galaxies.  A significant fraction ($> 50\%$) of the total energy lies in supersonic motions with Mach number ${\cal M}>1$ in passive haloes, compared to $<30\%$ in star-forming galaxy haloes.  For star-forming galaxies, we also found a modest difference between $z\approx 0.3-1.0$ and $z\approx 2-3$, at a 2.9-$\sigma$ level based on a 2-sample KS test.  This difference is most apparent for components with high thermal energy ratio $E_{\rm T}/E_{\rm tot} > 0.8$.

The distinction between passive and star-forming haloes remains when considering only high-density components based on the presence of \ion{Mg}{II}.
Figure \ref{fig:Tratio}\,C shows that \ion{Mg}{II}-selected components originate in gas with densities of $\log\,n_{\rm H}/{\rm cm}^{-3}\approx -3$ or higher. The red and cyan bands show these \ion{Mg}{II}-selected  high-density clumps associated with passive and star-forming galaxies, respectively.
The cumulative distributions of $E_{\rm T}/E_{\rm tot}$ for the \ion{Mg}{II}-selected components in Figure \ref{fig:Tratio}\,D consistently show a higher power in non-thermal energy in passive haloes, confirming that independent of cool gas density passive haloes appear to show a larger non-thermal line width than star-forming ones.

\subsection{The prevalence of multiphase gas at $\bm{d<100}$ kpc}
\label{sec:everywhere}

Previous studies have shown that the CGM is multiphase, including cold molecular gas, cool photoionized gas, warm-hot collisionally-ionized gas, and hot X-ray/$\gamma$-ray emitting plasma \citep[e.g.,][]{Savage:2014aa, Werk:2016aa, Bogdan:2017aa, Li:2018aa, Das:2020aa, Boettcher:2021aa, Karwin:2021aa}.
Theoretically, the volume-filling hot medium is formed via virial shocks and galactic feedback \citep[e.g.,][]{Cen:2006aa}, while lower temperature gas is confined within this volume-filling hot medium in the forms of clouds or filaments.
In observations, the hot and cool phases are best constrained with X-ray and UV studies, respectively.
However, direct comparisons between these two phases are difficult, because they are not necessarily co-spatial given the dramatic differences in temperature and ionization mechanisms.

Recent studies have revealed that kinematically aligned absorption lines from different ions may require more than one gas density to fully explain the observed relative abundances between low- and high-ionization species under a single photoionization model, motivating the need of considering multiphase gas (Figure \ref{fig:mpc_system}; \citealt{Zahedy:2021aa, Cooper:2021aa}).
These multiphase components typically have narrow absorption features for low ionization state ions (i.e., \ion{O}{II} and \ion{S}{II}) and broad absorption features for higher ions (i.e., \ion{O}{IV} and \ion{S}{V}), and can be characterized by a two-phase model \citep{Cooper:2021aa, Sameer:2021aa, Zahedy:2021aa}.

As illustrated in Section \ref{sec:photo}, multiphase components in our analysis are found based on a simultaneous presence of low and intermediate ions  (e.g., from \ion{H}{I} to \ion{O}{IV}) with kinematically aligned absorption profiles
that cannot be modeled by a single-density photoionization model and therefore require two phases.
The densities of the two phases typically differ by a factor of $10-30$ ($\log n_{\rm H}/\cc \approx -2$ for the high-density phase compared to $\log n_{\rm H}/\cc \approx-3.5$ for the low-density phase). Whether or not the two phases are physically associated has strong implications for the hydrodynamic state of the gas (see \S\S\ 4.4 and 5.2 below).

Among the 26 galaxies/galaxy groups at $d<200$ kpc from a QSO sightline in the combined sample, including CUBSz1, CUBS III, CUBS IV, and COS LRG, only 12 have data available for investigating the multiphase nature of the gas.  Most systems in the COS LRG sample are at sufficiently low redshift ($z \lesssim 0.5$) that existing FUV spectra do not cover \ion{O}{IV} (\citealt{Zahedy:2019aa}).  While the inferred gas densities are robust for the high-density clumps, no information is available for the presence or absence of additional low-density phases.
In addition, two LLSs in CUBS III have $\log\,(N_{\rm H{\small I}}/{\rm cm^{-2}}) \gtrsim 18$ sufficient to attenuate the background QSO light below $\lambda_{\rm abs} < 912$ \AA.
Consequently, \ion{O}{IV} cannot be detected even though it is covered in available COS/FUV spectra.
To constrain the presence or absence of a low-density phase in such systems requires coverage of \ion{C}{IV} \,$\lambda \lambda\,1548, 1550$ at longer wavelengths (e.g., \citealt{Cooper:2021aa}; see also Figure \ref{fig:mpc_system}).
Excluding these systems from the total sample leaves 12 galaxies/galaxy groups for which robust constraints on the multiphase nature of CGM absorbers can be obtained.
Out of these 12, five galaxies/galaxy groups exhibit kinematically aligned absorption features in seven components that are indicative of multiphase gas, leading to a detection rate of $\approx 40$\%.
However, all five galaxies/galaxy groups with multiphase components detected occur at $d<80$ kpc from the QSO sightlines.
Therefore, restricting the distance to $d<100$ kpc leads to a still higher detection rate of 80\%, with five of six galaxies/galaxy groups showing these multiphase components.
In contrast, zero out of six galaxies/galaxy groups at $d=100$-200 kpc exhibit traces of multiphase gas.  The incidence of multiphase components can be constrained to be $<15$\% at $d>100$ kpc (a single-sided 84\% upper limit).
The observed steep decline in the incidence of multiphase components from small to large distances indicates these different phases are most likely associated with the same host galaxies, rather than originating in cosmologically distinct regions but kinematically aligned by projection \citep[e.g.,][]{Ho:2020aa}.

To further assess whether the low- and high-density phases are co-spatial, instead of being due to line-of-sight projections of inner halo high-density gas and low-density outskirts \citep[e.g.,][]{Voit:2019aa}, we estimate the occurrence rate of low-density components by considering only galaxies in CUBSz1.  As described in \S\ \ref{sec:cubsz1}, the CUBSz1 sample selects galaxies
based on their close proximity to the QSO sightline
with no prior knowledge of whether or not an absorption feature is present.  This galaxy-centric sample is necessary for measuring the incidence of absorbers in galaxy halos.
We find that $40\pm16\%$ of galaxies/galaxy groups at $d=100-200$ kpc have low-density components ($\log n_{\rm H} \lesssim -3$) detected (i.e., six detections among 15 galaxy systems at $z\approx1$; CUBS VI, in preparation).
This detection rate is a factor of two lower than the detection rate of $\approx 80\%$ for low-density components at $d<100$ kpc.  The decline is in contrast to a constant detection rate expected from a flat surface mass density profile observed for the warm, low-density CGM around low-redshift galaxies \citep[e.g.,][]{Singh:2018}.
We therefore argue that the two phases in each multiphase absorption component are most likely physically associated and that multiphase gas is common in the CGM.

Before concluding, we briefly comment on a caveat in  photoionization analyses of the CGM.
In this work and CUBS III \citep{Zahedy:2021aa}, the adopted incident field is the HM05 UVB.
Considering the density difference between the two phases in the multiphase component ($\approx 1.5$ dex), increasing the incident radiation field at $\approx 50-200$ eV by 1.5 dex from the standard value in HM05 would lead to a 1.5 dex increase in the density required to explain the observed abundances of high-ionization species such as \ion{N}{IV} and \ion{O}{IV} (see Figure \ref{fig:mpc_system}). 
The observed wide range of ionization states may be reproduced by a single phase with $\log (n_{\rm H}/{\rm cm^{-3}}) \approx -2.5$ to $-2$.
However, \citet{Upton-Sanderbeck:2018aa} investigated several sources contributing to the extreme UVB currently not in HM05, and showed that AGN are still the dominant component at $\approx 100$ eV (see also \citealt{Khaire:2019aa}).
We also consider contributions of extreme UV photons from the host galaxy following \citet{Upton-Sanderbeck:2018aa}.
To raise the incident field at $\approx100$ eV by $> 1$ dex, it would require a close distance to the host galaxy, $\lesssim 5$ kpc with an escape fraction of $1-10\%$, which is much smaller than the impact parameters of multiphase components ($\approx 50$ kpc).
Therefore, we conclude that the variation of the incident field is unlikely to explain the simultaneous presence of a wide range of ionization state species.

\subsection{Pressure balance in the multiphase CGM}
\label{sec:mpc}

A long-standing question is whether different phases in the multiphase CGM are in pressure balance, which is typically assumed in physical models \citep[e.g.,][]{Faerman:2017aa, Qu:2018aa, Voit:2019aa} and found in numerical simulations \citep[e.g.,][]{vandeVoort:2012aa, Ji:2019aa, Fielding:2020aa}.
In \S\ \ref{sec:everywhere}, we argue that the kinematically-aligned multiphase gas arises in the same host haloes.  Here we examine the pressure balance between different phases in these multiphase absorbers under the assumption that these different phases are physically cospatial.

\begin{figure*}
\begin{center}
\includegraphics[width=0.49\textwidth]{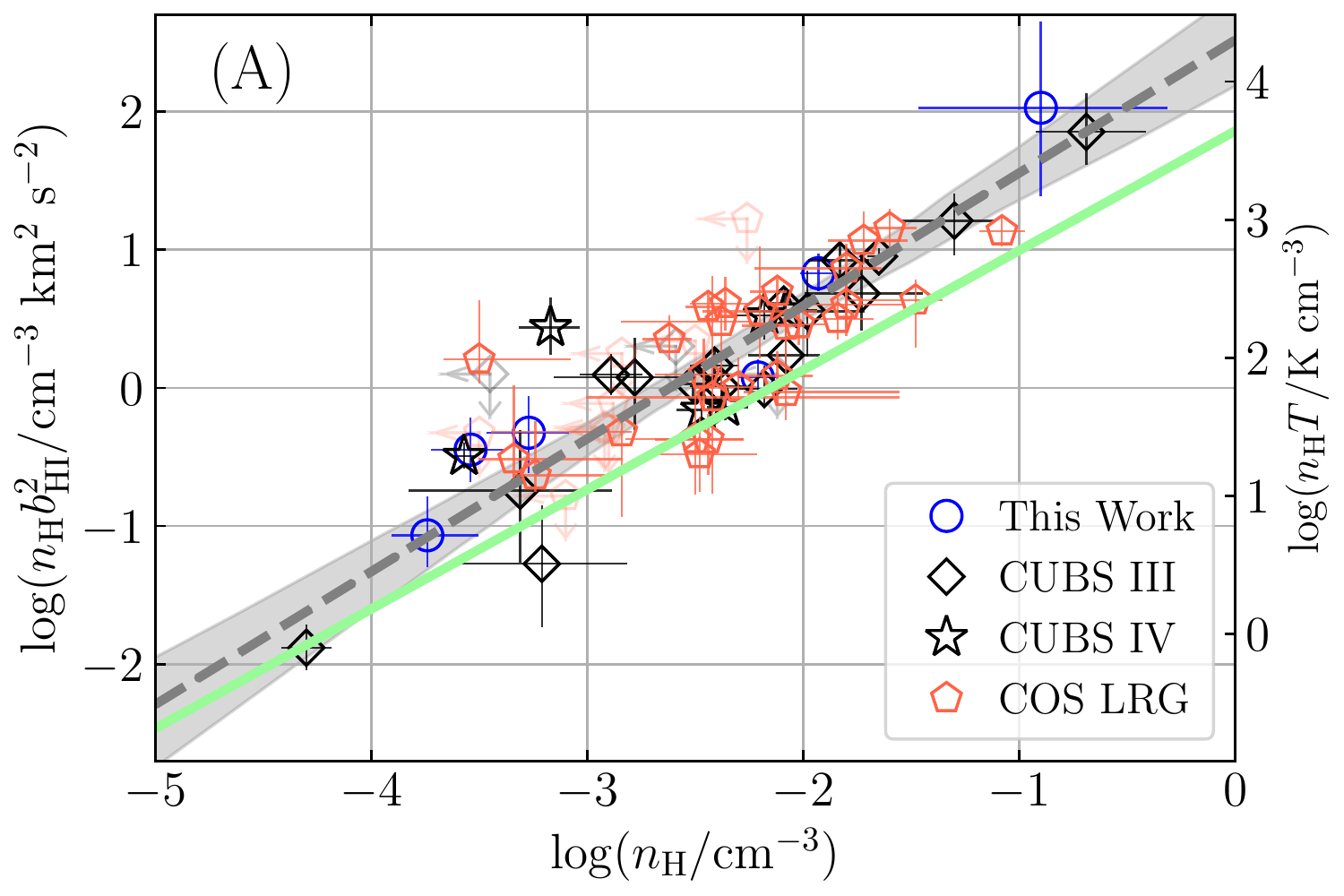}
\hskip 0.1cm
\includegraphics[width=0.49\textwidth]{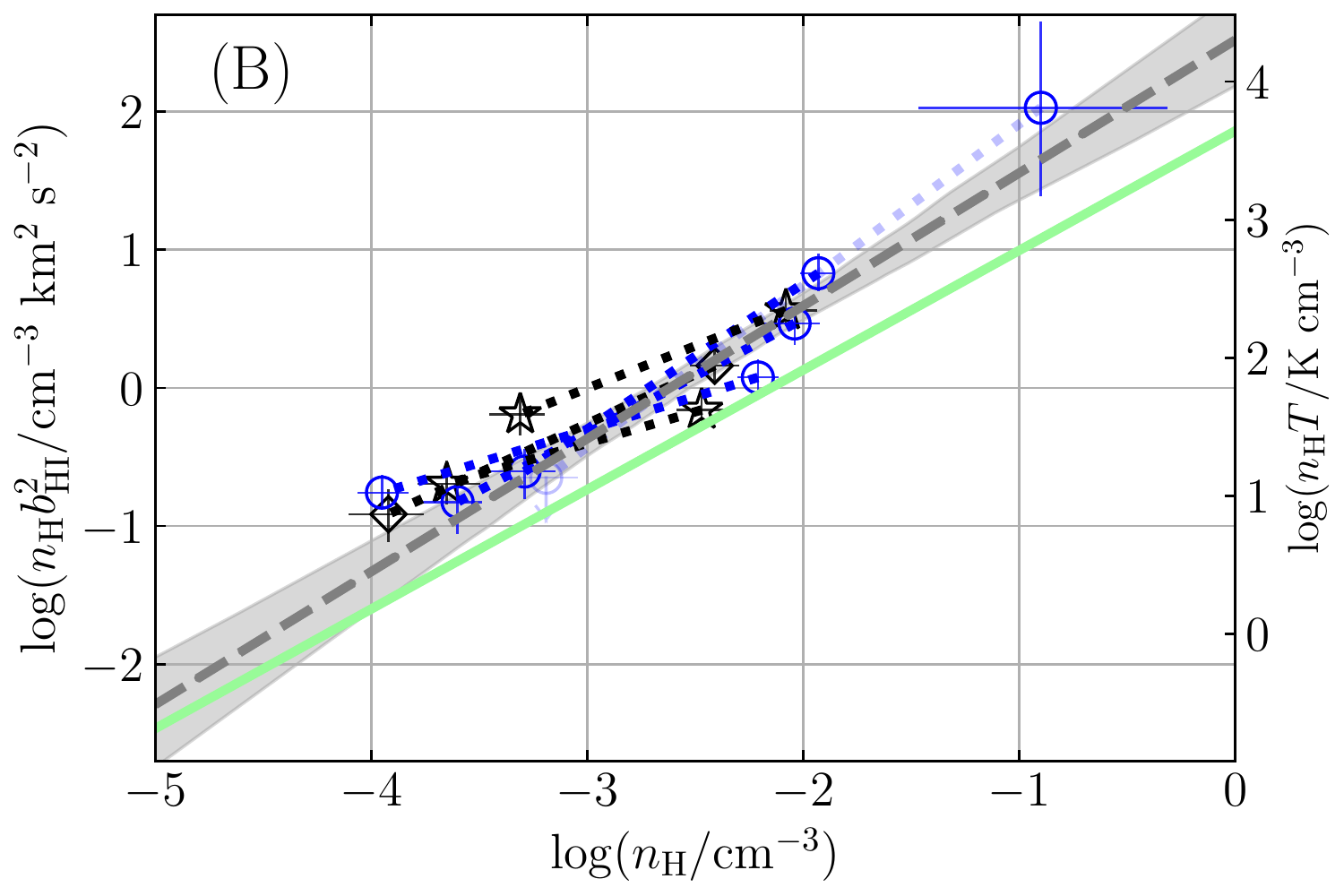}
\vskip 0.2cm
\includegraphics[width=0.49\textwidth]{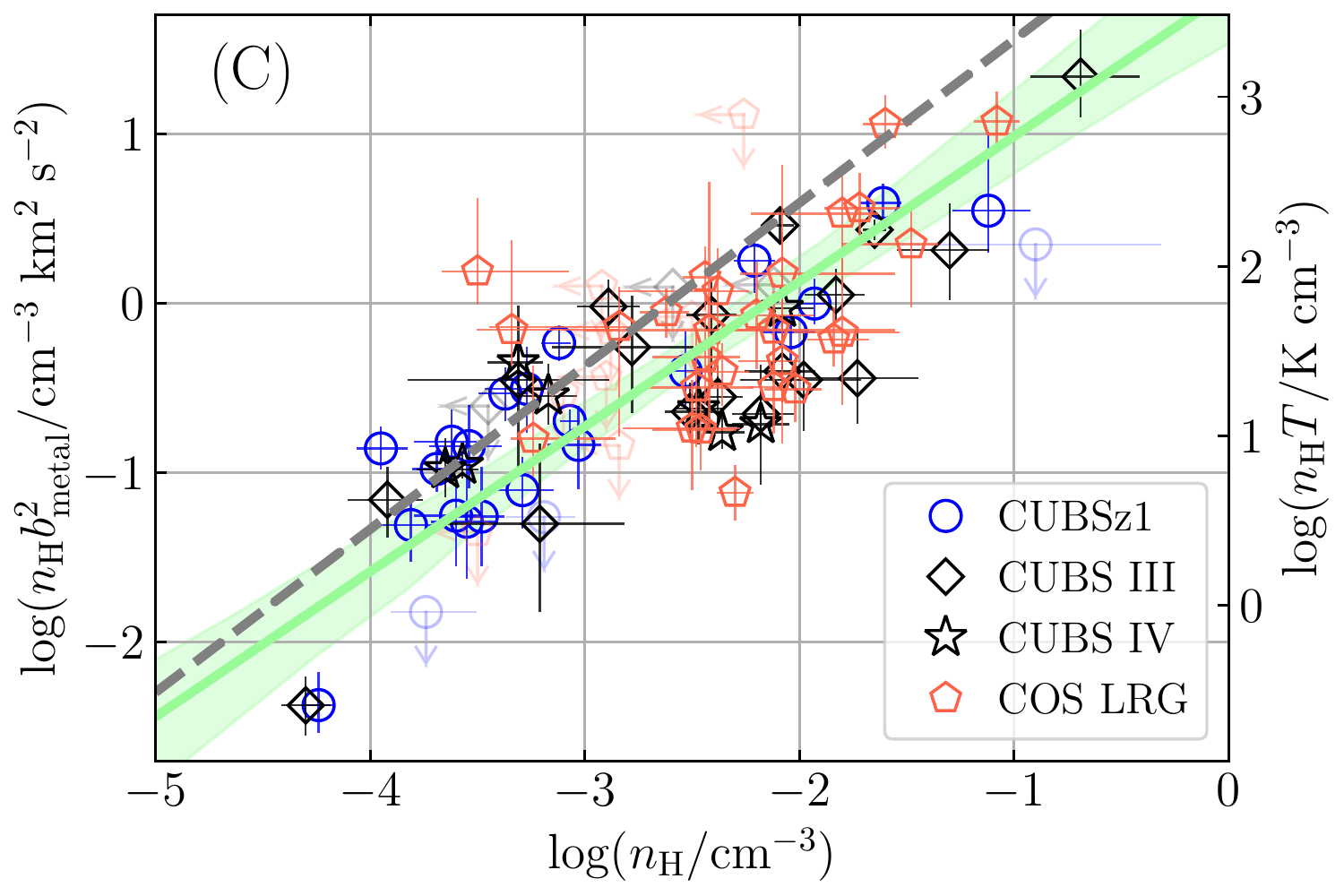}
\hskip 0.1cm
\includegraphics[width=0.49\textwidth]{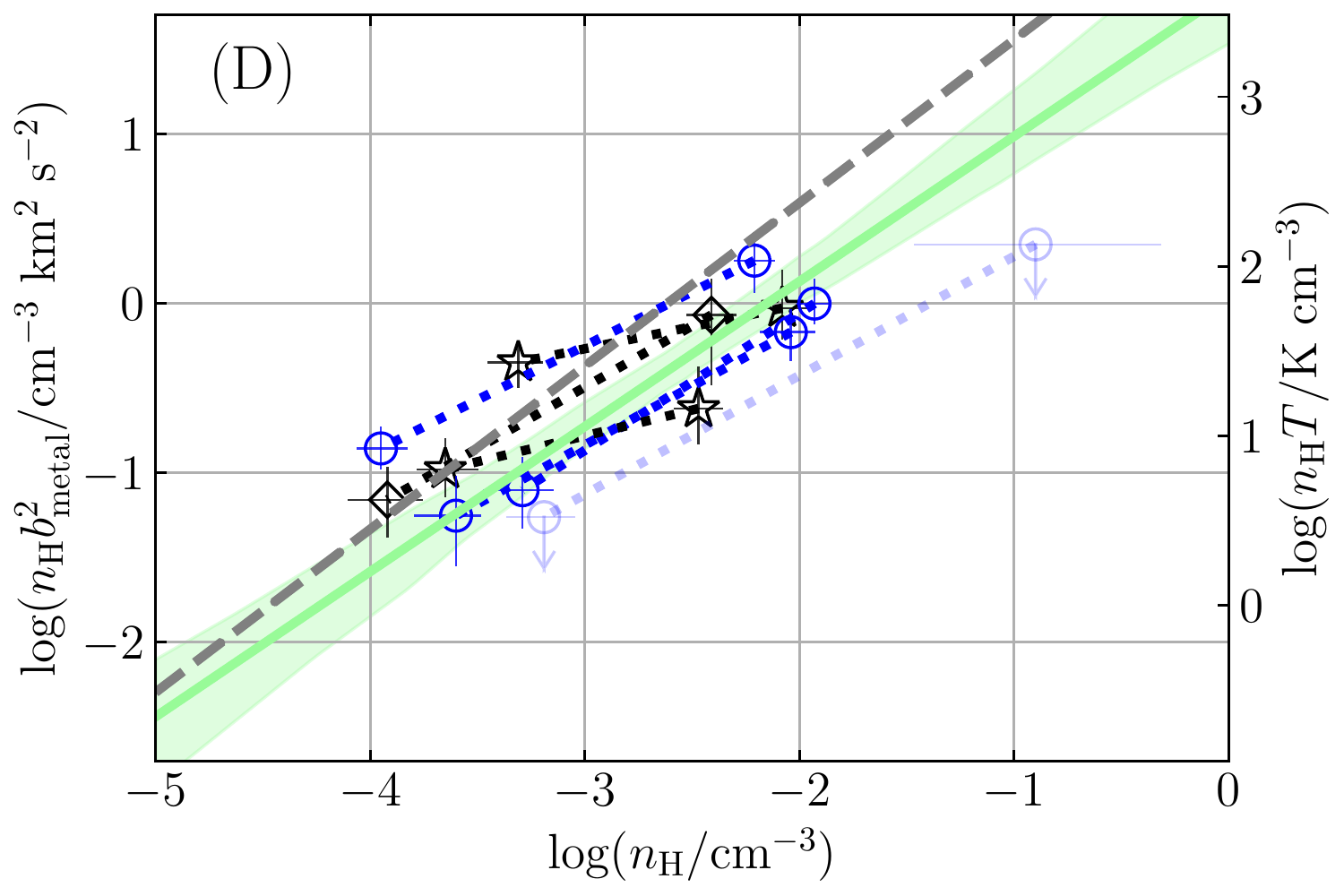}
\end{center}
 \vspace{-0.3cm}
\caption{Gas pressure determined for cool photoionized CGM.  Panel (A) shows the total pressure calculated based on $n_{\rm H}b_{\rm H{\small I}}^2$. The gray dashed line and shaded band represent the best-fit power law of $\log\,(n_{\rm H}b_{\rm H{\small I}}^2/\cmjj~{\rm km^2~s^{-2}}) = (0.96\pm0.07) \log\,(n_{\rm H}/\cmjj) + (2.52 \pm 0.17)$ and the 95\% confidence interval.
The light green line is the best-fit relation between $\log\,(n_{\rm H}b_{\rm metal}^2/\cmjj~{\rm km^2~s^{-2}})$ and $n_{\rm H}$ shown in panels (C) and (D).
Panel (B) shows the derived total pressure using the scaled $b_{\rm HI}$ from $b_{\rm metal}$ for the high- and low-density phases in multiphase components.
The two phases from each multiphase component are connected by a dotted line.
All multiphase components have consistent slopes with the full sample, showing higher dynamic pressures at higher densities.
Panel (C) shows turbulent pressure determined based on $n_{\rm H}b_{\rm metal}^2$.
The metal $b$ values are determined mainly for oxygen and nitrogen, which are dominated by non-thermal broadening (Figure \ref{fig:T_bNT}).
The correlation can be approximated by a power law following $\log (n_{\rm H}b_{\rm metal}^2/\cmjj~{\rm km^2~s^{-2}}) = (0.86\pm0.07) \log (n_{\rm H}/\cmjj) + (1.85 \pm 0.19)$, where $\gamma = 0.86\pm0.07$ is the empirical polytropic index (light green lines and shaded region).
The gray dashed line is the best-fit relation between $\log\,(n_{\rm H}b_{\rm H{\small I}}^2/\cmjj~{\rm km^2~s^{-2}})$ and $n_{\rm H}$ from Panel (A).
Panel (D) shows the comparison of turbulent pressures between differennt phases.  This exercise shows that both the total and turbulent pressure between the low- and high-density phases differ by a factor of 10 in the multiphase components.
}
\label{fig:mpc_pressure}
\end{figure*}

In Figure \ref{fig:mpc_pressure}, we show the pressure-density ($P$\,-\,$n$) relation of the photoionized gas, highlighting the multiphase components (points connected by dotted lines in panels B and D).
In these figures, we consider a combination of observed parameters to measure this relation with the lowest degree of uncertainty.
Here, we adopt $b_{\rm HI}$ as a tracer of the total pressure (Figure \ref{fig:mpc_pressure}\,A), because the motions of hydrogen particles dominate the pressure force, $P_{\rm gas}\propto n_{\rm H} b_{\rm HI}^2$. 
For the sub-sample with detected \ion{H}{I}, we find that  $n_{\rm H} b_{\rm HI}^2$ depends on the gas density with a power-law slope of $0.96\pm0.07$.  In comparison, the density dependence of thermal pressure expected from the PIE model follows $P_{\rm T}/k_{\rm B}= n_{\rm H}T \propto n_{\rm H}^{0.83\pm 0.07}$ (see Figure \ref{fig:T_bNT}\,A).  Therefore, we find that both total and thermal pressures display a consistent density dependence to within $1\,\sigma$, because of the predominance of thermal energy in cool CGM.

As described in Section \ref{sec:scaling}, broad \ion{H}{I} features from the low-density phase are obscured by the stronger \ion{H}{I} lines from the high-density phase in multiphase components and therefore $b_{\rm HI, broad}$ cannot be directly measured.  We apply the empirical $b_{\rm HI}-b_{\rm metal}$ relation displayed in Figure \ref{fig:T_bNT}\,D to infer $b_{\rm HI, broad}$ for these components in order to estimate the total pressure in Figure \ref{fig:mpc_pressure}\,B.
Here, we assume the low-density phases in multiphase components are similar to single-phase low-density clouds, which determine the scaling relationships shown in Figure \ref{fig:T_bNT}.
The derived total pressures of multiphase components follow the slope of the \ion{H}{I}-detected sub-sample, indicating that the high-density gas also experiences a higher pressure than the low-density gas. These empirical constraints on gas pressure in multi-phase absorbers suggest that the gas is either not in pressure equilibrium or additional non-thermal sources of pressure must be present. 

One caveat is the assumption described above that $b_{\rm HI, broad}$ follows the general  $b_{\rm HI}-b_{\rm metal}$ relation shown in Figure \ref{fig:T_bNT}\,D. To demonstrate the robustness of the inferred pressure imbalance, we consider the limiting case in which $b_{\rm metal}$ for the low-density phase was dominated by thermal motions. We note that the data shown in Figure \ref{fig:T_bNT}B disfavor this possibility, but we consider it as strict upper bound to the gas pressure in the low-density phase. Under these very conservative assumptions, there would still be four systems (out of seven total) with discrepant pressures between the high- and low-density phases.

We also estimate turbulent pressure using the observed line widths of metal lines following $P_{\rm turb}\propto n_{\rm H} b_{\rm metal}^2$ (Figure \ref{fig:mpc_pressure}\,C).  Recall that non-thermal
motions dominate the line widths of metal lines at all densities (Figure  \ref{fig:T_bNT}\,B).  The approximated turbulent pressure depends on the gas density with a power-law slope of $0.86\pm0.07$ (Kendall $\tau=0.76$), confirming that turbulent pressure in high-density gas is also higher than in low-density regions (Figure \ref{fig:mpc_pressure}\,D).

It is clear from Figure \ref{fig:mpc_pressure} that the derived pressures (both total and turbulent) differ by a factor of 10 between the low- and high-density phases in the multiphase components that occur at $d=50$-80 kpc in the combined $z\lesssim 1$ sample.
Therefore, extra pressure contributions are needed to balance the thermodynamic pressure in these multiphase components, or non-equilibrium processes are required to maintain large pressure fluctuations at $d\lesssim 100$ kpc in the CGM.

\section{Discussion}

The results presented in \S\ \ref{sec:results} provide a detailed characterization of the thermodynamic properties of the cool CGM around $z\lesssim 1$ galaxies, covering a broad range in star formation history.  Our analysis has for the  first time uncovered a clear distinction in the significance of non-thermal  broadening between passive and star-forming haloes.  In addition, we show that while multiphase gas is common in the inner 100 kpc of the CGM at $z\lesssim 1$, there exists a large pressure difference between the low- and high-density phases.  Here we discuss implications of these findings.

\subsection{Connections between internal non-thermal motions of cool clumps in the CGM and star formation in galaxies}

Previous studies have shown that massive quiescent haloes contain a significant amount of cool gas in their CGM \citep[e.g.,][]{Gauthier2009,Gauthier2010} with an estimated cool gas mass of $M_{\rm cool}\approx 10^{10}\,\msun$ at $d\lesssim 160$ kpc \citep{Zahedy:2019aa}.  With little ongoing star formation detected in these passive galaxies, additional mechanisms are needed to prevent these cold clumps from continuing to cool to trigger star formation in the central galaxies.  

We have demonstrated in Section \ref{sec:T_NT} that passive haloes harbor cool clouds with a higher power in their internal non-thermal motions than star forming galaxies (Figure \ref{fig:Tratio}).  This is in contrast to a suppressed bulk velocity dispersion observed for cool gas in these passive haloes \citep[][]{Huang:2016}, which is not seen around star-forming galaxies \citep{Chen:2010aa,Huang:2021aa}\footnote{See also \cite{Tumlinson:2013aa} for the velocity distributions observed for blue and red galaxies in the COS-Halos sample.  Comparisons of the observed velocity distributions and the expected virial motions based on the halos mass will provide an independent evaluation for the reported differential motions.}.  Together, these observations show that while the intercloud bulk velocities are suppressed relative to the expected virial motions in passive haloes, the internal cloud energetics are enhanced in these passive halos in comparison to cool gas detected in star-forming haloes.  

To date, the presence of cool gas in massive halos of quenched galaxies and their sub-virial velocities have presented a perplexing challenge to explain how such gas does not drive future star formation. Thus, the finding of increased non-thermal broadening leads to a natural question of whether the source of the broadening contributes to or results from processes which drive the cessation of star formation in these systems.

A natural candidate for the enhanced non-thermal broadening observed in passive haloes is turbulent motions.\footnote{Here turbulent motions refer to internal velocity field of individual absorbing components, not the bulk motions of absorbing components within the galaxy halo.} Possible mechanisms for driving the increased turbulence in passive haloes include feedback from active galactic nuclei \citep[AGN; e.g.,][]{Li:2015aa, Werner:2019aa}, stellar winds from evolved asymptotic giant branch (AGB) stars, and energy input from Type Ia supernovae \citep[e.g.,][]{Conroy:2015aa, Li:2020aa}.  In addition, the suppressed intercloud bulk velocity in passive haloes suggests that ram-pressure drag forces imposed on the cool clouds by the ambient hot medium may be effective.  The differential velocities between cool clumps and the hot halo may trigger the Kelvin-Helmholtz instability to develop \citep[e.g.,][]{Maller:2004,Afruni:2019}, providing an additional source of turbulence in these passive haloes.
In this scenario, a  key ingredient is the presence of a hot halo.  While it is still unclear how the x-ray properties of the hot CGM change between passive and star-forming galaxies, recent x-ray observations have shown that more massive galaxies host more massive hot CGM \citep{Chadayammuri:2022aa, Comparat:2022aa}.
Because passive galaxies in the COS-LRG sample are also more massive than the star-forming galaxies in the CUBSz1 sample,
the effects of hot gas may have a more significant impact on cool clouds around these quenched galaxies.

While large-scale bulk flows may also contribute to non-thermal broadening, we consider this a less likely scenario for these spectrally-resolved components with relatively narrow line widths and a median size of $\approx\!100$ pc \citep[typically 10-1000 pc; e.g.,][]{Rauch2001bb, Keeney:2017aa, Rudie:2019aa, Zahedy:2021aa}.  The implied internal velocity gradient across individual clumps would exceed $100\,\kms\, {\rm kpc^{-1}}$.
In addition, numerical simulations have shown that large-scale infall can result in absorbing clouds, which reside in distinct haloes to appear as blended components due to projection effects along the line-of-sight velocity axis \citep[e.g.,][]{Turner:2017aa,Ho:2020aa}.
We note however that the velocity dispersions of galaxy-absorber pairs have been found to be either comparable to or much smaller than the projected virial velocities of the host haloes at small projected distances of $d\lesssim 100$ kpc  \citep[e.g.,][]{Chen:2010aa, Tumlinson:2011aa,Huang:2021aa}.  Therefore, while we cannot rule out possible contamination due to line-of-sight projections, we consider it an unlikely scenario for explaining the observed dominance of non-thermal line widths in the majority of passive haloes.  

Combining all empirical constraints available for the cool CGM at $z\lesssim 1$, we therefore find it particularly interesting that the cool CGM surrounding passive galaxies has a higher non-thermal contribution to its total energy than the cool CGM in star-forming galaxies (Figure \ref{fig:Tratio}\,B).
We postulate that the observed non-thermal line widths may be attributed to turbulence in the cool CGM, and that turbulent energy appears to be more prominent in massive, quenched galaxies.
Given the possibility that these non-thermal motions may provide a new window into the physics of star-formation quenching, future CGM studies should work to further quantify and understand their predominance and their relationship to other galaxy properites.

\subsection{Implications of large pressure fluctuations in the multiphase CGM}
\label{sec:imbalance}
The observed large pressure difference by a factor of 10-30 between low- and high-density phases in the multiphase components at $d=50$-80 kpc (Figure \ref{fig:mpc_pressure}) also implies a volatile condition in the cool CGM at $z\lesssim 1$\footnote{Note that the lower bound of this impact parameter range is limited by the galaxy sample (i.e., no sight lines projected within 50 kpc are available in the combined galaxy sample), rather than non-detections of multiphase components in the inner halo.}.
Such a large pressure variation is unexpected for the global pressure variation.
In both analytic and simulation models, the global pressure profile can be approximated by a power law with slope of 1\,-\,1.5 \citep[i.e., ][]{Faerman:2017aa, Voit:2019aa, Ji:2019aa}.
Therefore, a factor of 10 difference in pressure can be translated to a factor of 5\,-\,10 difference in distances.
If the high-density (i.e, high-pressure) phase in the detected multiphase component occurs at radius of 50\,-\,80 kpc (i.e.,\ the impact parameter), the low-density phase would be at a distance of 250\,-\,800 kpc.
However, in the outskirts of CGM ($d>200$ kpc), the detection rate of absorption features is low ($< 10\%$; \citealt{Liang:2014aa, Johnson:2015aa}), which cannot account for the high detection rate  ($\approx 80\%$; Section \ref{sec:everywhere}) of low-density phases in the multiphase components at $d<100$ kpc.
Therefore, the observed pressure imbalance likely represent the local pressure variation instead of the global variation.

Numerical simulations suggest local variations in pressure of $\approx 20\%$ ($\approx 0.2$ dex) in the mixing boundary layer between the cool cloud and the hot ambient medium \citep{Ji:2019aa, Fielding:2020aa}.
Therefore, direct mixing between the cool gas and the hot ambient medium alone cannot fully explain the observed pressure variation in the multiphase components.  On the other hand, large pressure fluctuations by more than a factor of 10 can occur in simulated inner haloes where the CGM undergoes rapid radiative cooling (see examples in \citealt{Stern:2021aa}), but these large pressure fluctuations appear to be confined in the inner halo at $d\lesssim 50$ kpc.

One possible scenario is that high-density gas originates in clumps in the low-density photoionized medium, which may allow for large local pressure variation \citep{McCourt:2018aa, Sparre:2019aa, Gronke:2022aa}.
Although the volume filling factor is small, the covering factor of the high-density clumps in the low-density clouds can be high \citep{Liang:2020aa}.   

These high density clumps in the low density clouds are dynamically unstable if the multiphase medium is not in pressure equilibrium.
If high-density clumps are ejected from the galaxy, the dynamic timescale of the multiphase components at $d\approx 50$ kpc, which is about the minimum $d$ of the detected multiphase components, is
$\approx 50~ {\rm kpc} / 300\,{\kms} \approx150~{\rm Myr}$, which is much longer than the sound crossing timescale ($\approx 0.1~{\rm kpc}/ 10 {\kms}\approx 10 ~{\rm Myr}$) for a typical size of $100$ pc for the high density phase \citep{Zahedy:2021aa}.
Consequently, pressure imbalance between multiple phases is not expected to be detected beyond 5 kpc (balancing the dynamical timescale and the sound crossing timescale), but multiphase components are commonly detected at $d=50$-80 kpc in our sample.

Recent cosmological simulations (box size of $\gtrsim 100$ kpc) reveal that magnetic or cosmic-ray (CR) pressure could provide additional pressure balance to support cold clumps that are out of equilibrium, and change the morphology and thermodynamic properties of the cool CGM in CR pressure dominated haloes \citep[e.g.,][]{Ji:2019aa, Nelson:2020aa, Butsky:2020aa}.
However, in these simulations, CR pressure is higher in the cool CGM ($\log\,T/{\rm K} \approx 4-4.5$) than in the hot medium ($\log\,T/{\rm K}\approx 6$), whereas our observations suggest additional pressure is needed in the low-density phase.
This does not mean that CR pressure is ruled out as a candidate for balancing gas pressures in the cool CGM, because of uncertainties both in the propagation theory of cosmic rays and in the implementations  of numerical simulations. 
For example, \citet{Commercon:2019aa} showed that CR pressure in the interstellar medium is higher in the low-density gas than in the high-density gas where the CR diffusion coefficient is large in zoomed-in simulations (box size of $\approx 10$ pc).
Continuing effort to increase the sample size for CGM clouds with empirical constraints on the detailed component structures will provide critical tests for these different models.

\section{Summary and conclusions}

In this work, we have carried out a systematic investigation of the thermodynamic properties of the cool CGM at $z\lesssim 1$, by combining a new CGM sample at $z\approx 1$ from the CUBS program and published samples from the literature.  We analyze high S/N and high spectral resolution absorption spectroscopy, which provides a strong constraint on the thermodynamic properties of the cool CGM at $z \lesssim 1$.
Combining with a deep galaxy survey, we investigate possible dependence of the observed thermodynamic properties of the cool CGM on the star formation history in the host galaxies.
Our key findings are summarized below:
\begin{itemize}
    \item Although the density of the cool CGM varies over three decades $\log\,(n_{\rm H}/{\rm cm^{-3}}) = -4$ to $-1$, the temperature is narrow, $\log\,(T/{\rm K}) \approx 4.3\pm 0.3$), which is consistent with expectation from photoionization equilibrium models (Figure \ref{fig:T_bNT}).
    \item More than 30\% of the cool CGM at $z\lesssim 1$ exhibit line widths driven by non-thermal motions, in comparison to $< 20$\% found at $z\approx 2$-3 (Figure \ref{fig:Tratio} A and B). 
    \item Non-thermal motions contribute substantially more to the total internal energy of the gas within the cool CGM around passive galaxies than around star-forming galaxies at $z\lesssim 1$ (Figure \ref{fig:Tratio} C and D).
    \item The kinematically-aligned multiphase component in the CGM is not in pressure equilibrium. The two phases with densities that differ by a factor of 30 have pressures that differ by a factor of 10 (Figure \ref{fig:mpc_pressure} and Section \ref{sec:imbalance}).
\end{itemize}

The predominance of non-thermal broadening in gas surrounding passive galaxies when compared to those that are star-forming halos highlights a new critical window into the relationship between the detailed properties of gas within the CGM and their central galaxies.  While the large discrepancies in pressure inferred for seemingly co-moving and presumably co-spatial multi-phase absorbers suggests that the CGM is either highly dynamic and out of equilibrium or that new non-thermal sources of pressure need to be incorporated into models of the CGM.

In conclusion, the thermodynamic properties of the CGM provide unique insights into connection between the star formation history in galaxies and their CGM.
Further investigation will quantify the dependence of the thermodynamic properties of gas within the CGM on galaxy type and other galactic properties (e.g., the stellar mass and star formation rate).

\section*{Acknowledgements}
The authors thank the referee Benjamin Oppenheimer for a timely and helpful report.
ZQ acknowledges partial support from HST-GO-15163.001A, NSF AST-1715692 grants, and NASA ADAP grant 80NSSC22K0481.
HWC, and MCC acknowledge partial support from HST-GO-15163.001A and NSF AST-1715692 grants.
GCR acknowledges partial support from HST-GO-15163.015A.
FSZ acknowledges the support of a Carnegie Fellowship from the Observatories of the Carnegie Institution for Science.
SDJ acknowleges partial support from HST-GO-15280.009A.
EB acknowledges partial support from NASA under award No. 80GSFC21M0002.
KLC acknowledges partial support from NSF AST-1615296.
SC gratefully acknowledges support from the European Research Council (ERC) under the European Union’s Horizon 2020 research and innovation programme grant agreement No 864361.
CAFG was supported by NSF through grants AST-1715216, AST-2108230, and CAREER award AST-1652522; by NASA through grants 17-ATP17-006 7 and 21-ATP21-0036; by STScI through grants HST-AR-16124.001-A and HST-GO-16730.016-A; by CXO through grant TM2-23005X; and by the Research Corporation for Science Advancement through a Cottrell Scholar Award.
This work is based on observations made with ESO Telescopes at the Paranal Observatory under program ID 0104.A-0147(A), observations made with the 6.5m Magellan Telescopes located at Las Campanas Observatory, and spectroscopic data gathered under the HST-GO-15163.01A program using the NASA/ESA Hubble Space Telescope operated by the Space Telescope Science Institute and the Association of Universities for Research in Astronomy, Inc., under NASA contract NAS 5-26555. This research has made use of the services of the ESO Science Archive Facility and the Astrophysics Data Service (ADS)\footnote{\url{https://ui.adsabs.harvard.edu}}. The analysis in this work was greatly facilitated by the following \texttt{python} packages:  \texttt{Numpy} \citep{Numpy}, \texttt{Scipy} \citep{Scipy}, \texttt{Astropy} \citep{astropy:2013,astropy:2018}, \texttt{Matplotlib} \citep{Matplotlib}, and \texttt{emcee} \citep{Foreman-Mackey:2013aa}.

\section*{Data Availability}
The data underlying this article will be shared on reasonable request to the corresponding author.

\begin{table*}
\caption{Summary of density and pressure properties}
\label{tab:sample}
\begin{tabular}{llcccccccc}
\hline
\hline
& & $\log\,n_{\rm H}/$ & $b_{{\rm HI}}$ & $b_{c,{\rm metal}}^\ddagger$ & Metal & log$T/$ & $b_{\rm NT}$ & $\log P_{\rm T}/k_{\rm B}/$ & $\log P_{\rm NT}/k_{\rm B}/$\\
Sample & ID$^\dagger$ & cm$^{-3}$ & (km s$^{-1}$) & (km s$^{-1}$) & Ions & K & (km s$^{-1}$) & K cm$^{-3}$ & K cm$^{-3}$\\
\hline
CUBSz1 & s2c1 & $-3.74_{-0.16}^{+0.23}$ & $21.7_{-3.8}^{+3.7}$ & $<9.1$ & O {\scriptsize III}-{\scriptsize V} & $4.41_{-0.20}^{+0.13}$ & $<6.9$ & $0.67_{-0.26}^{+0.26}$ & $<-0.3$\\
CUBSz1 & s2c2 & $-3.27_{-0.19}^{+0.18}$ & $29.7_{-7.1}^{+6.2}$ & $24.1_{-4.5}^{+4.5}$ & O {\scriptsize III}-{\scriptsize IV} & $<4.6$ & $22.4_{-4.8}^{+4.6}$ & $<1.3$ & $1.21_{-0.27}^{+0.25}$\\
CUBSz1 & s2c3 & $-2.53_{-0.11}^{+0.10}$ & $...$ & $11.6_{-2.4}^{+2.7}$ & Mg {\scriptsize II}, O {\scriptsize III}-{\scriptsize IV} & $<5.2$ & $<10.6$ & $<2.7$ & $<1.3$\\
CUBSz1 & s2c4l & $-0.90_{-0.56}^{+0.58}$ & $29.1_{-9.8}^{+6.9}$ & $<4.2$ & Mg {\scriptsize II} & $4.34_{-0.26}^{+0.24}$ & $<3.2$ & $3.44_{-0.62}^{+0.63}$ & $<1.9$\\
CUBSz1 & s2c4h & $-3.19_{-0.17}^{+0.14}$ & ... & $<9.2$ & O {\scriptsize III}-{\scriptsize V} & $<5.1$ & $<7.7$ & $<1.9$ & $<0.4$\\
CUBSz1 & s5c2 & $-3.54_{-0.18}^{+0.15}$ & $35.2_{-5.4}^{+6.9}$ & $22.3_{-4.1}^{+4.7}$ & O {\scriptsize III}-{\scriptsize IV} & $4.72_{-0.29}^{+0.24}$ & $19.8_{-4.7}^{+5.1}$ & $1.18_{-0.34}^{+0.28}$ & $0.84_{-0.27}^{+0.27}$\\
CUBSz1 & s6c1l & $-2.21_{-0.09}^{+0.09}$ & $13.9_{-1.3}^{+1.3}$ & $17.0_{-3.1}^{+2.3}$ & O {\scriptsize II}, S {\scriptsize II} & $4.00_{-0.15}^{+0.10}$ & $7.5_{-0.4}^{+0.5}$ & $1.79_{-0.17}^{+0.13}$ & $1.32_{-0.10}^{+0.11}$\\
CUBSz1 & s6c1h & $-3.95_{-0.11}^{+0.12}$ & ... & $35.2_{-1.5}^{+1.1}$ & O {\scriptsize IV}, N {\scriptsize IV} & $<4.5$ &  $34.8_{-1.6}^{+1.2}$ & $<0.6$ & $0.92_{-0.12}^{+0.12}$\\
CUBSz1 & s6c2l & $-1.93_{-0.08}^{+0.07}$ & $24.0_{-2.6}^{+3.1}$ & $9.2_{-0.9}^{+1.3}$ & N {\scriptsize II}, O {\scriptsize II} & $4.57_{-0.11}^{+0.11}$ & $6.3_{-0.7}^{+0.7}$ & $2.64_{-0.14}^{+0.13}$ & $1.45_{-0.13}^{+0.12}$\\
CUBSz1 & s6c2h & $-3.29_{-0.18}^{+0.14}$ & ... & $12.4_{-1.8}^{+1.8}$ & N {\scriptsize IV},  O {\scriptsize IV}-{\scriptsize V} & $<5.1$ & $9.8_{-4.3}^{+2.7}$ & $<1.8$ & $0.47_{-0.42}^{+0.28}$\\
CUBSz1 & s6c3 & $-3.12_{-0.08}^{+0.05}$ & ... & $27.7_{-2.0}^{+2.7}$ & O {\scriptsize III}-{\scriptsize V} & $<5.0$ & $25.7_{-3.0}^{+3.0}$ & $<1.9$ & $1.48_{-0.13}^{+0.11}$\\
CUBSz1 & s12c1 & $-3.62_{-0.17}^{+0.11}$ & ... & $25.2_{-3.6}^{+4.2}$ & N {\scriptsize IV}, O {\scriptsize IV}-{\scriptsize V} & $<5.4$ & $<24.0$ & $<1.8$ & $<0.9$\\
CUBSz1 & s12c2 & $-3.69_{-0.11}^{+0.08}$ & ... & $22.7_{-1.8}^{+1.8}$ & O {\scriptsize III}-{\scriptsize V} & $<5.2$ & $<21.0$ & $<1.5$ & $<0.7$\\
CUBSz1 & s13c1 & $-1.61_{-0.10}^{+0.08}$ & ... & $12.6_{-0.9}^{+1.0}$ & C {\scriptsize II}, O {\scriptsize II} & $4.42_{-0.23}^{+0.18}$ & $6.6_{-0.7}^{+0.5}$ & $2.81_{-0.25}^{+0.20}$ & $1.81_{-0.14}^{+0.10}$ \\
CUBSz1 & s13c2 & $-1.12_{-0.16}^{+0.19}$ & ... & $6.8_{-1.4}^{+3.3}$ & C {\scriptsize II}, O {\scriptsize II} & $4.03_{-0.32}^{+0.24}$ & $2.4_{-1.4}^{+1.0}$ & $2.91_{-0.36}^{+0.31}$ & $1.42_{-0.53}^{+0.41}$\\
CUBSz1 & s15c1 & $-3.07_{-0.04}^{+0.04}$ & ... & $15.4_{-0.6}^{+0.8}$ & C {\scriptsize II}, O {\scriptsize IV}, Mg {\scriptsize II} & $4.60_{-0.43}^{+0.26}$ & $14.4_{-1.2}^{+1.0}$ & $1.53_{-0.43}^{+0.26}$ & $1.03_{-0.08}^{+0.07}$\\
CUBSz1 & s15c2 & $-3.81_{-0.16}^{+0.14}$ & ... & $17.8_{-2.8}^{+1.5}$ & N {\scriptsize IV}, O {\scriptsize IV}-{\scriptsize V} & $<5.5$ & $<19.1$ & $<0.8$ & $0.31_{-0.30}^{+0.22}$\\
CUBSz1 & s15c3l & $-2.04_{-0.14}^{+0.11}$ & ... & $8.6_{-0.9}^{+0.7}$ & C {\scriptsize II}, S {\scriptsize III}, N {\scriptsize III} & $4.65_{-0.19}^{+0.09}$ & $<4.8$ & $2.61_{-0.24}^{+0.14}$ & $<1.1$\\
CUBSz1 & s15c3h & $-3.60_{-0.19}^{+0.11}$ & ... & $14.9_{-3.8}^{+3.4}$ & N {\scriptsize IV}, O {\scriptsize V} & $<5.4$ & $<14.0$ & $<1.8$ & $<0.5$\\
CUBSz1 & s17c2 & $-3.48_{-0.13}^{+0.10}$ & ... & $12.9_{-3.8}^{+4.0}$ & O {\scriptsize III}-{\scriptsize V} & $<5.1$ & $<13.2$ & $<1.6$ & $<0.5$\\
CUBSz1 & s17c4 & $-3.03_{-0.14}^{+0.10}$ & ... & $12.5_{-3.1}^{+2.9}$ & O {\scriptsize III}-{\scriptsize IV} & $<5.0$ & $<12.2$ & $<2.0$ & $<0.9$\\
CUBSz1 & s18c1 & $-3.37_{-0.12}^{+0.10}$ & ... & $26.2_{-2.9}^{+3.2}$ & O {\scriptsize III}-{\scriptsize V} & $<5.4$ & $<25.0$ & $<2.0$ & $<1.2$\\
CUBSz1 & s18c2 & $-3.55_{-0.16}^{+0.14}$ & ... & $13.5_{-4.5}^{+4.8}$ & O {\scriptsize III}-{\scriptsize IV} & $<5.3$ & $<12.4$ & $<1.8$ & $<0.4$\\
CUBSz1 & s20c1 & $-4.24_{-0.05}^{+0.06}$ & ... & $8.6_{-1.5}^{+1.8}$ & C {\scriptsize IV}, O {\scriptsize IV}-{\scriptsize V} & $<4.9$ & $<7.8$ & $<0.7$ & $<-0.7$\\
\hline
CUBS III & s1c1 & $-2.89_{-0.14}^{+0.14}$ & $31.1_{-0.8}^{+1.2}$ & $27.2_{-1.9}^{+2.0}$ & C {\scriptsize III} & $4.23_{-0.27}^{+0.16}$ & $26.6_{-2.0}^{+2.0}$ & $1.34_{-0.30}^{+0.21}$ & $1.74_{-0.15}^{+0.15}$\\
CUBS III & s1c2 & $-1.83_{-0.14}^{+0.13}$ & $23.8_{-0.7}^{+0.5}$ & $8.7_{-0.7}^{+0.7}$ & C {\scriptsize II}, Mg {\scriptsize II} & $4.49_{-0.03}^{+0.03}$ & $7.3_{-0.9}^{+0.8}$ & $2.66_{-0.14}^{+0.13}$ & $1.68_{-0.18}^{+0.16}$\\
CUBS III & s1c3 & $-1.65_{-0.05}^{+0.05}$ & $20.0_{-0.3}^{+0.3}$ & $11.0_{-0.3}^{+0.3}$ & N {\scriptsize II}, Mg {\scriptsize II}, Si {\scriptsize II} & $4.25_{-0.03}^{+0.02}$ & $10.5_{-0.4}^{+0.3}$ & $2.60_{-0.06}^{+0.05}$ & $2.17_{-0.06}^{+0.06}$\\
CUBS III & s1c4 & $-3.21_{-0.44}^{+0.39}$ & $9.3_{-1.0}^{+1.5}$ & $9.0_{-2.7}^{+2.7}$ & Si {\scriptsize III} & $<3.6$ & $8.9_{-3.0}^{+2.8}$ & $<0.4$ & $0.47_{-0.53}^{+0.48}$\\
CUBS III & s1c6 & $-3.31_{-0.51}^{+0.42}$ & $19.2_{-1.7}^{+2.3}$ & $26.8_{-3.0}^{+3.5}$ & C {\scriptsize III} & $<3.7$ & $19.2_{-1.7}^{+2.3}$ & $<0.8$ & $1.04_{-0.52}^{+0.43}$\\
CUBS III & s2c1 & $-2.09_{-0.06}^{+0.06}$ & $22.2_{-0.3}^{+0.4}$ & $18.8_{-1.6}^{+1.4}$ & \ion{N}{III}, O {\scriptsize II} & $3.92_{-0.23}^{+0.16}$ & $19.1_{-1.6}^{+1.3}$ & $1.83_{-0.24}^{+0.17}$ & $2.25_{-0.09}^{+0.08}$\\
CUBS III & s2c2 & $-2.51_{-0.11}^{+0.12}$ & $18.6_{-0.6}^{+0.6}$ & $8.6_{-1.0}^{+1.1}$ & C {\scriptsize II}, O {\scriptsize II} & $4.25_{-0.05}^{+0.04}$ & $7.5_{-1.3}^{+1.3}$ & $1.74_{-0.12}^{+0.13}$ & $1.02_{-0.19}^{+0.19}$\\
CUBS III & s2c3l & $-2.41_{-0.11}^{+0.11}$ & $19.3_{-0.6}^{+0.6}$ & $14.8_{-6.7}^{+3.0}$ & O {\scriptsize II} & $4.18_{-0.31}^{+0.14}$ & $11.2_{-6.1}^{+4.7}$ & $1.94_{-0.11}^{+0.11}$ & $<0.5$\\
CUBS III & s2c3h & $-3.92_{-0.18}^{+0.16}$ & ... & $24.0_{-3.3}^{+2.8}$ & O {\scriptsize IV}, Si {\scriptsize III} & $<5.8$ & $22.2_{-9.0}^{+3.6}$ & $<1.9$ & $0.55_{-0.40}^{+0.21}$\\
CUBS III & s2c4 & $<-2.12$ & $12.9_{-2.8}^{+2.8}$ & $13.0_{-0.8}^{+0.8}$ & C {\scriptsize III} & $<3.8$ & $12.9_{-0.8}^{+0.8}$ & $<1.7$ & $<1.9$\\
CUBS III & s2c5 & $<-2.59$ & $27.9_{-2.8}^{+3.7}$ & $22.0_{-2.0}^{+2.2}$ & C {\scriptsize III} & $4.38_{-0.31}^{+0.21}$ & $21.2_{-2.4}^{+2.3}$ & $<1.8$ & $<1.8$\\
CUBS III & s2c6 & $-4.30_{-0.11}^{+0.11}$ & $16.2_{-2.0}^{+2.1}$ & $9.2_{-1.4}^{+1.3}$ & C {\scriptsize III}, O {\scriptsize IV} & $4.01_{-0.32}^{+0.18}$ & $8.8_{-1.8}^{+1.7}$ & $-0.29_{-0.34}^{+0.21}$ & $-0.63_{-0.21}^{+0.20}$\\
CUBS III & s3c2 & $-2.78_{-0.37}^{+0.28}$ & $26.8_{-0.6}^{+0.6}$ & $18.2_{-2.2}^{+2.2}$ & C {\scriptsize III}, N {\scriptsize III} & $4.38_{-0.13}^{+0.09}$ & $17.8_{-2.6}^{+2.6}$ & $1.60_{-0.39}^{+0.29}$ & $1.50_{-0.39}^{+0.31}$\\
CUBS III & s3c3 & $-0.69_{-0.23}^{+0.27}$ & $18.7_{-0.2}^{+0.2}$ & $10.3_{-0.3}^{+0.3}$ & C {\scriptsize II}, Mg {\scriptsize II} & $4.19_{-0.02}^{+0.02}$ & $9.8_{-0.4}^{+0.4}$ & $3.50_{-0.23}^{+0.27}$ & $3.07_{-0.23}^{+0.27}$\\
CUBS III & s3c4 & $-1.98_{-0.29}^{+0.26}$ & $18.5_{-2.8}^{+2.7}$ & $5.8_{-0.3}^{+0.4}$ & C {\scriptsize II}, Mg {\scriptsize II} & $4.34_{-0.17}^{+0.15}$ & $4.1_{-0.9}^{+0.7}$ & $2.36_{-0.34}^{+0.30}$ & $1.03_{-0.35}^{+0.30}$\\
CUBS III & s4c1 & $-2.18_{-0.13}^{+0.15}$ & $12.2_{-0.6}^{+0.8}$ & $5.8_{-0.3}^{+0.3}$ & C {\scriptsize II}, Mg {\scriptsize II} & $3.88_{-0.06}^{+0.06}$ & $5.2_{-0.2}^{+0.2}$ & $1.70_{-0.14}^{+0.16}$ & $1.03_{-0.13}^{+0.15}$\\
CUBS III & s4c2 & $<-3.45$ & $59.6_{-3.0}^{+3.3}$ & $<26.4$ & C {\scriptsize III} & $5.28_{-0.16}^{+0.07}$ & $<26.3$ & $<1.8$ & $<1.2$\\
CUBS III & s4c3 & $-1.73_{-0.26}^{+0.28}$ & $16.1_{-0.5}^{+0.6}$ & $4.4_{-0.2}^{+0.2}$ & N {\scriptsize II}, Mg {\scriptsize II} & $3.93_{-0.06}^{+0.06}$ & $3.6_{-0.3}^{+0.3}$ & $2.20_{-0.27}^{+0.29}$ & $1.16_{-0.27}^{+0.29}$\\
CUBS III & s4c4 & $-2.08_{-0.17}^{+0.15}$ & $14.4_{-0.7}^{+1.0}$ & $6.9_{-0.3}^{+0.3}$ & Mg {\scriptsize II}, S {\scriptsize II} & $4.10_{-0.06}^{+0.06}$ & $4.9_{-0.2}^{+0.2}$ & $2.02_{-0.18}^{+0.16}$ & $1.08_{-0.17}^{+0.15}$\\
CUBS III & s4c5 & $-1.30_{-0.24}^{+0.18}$ & $18.0_{-1.0}^{+1.0}$ & $6.4_{-1.2}^{+1.5}$ & N {\scriptsize II}, Mg {\scriptsize II}, S {\scriptsize II} & $4.28_{-0.05}^{+0.04}$ & $<2.2$ & $2.98_{-0.25}^{+0.18}$ & $<1.2$\\
CUBS III & s4c6 & $-2.38_{-0.14}^{+0.11}$ & $15.7_{-1.2}^{+1.5}$ & $8.2_{-1.1}^{+1.3}$ & N {\scriptsize II}, C {\scriptsize II}, Mg {\scriptsize II} & $4.15_{-0.09}^{+0.08}$ & $5.5_{-0.4}^{+0.4}$ & $1.77_{-0.17}^{+0.14}$ & $0.88_{-0.15}^{+0.13}$\\
\hline
CUBS IV & s1c1l & $-2.08_{-0.07}^{+0.14}$ & $20.9_{-1.3}^{+1.4}$ & $10.6_{-2.0}^{+2.1}$ & Mg {\scriptsize II}, S {\scriptsize II}& $4.31_{-0.12}^{+0.09}$ & $9.9_{-2.5}^{+2.5}$ & $2.03_{-0.14}^{+0.17}$ & $1.49_{-0.23}^{+0.26}$\\
CUBS IV & s1c1h & $-3.31_{-0.14}^{+0.11}$ & ... & $30.2_{-1.5}^{+1.5}$ & C {\scriptsize III}, O {\scriptsize III}-{\scriptsize IV} & $<5.4$ & $29.5_{-5.3}^{+1.9}$ & $<1.9$ & $1.21_{-0.21}^{+0.12}$\\
CUBS IV & s1c2 & $-2.18_{-0.13}^{+0.13}$ & $22.5_{-2.7}^{+2.4}$ & $5.4_{-2.0}^{+2.0}$ & C {\scriptsize II}-{\scriptsize III}, Mg {\scriptsize II}& $4.48_{-0.10}^{+0.09}$ & $<3.8$ & $2.10_{-0.16}^{+0.16}$ & $<0.6$\\
CUBS IV & s1c3 & $-3.57_{-0.05}^{+0.05}$ & $34.5_{-2.6}^{+3.0}$ & $20.3_{-1.2}^{+1.2}$ & C {\scriptsize III}, O {\scriptsize IV} & $4.68_{-0.12}^{+0.11}$ & $19.1_{-1.5}^{+1.4}$ & $0.91_{-0.13}^{+0.12}$ & $0.57_{-0.08}^{+0.08}$\\
CUBS IV & s1c4l & $-2.47_{-0.11}^{+0.11}$ & $14.3_{-0.5}^{+0.5}$ & $8.4_{-1.7}^{+2.1}$ & C {\scriptsize II}, O {\scriptsize II} & $4.07_{-0.04}^{+0.04}$ & $3.7_{-1.3}^{+1.1}$ & $1.40_{-0.12}^{+0.12}$ & $0.25_{-0.32}^{+0.28}$\\
CUBS IV & s1c4h & $-3.65_{-0.13}^{+0.15}$ & ... & $21.6_{-2.2}^{+2.3}$ & N {\scriptsize IV}, O {\scriptsize IV} & $<5.2$ & $20.7_{-4.3}^{+2.6}$ & $<1.4$ & $0.56_{-0.22}^{+0.19}$\\
CUBS IV & s2c1 & $-2.36_{-0.08}^{+0.10}$ & $12.9_{-0.6}^{+0.6}$ & $6.3_{-0.3}^{+0.2}$ & N {\scriptsize II}-{\scriptsize III}, O {\scriptsize II} & $3.91_{-0.03}^{+0.03}$ & $5.6_{-0.5}^{+0.5}$ & $1.35_{-0.09}^{+0.10}$ & $0.72_{-0.11}^{+0.13}$\\
CUBS IV & s2c2 & $-3.17_{-0.14}^{+0.13}$ & $63.7_{-9.3}^{+11.8}$ & $20.5_{-2.1}^{+3.0}$ & N {\scriptsize III}-{\scriptsize IV}, O {\scriptsize III}& $5.37_{-0.14}^{+0.15}$ & $12.7_{-6.2}^{+6.9}$ & $2.00_{-0.20}^{+0.20}$ & $0.62_{-0.45}^{+0.49}$\\
\hline
COS LRG & s1c1 & $-2.84_{-0.60}^{+0.04}$ & $18.2_{-1.7}^{+1.7}$ & $22.4_{-5.8}^{+5.9}$ & C {\scriptsize III}, Si {\scriptsize III} & $<4.3$ & $18.2_{-1.7}^{+1.7}$ & $<1.5$ & $1.46_{-0.61}^{+0.09}$\\
COS LRG & s1c2 & $-2.08_{-0.19}^{+0.52}$ & $10.6_{-0.6}^{+0.7}$ & $13.4_{-2.0}^{+5.6}$ & C {\scriptsize III} & $<3.8$ & $10.6_{-0.6}^{+0.7}$ & $<1.7$ & $1.75_{-0.20}^{+0.52}$\\
COS LRG & s1c3 & $-2.44_{-0.24}^{+0.16}$ & $10.8_{-0.9}^{+1.0}$ & $11.5_{-1.8}^{+2.3}$ & C {\scriptsize II}, N {\scriptsize II} & $<3.6$ & $10.8_{-0.9}^{+1.0}$ & $<1.2$ & $1.41_{-0.25}^{+0.18}$\\
COS LRG & s1c4 & $-2.38_{-0.46}^{+0.14}$ & $26.9_{-1.9}^{+2.6}$ & $16.8_{-3.6}^{+3.9}$ & Si {\scriptsize II}-{\scriptsize III} & $4.48_{-0.20}^{+0.14}$ & $16.5_{-4.1}^{+4.0}$ & $2.10_{-0.50}^{+0.20}$ & $1.84_{-0.51}^{+0.25}$\\
COS LRG & s1c5 & $-3.50_{-0.16}^{+0.42}$ & $71.3_{-3.1}^{+3.7}$ & $69.8_{-8.6}^{+6.7}$ & C {\scriptsize III}, Si {\scriptsize III} & $<5.0$ & $63.7_{-8.3}^{+5.6}$ & $<1.5$ & $1.89_{-0.20}^{+0.43}$\\
COS LRG & s2c3 & $-1.80_{-0.42}^{+0.16}$ & $21.5_{-1.5}^{+1.6}$ & $14.6_{-2.6}^{+2.6}$ & Mg {\scriptsize II}, C {\scriptsize II}, Si {\scriptsize II} & $4.22_{-0.22}^{+0.14}$ & $13.7_{-2.9}^{+2.9}$ & $2.42_{-0.47}^{+0.21}$ & $2.26_{-0.46}^{+0.24}$\\
\hline
\hline
\end{tabular}
\end{table*}

\begin{table*}
\contcaption{Summary of density and pressure properties}
\begin{tabular}{llcccccccc}
\hline
\hline
& & $\log\,n_{\rm H}/$ & $b_{{\rm HI}}$ & $b_{c,{\rm metal}}^\ddagger$ & Metal & log$T/$ & $b_{\rm NT}$ & $\log P_{\rm T}/k_{\rm B}/$ & $\log P_{\rm NT}/k_{\rm B}/$\\
Sample & ID$^\dagger$ & cm$^{-3}$ & (km s$^{-1}$) & (km s$^{-1}$) & Ions & K & (km s$^{-1}$) & K cm$^{-3}$ & K cm$^{-3}$\\
\hline
COS LRG & s2c4 & $-1.48_{-0.32}^{+0.12}$ & $11.4_{-1.2}^{+1.0}$ & $8.2_{-1.7}^{+2.0}$ & Mg {\scriptsize II}, Si {\scriptsize II}& $3.62_{-0.32}^{+0.18}$ & $7.8_{-1.7}^{+1.8}$ & $2.14_{-0.45}^{+0.22}$ & $2.09_{-0.37}^{+0.23}$\\
COS LRG & s3c1 & $-3.24_{-0.10}^{+0.38}$ & $20.1_{-1.0}^{+1.1}$ & $16.6_{-3.7}^{+3.7}$ & N {\scriptsize III}, Si {\scriptsize III} & $4.01_{-0.41}^{+0.22}$ & $15.6_{-3.6}^{+2.9}$ & $0.77_{-0.42}^{+0.44}$ & $0.93_{-0.22}^{+0.41}$\\
COS LRG & s3c2 & $-1.60_{-0.10}^{+0.12}$ & $23.9_{-2.0}^{+1.7}$ & $21.3_{-2.4}^{+2.7}$ & C {\scriptsize II}, N {\scriptsize II}-{\scriptsize III}, O {\scriptsize I} & $4.00_{-0.44}^{+0.24}$ & $20.2_{-2.2}^{+2.2}$ & $2.40_{-0.45}^{+0.27}$ & $2.79_{-0.14}^{+0.15}$\\
COS LRG & s3c3 & $-1.08_{-0.10}^{+0.10}$ & $12.8_{-0.3}^{+0.4}$ & $11.9_{-1.6}^{+1.9}$ & C {\scriptsize II}, N {\scriptsize II} & $<3.5$ & $11.8_{-1.8}^{+2.1}$ & $<2.4$ & $2.85_{-0.17}^{+0.18}$\\
COS LRG & s4c1 & $<-3.50$ & $38.7_{-1.7}^{+6.0}$ & $11.9_{-2.1}^{+7.3}$ & C {\scriptsize III} & $4.97_{-0.10}^{+0.10}$ & $<14.2$ & $<1.5$ & $<0.6$\\
COS LRG & s4c2 & $<-2.84$ & $35.0_{-6.2}^{+11.0}$ & $10.0_{-3.5}^{+6.3}$ & C {\scriptsize III} & $4.93_{-0.21}^{+0.19}$ & $<10.4$ & $<2.1$ & $<1.0$\\
COS LRG & s6c1 & $-3.34_{-0.16}^{+0.50}$ & $25.8_{-2.4}^{+5.1}$ & $39.0_{-5.6}^{+6.8}$ & C {\scriptsize III}, Si {\scriptsize III} & $<4.2$ & $25.8_{-2.4}^{+5.1}$ & $<0.9$ & $1.27_{-0.18}^{+0.53}$\\
COS LRG & s6c2 & $-2.42_{-0.58}^{+0.86}$ & $15.0_{-6.5}^{+2.6}$ & $13.5_{-1.1}^{+1.6}$ & Mg {\scriptsize II} & $<3.9$ & $12.9_{-0.9}^{+1.4}$ & $<1.5$ & $1.58_{-0.58}^{+0.87}$\\
COS LRG & s6c3 & $-1.72_{-0.16}^{+0.20}$ & $24.7_{-1.1}^{+0.9}$ & $13.8_{-0.4}^{+0.4}$ & Mg {\scriptsize II}, Si {\scriptsize II} & $4.41_{-0.06}^{+0.05}$ & $13.2_{-0.5}^{+0.4}$ & $2.69_{-0.17}^{+0.21}$ & $2.30_{-0.16}^{+0.20}$\\
COS LRG & s6c4 & $<-3.10$ & $14.4_{-1.6}^{+4.1}$ & $19.3_{-5.3}^{+4.3}$ & C {\scriptsize III} & $<4.1$ & $14.4_{-1.6}^{+4.1}$ & $<1.0$ & $<1.0$\\
COS LRG & s7c1 & $-2.12_{-0.12}^{+0.04}$ & $25.6_{-0.9}^{+2.3}$ & $6.5_{-1.8}^{+1.8}$ & C {\scriptsize III}, Mg {\scriptsize II} & $4.64_{-0.06}^{+0.06}$ & $<4.2$ & $2.52_{-0.13}^{+0.07}$ & $<0.9$\\
COS LRG & s7c2 & $-2.08_{-0.20}^{+0.08}$ & $18.4_{-1.4}^{+1.3}$ & $7.4_{-3.7}^{+6.4}$ & Mg {\scriptsize II} & $4.22_{-0.22}^{+0.11}$ & $7.8_{-4.8}^{+5.3}$ & $2.14_{-0.30}^{+0.14}$ & $1.49_{-0.57}^{+0.60}$\\
COS LRG & s9c1 & $-2.44_{-0.10}^{+0.06}$ & $32.6_{-1.7}^{+2.2}$ & $19.8_{-3.6}^{+3.7}$ & C {\scriptsize II}, Si {\scriptsize II} & $4.68_{-0.14}^{+0.10}$ & $17.6_{-4.6}^{+4.2}$ & $2.24_{-0.17}^{+0.12}$ & $1.83_{-0.25}^{+0.22}$\\
COS LRG & s9c2 & $-2.62_{-0.12}^{+0.10}$ & $30.7_{-2.9}^{+4.6}$ & $19.2_{-1.7}^{+1.9}$ & N {\scriptsize III}, Mg {\scriptsize II} & $4.62_{-0.18}^{+0.16}$ & $18.5_{-1.9}^{+2.0}$ & $2.00_{-0.22}^{+0.19}$ & $1.70_{-0.15}^{+0.14}$\\
COS LRG & s9c3 & $-2.12_{-0.18}^{+0.16}$ & $12.7_{-1.0}^{+0.9}$ & $9.6_{-1.6}^{+1.7}$ & C {\scriptsize II}-{\scriptsize III}, Mg {\scriptsize II} & $3.68_{-0.31}^{+0.17}$ & $9.2_{-1.7}^{+1.7}$ & $1.56_{-0.36}^{+0.23}$ & $1.59_{-0.24}^{+0.23}$\\
COS LRG & s9c4 & $<-2.26$ & $55.2_{-4.9}^{+7.7}$ & $48.5_{-26.1}^{+18.5}$ & C {\scriptsize III} & $<5.3$ & $<50.0$ & $<3.0$ & $<2.9$\\
COS LRG & s10c2 & $-2.02_{-0.14}^{+0.12}$ & $17.4_{-0.4}^{+0.4}$ & $5.7_{-0.8}^{+0.8}$ & C {\scriptsize III}, Mg {\scriptsize II} & $4.24_{-0.03}^{+0.03}$ & $4.2_{-1.3}^{+1.1}$ & $2.22_{-0.14}^{+0.12}$ & $1.01_{-0.30}^{+0.26}$\\
COS LRG & s10c4 & $-1.80_{-0.22}^{+0.26}$ & $15.9_{-0.3}^{+0.4}$ & $6.5_{-2.7}^{+2.8}$ & C {\scriptsize III}, Mg {\scriptsize II} & $4.15_{-0.09}^{+0.04}$ & $<8.0$ & $2.35_{-0.24}^{+0.26}$ & $<1.8$\\
COS LRG & s10c5 & $-2.48_{-0.26}^{+0.26}$ & $10.0_{-0.4}^{+0.4}$ & $9.8_{-2.6}^{+2.7}$ & C {\scriptsize III}, Mg {\scriptsize II} & $4.15_{-0.09}^{+0.06}$ & $4.7_{-3.0}^{+3.4}$ & $1.67_{-0.28}^{+0.27}$ & $0.65_{-0.61}^{+0.68}$\\
COS LRG & s10c6 & $<-2.88$ & $24.2_{-0.9}^{+0.9}$ & $22.8_{-9.3}^{+5.8}$ & C {\scriptsize III}  & $<4.4$ & $23.4_{-4.5}^{+1.2}$ & $<1.5$ & $<1.6$\\
COS LRG & s12c1 & $-1.84_{-0.12}^{+0.16}$ & $14.8_{-1.4}^{+0.9}$ & $6.5_{-0.7}^{+0.7}$ & Mg {\scriptsize I}-{\scriptsize II} & $4.04_{-0.10}^{+0.08}$ & $5.9_{-0.8}^{+0.8}$ & $2.20_{-0.16}^{+0.18}$ & $1.48_{-0.17}^{+0.20}$\\
COS LRG & s12c2 & $<-2.92$ & $20.8_{-2.5}^{+2.5}$ & $32.4_{-15.5}^{+15.5}$ & C {\scriptsize III} & $<4.6$ & $20.8_{-2.5}^{+2.5}$ & $<1.7$ & $<1.5$\\
COS LRG & s12c3 & $<-2.50$ & $26.3_{-2.7}^{+3.8}$ & $16.2_{-6.7}^{+16.9}$ & C {\scriptsize III} & $4.49_{-0.38}^{+0.19}$ & $15.2_{-8.6}^{+8.1}$ & $<2.0$ & $<1.6$\\
COS LRG & s12c4 & $<-2.90$ & $19.9_{-3.6}^{+5.6}$ & $17.1_{-3.9}^{+8.8}$ & C {\scriptsize III} & $<4.5$ & $16.3_{-5.0}^{+5.0}$ & $<1.6$ & $<1.3$\\
COS LRG & s14c1 & $-2.36_{-0.16}^{+0.12}$ & $30.5_{-3.2}^{+5.1}$ & $9.5_{-1.2}^{+1.2}$ & N {\scriptsize II}, Mg {\scriptsize II} & $4.81_{-0.12}^{+0.11}$ & $5.5_{-2.8}^{+2.2}$ & $2.42_{-0.21}^{+0.17}$ & $0.98_{-0.41}^{+0.33}$\\
COS LRG & s14c2 & $-2.30_{-0.08}^{+0.08}$ & $14.2_{-1.4}^{+3.1}$ & $3.9_{-0.6}^{+0.6}$ & Mg {\scriptsize II} & $4.18_{-0.13}^{+0.12}$ & $2.1_{-1.3}^{+1.1}$ & $1.88_{-0.15}^{+0.14}$ & $0.13_{-0.54}^{+0.46}$\\
COS LRG & s16c1 & $-2.20_{-0.26}^{+0.26}$ & $23.8_{-5.5}^{+10.5}$ & $11.6_{-2.3}^{+2.3}$ & C {\scriptsize III}, Mg {\scriptsize II} & $4.53_{-1.04}^{+0.35}$ & $9.5_{-3.7}^{+5.0}$ & $2.33_{-1.07}^{+0.44}$ & $1.54_{-0.43}^{+0.53}$\\
COS LRG & s16c2 & $-2.50_{-0.32}^{+0.22}$ & $11.6_{-3.0}^{+5.5}$ & $7.6_{-1.4}^{+1.5}$ & C {\scriptsize II}, Mg {\scriptsize II} & $4.11_{-0.40}^{+0.29}$ & $6.4_{-2.0}^{+1.8}$ & $1.61_{-0.51}^{+0.36}$ & $0.89_{-0.42}^{+0.33}$\\
COS LRG & s16c3 & $-2.46_{-0.22}^{+0.16}$ & $19.0_{-2.2}^{+4.3}$ & $7.2_{-0.7}^{+0.7}$ & Mg {\scriptsize II} & $4.43_{-0.17}^{+0.16}$ & $5.6_{-1.4}^{+1.1}$ & $1.97_{-0.28}^{+0.23}$ & $0.82_{-0.31}^{+0.23}$\\
\hline
\hline
\end{tabular}

\begin{flushleft}
$^\dagger$ The absorber ID is defined using the system number and component number. The ``s1c1'' label represents the first component in the first galaxy-absorption system.  For four components in the CUBSz1 sample, one component in CUBS III, and two in CUBS IV, an additional letter ``l'' (or ``h'') at the end of the ID represents the low-ionization/high density (high-ionization/low density) state of the multiphase gas.\\
$^\ddagger$ Mean line width determined from a combination of available metal-lines listed in Column (6), which include C, N, O, Mg, Si, or S ions originating in the same density phase.
\end{flushleft}
\end{table*}

\bibliographystyle{mnras}
\bibliography{ms}

\begin{thebibliography}{}
\makeatletter
\relax
\def\mn@urlcharsother{\let\do\@makeother \do\$\do\&\do\#\do\^\do\_\do\%\do\~}
\def\mn@doi{\begingroup\mn@urlcharsother \@ifnextchar [ {\mn@doi@}
  {\mn@doi@[]}}
\def\mn@doi@[#1]#2{\def\@tempa{#1}\ifx\@tempa\@empty \href
  {http://dx.doi.org/#2} {doi:#2}\else \href {http://dx.doi.org/#2} {#1}\fi
  \endgroup}
\def\mn@eprint#1#2{\mn@eprint@#1:#2::\@nil}
\def\mn@eprint@arXiv#1{\href {http://arxiv.org/abs/#1} {{\tt arXiv:#1}}}
\def\mn@eprint@dblp#1{\href {http://dblp.uni-trier.de/rec/bibtex/#1.xml}
  {dblp:#1}}
\def\mn@eprint@#1:#2:#3:#4\@nil{\def\@tempa {#1}\def\@tempb {#2}\def\@tempc
  {#3}\ifx \@tempc \@empty \let \@tempc \@tempb \let \@tempb \@tempa \fi \ifx
  \@tempb \@empty \def\@tempb {arXiv}\fi \@ifundefined
  {mn@eprint@\@tempb}{\@tempb:\@tempc}{\expandafter \expandafter \csname
  mn@eprint@\@tempb\endcsname \expandafter{\@tempc}}}

\bibitem[\protect\citeauthoryear{{Afruni}, {Fraternali}  \&
  {Pezzulli}}{{Afruni} et~al.}{2019}]{Afruni:2019}
{Afruni} A.,  {Fraternali} F.,   {Pezzulli} G.,  2019, \mn@doi [\aap]
  {10.1051/0004-6361/201835002}, \href
  {https://ui.adsabs.harvard.edu/abs/2019A&A...625A..11A} {625, A11}

\bibitem[\protect\citeauthoryear{{Astropy Collaboration} et~al.,}{{Astropy
  Collaboration} et~al.}{2013}]{astropy:2013}
{Astropy Collaboration} et~al., 2013, \mn@doi [\aap]
  {10.1051/0004-6361/201322068}, \href
  {http://adsabs.harvard.edu/abs/2013A%26A...558A..33A} {558, A33}

\bibitem[\protect\citeauthoryear{{Bacon} et~al.,}{{Bacon}
  et~al.}{2010}]{Bacon2010}
{Bacon} R.,  et~al., 2010, in {McLean} I.~S.,  {Ramsay} S.~K.,   {Takami} H.,
  eds,  Society of Photo-Optical Instrumentation Engineers (SPIE) Conference
  Series Vol. 7735, Ground-based and Airborne Instrumentation for Astronomy
  III. p. 773508, \mn@doi{10.1117/12.856027}

\bibitem[\protect\citeauthoryear{{Bernstein}, {Shectman}, {Gunnels},
  {Mochnacki}  \& {Athey}}{{Bernstein} et~al.}{2003}]{Bernstein2003}
{Bernstein} R.,  {Shectman} S.~A.,  {Gunnels} S.~M.,  {Mochnacki} S.,   {Athey}
  A.~E.,  2003, in {Iye} M.,  {Moorwood} A.~F.~M.,  eds,  Society of
  Photo-Optical Instrumentation Engineers (SPIE) Conference Series Vol. 4841,
  Instrument Design and Performance for Optical/Infrared Ground-based
  Telescopes. pp 1694--1704, \mn@doi{10.1117/12.461502}

\bibitem[\protect\citeauthoryear{{Boettcher} et~al.,}{{Boettcher}
  et~al.}{2021}]{Boettcher:2021aa}
{Boettcher} E.,  et~al., 2021, \mn@doi [\apj] {10.3847/1538-4357/abf0a0}, 913,
  18

\bibitem[\protect\citeauthoryear{{Bogd{\'a}n}, {Bourdin}, {Forman}, {Kraft},
  {Vogelsberger}, {Hernquist}  \& {Springel}}{{Bogd{\'a}n}
  et~al.}{2017}]{Bogdan:2017aa}
{Bogd{\'a}n} {\'A}.,  {Bourdin} H.,  {Forman} W.~R.,  {Kraft} R.~P.,
  {Vogelsberger} M.,  {Hernquist} L.,   {Springel} V.,  2017, \mn@doi [\apj]
  {10.3847/1538-4357/aa9523}, 850, 98

\bibitem[\protect\citeauthoryear{{Bregman}}{{Bregman}}{2007}]{Bregman:2007aa}
{Bregman} J.~N.,  2007, \mn@doi [\araa]
  {10.1146/annurev.astro.45.051806.110619}, 45, 221

\bibitem[\protect\citeauthoryear{{Butsky}, {Fielding}, {Hayward}, {Hummels},
  {Quinn}  \& {Werk}}{{Butsky} et~al.}{2020}]{Butsky:2020aa}
{Butsky} I.~S.,  {Fielding} D.~B.,  {Hayward} C.~C.,  {Hummels} C.~B.,  {Quinn}
  T.~R.,   {Werk} J.~K.,  2020, \apj, 903, 77

\bibitem[\protect\citeauthoryear{{Cen} \& {Ostriker}}{{Cen} \&
  {Ostriker}}{2006}]{Cen:2006aa}
{Cen} R.,  {Ostriker} J.~P.,  2006, \mn@doi [\apj] {10.1086/506505}, 650, 560

\bibitem[\protect\citeauthoryear{{Chadayammuri}, {Bogdan}, {Oppenheimer},
  {Kraft}, {Forman}  \& {Jones}}{{Chadayammuri}
  et~al.}{2022}]{Chadayammuri:2022aa}
{Chadayammuri} U.,  {Bogdan} A.,  {Oppenheimer} B.,  {Kraft} R.,  {Forman} W.,
   {Jones} C.,  2022, arXiv e-prints, \href
  {https://ui.adsabs.harvard.edu/abs/2022arXiv220301356C} {p. arXiv:2203.01356}

\bibitem[\protect\citeauthoryear{{Chen}, {Helsby}, {Gauthier}, {Shectman},
  {Thompson}  \& {Tinker}}{{Chen} et~al.}{2010}]{Chen:2010aa}
{Chen} H.-W.,  {Helsby} J.~E.,  {Gauthier} J.-R.,  {Shectman} S.~A.,
  {Thompson} I.~B.,   {Tinker} J.~L.,  2010, \mn@doi [\apj]
  {10.1088/0004-637X/714/2/1521}, \href
  {https://ui.adsabs.harvard.edu/abs/2010ApJ...714.1521C} {714, 1521}

\bibitem[\protect\citeauthoryear{{Chen}, {Zahedy}, {Johnson}, {Pierce},
  {Huang}, {Weiner}  \& {Gauthier}}{{Chen} et~al.}{2018}]{Chen:2018aa}
{Chen} H.-W.,  {Zahedy} F.~S.,  {Johnson} S.~D.,  {Pierce} R.~M.,  {Huang}
  Y.-H.,  {Weiner} B.~J.,   {Gauthier} J.-R.,  2018, \mn@doi [\mnras]
  {10.1093/mnras/sty1541}, 479, 2547

\bibitem[\protect\citeauthoryear{{Chen} et~al.,}{{Chen}
  et~al.}{2020}]{Chen:2020aa}
{Chen} H.-W.,  et~al., 2020, \mn@doi [\mnras] {10.1093/mnras/staa1773}, 497,
  498

\bibitem[\protect\citeauthoryear{{Commer{\c{c}}on}, {Marcowith}  \&
  {Dubois}}{{Commer{\c{c}}on} et~al.}{2019}]{Commercon:2019aa}
{Commer{\c{c}}on} B.,  {Marcowith} A.,   {Dubois} Y.,  2019, \aap, 622, A143

\bibitem[\protect\citeauthoryear{{Comparat} et~al.,}{{Comparat}
  et~al.}{2022}]{Comparat:2022aa}
{Comparat} J.,  et~al., 2022, arXiv e-prints, p. arXiv:2201.05169

\bibitem[\protect\citeauthoryear{{Conroy}, {van Dokkum}  \&
  {Kravtsov}}{{Conroy} et~al.}{2015}]{Conroy:2015aa}
{Conroy} C.,  {van Dokkum} P.~G.,   {Kravtsov} A.,  2015, \mn@doi [\apj]
  {10.1088/0004-637X/803/2/77}, \href
  {https://ui.adsabs.harvard.edu/abs/2015ApJ...803...77C} {803, 77}

\bibitem[\protect\citeauthoryear{{Cooke} \& {Fumagalli}}{{Cooke} \&
  {Fumagalli}}{2018}]{Cooke:2018aa}
{Cooke} R.~J.,  {Fumagalli} M.,  2018, Nature Astronomy, 2, 957

\bibitem[\protect\citeauthoryear{{Cooper} et~al.,}{{Cooper}
  et~al.}{2021}]{Cooper:2021aa}
{Cooper} T.~J.,  et~al., 2021, \mn@doi [\mnras] {10.1093/mnras/stab2869}, 508,
  4359

\bibitem[\protect\citeauthoryear{{Correa}, {Schaye}, {Wyithe}, {Duffy},
  {Theuns}, {Crain}  \& {Bower}}{{Correa} et~al.}{2018}]{Correa:2018aa}
{Correa} C.~A.,  {Schaye} J.,  {Wyithe} J. S.~B.,  {Duffy} A.~R.,  {Theuns} T.,
   {Crain} R.~A.,   {Bower} R.~G.,  2018, \mn@doi [\mnras]
  {10.1093/mnras/stx2332}, \href
  {https://ui.adsabs.harvard.edu/abs/2018MNRAS.473..538C} {473, 538}

\bibitem[\protect\citeauthoryear{{Das}, {Mathur}  \& {Gupta}}{{Das}
  et~al.}{2020}]{Das:2020aa}
{Das} S.,  {Mathur} S.,   {Gupta} A.,  2020, \mn@doi [\apj]
  {10.3847/1538-4357/ab93d2}, 897, 63

\bibitem[\protect\citeauthoryear{{Donahue} \& {Voit}}{{Donahue} \&
  {Voit}}{2022}]{Donahue2022}
{Donahue} M.,  {Voit} G.~M.,  2022, arXiv e-prints, \href
  {https://ui.adsabs.harvard.edu/abs/2022arXiv220408099D} {p. arXiv:2204.08099}

\bibitem[\protect\citeauthoryear{{Dressler} et~al.,}{{Dressler}
  et~al.}{2011}]{Dressler2011}
{Dressler} A.,  et~al., 2011, \mn@doi [\pasp] {10.1086/658908}, \href
  {https://ui.adsabs.harvard.edu/abs/2011PASP..123..288D} {123, 288}

\bibitem[\protect\citeauthoryear{{Faerman}, {Sternberg}  \& {McKee}}{{Faerman}
  et~al.}{2017}]{Faerman:2017aa}
{Faerman} Y.,  {Sternberg} A.,   {McKee} C.~F.,  2017, \mn@doi [\apj]
  {10.3847/1538-4357/835/1/52}, 835, 52

\bibitem[\protect\citeauthoryear{{Faucher-Gigu{\`e}re}}{{Faucher-Gigu{\`e}re}}{2020}]{Faucher-Giguere:2020aa}
{Faucher-Gigu{\`e}re} C.-A.,  2020, \mn@doi [\mnras] {10.1093/mnras/staa302},
  493, 1614

\bibitem[\protect\citeauthoryear{{Faucher-Gigu{\`e}re}, {Kere{\v{s}}}  \&
  {Ma}}{{Faucher-Gigu{\`e}re} et~al.}{2011}]{Faucher-Giguere:2011aa}
{Faucher-Gigu{\`e}re} C.-A.,  {Kere{\v{s}}} D.,   {Ma} C.-P.,  2011, \mn@doi
  [\mnras] {10.1111/j.1365-2966.2011.19457.x}, \href
  {https://ui.adsabs.harvard.edu/abs/2011MNRAS.417.2982F} {417, 2982}

\bibitem[\protect\citeauthoryear{{Ferland} et~al.,}{{Ferland}
  et~al.}{2017}]{Ferland:2017RMxAA}
{Ferland} G.~J.,  et~al., 2017, \rmxaa, \href
  {https://ui.adsabs.harvard.edu/abs/2017RMxAA..53..385F} {53, 385}

\bibitem[\protect\citeauthoryear{{Fielding}, {Ostriker}, {Bryan}  \&
  {Jermyn}}{{Fielding} et~al.}{2020}]{Fielding:2020aa}
{Fielding} D.~B.,  {Ostriker} E.~C.,  {Bryan} G.~L.,   {Jermyn} A.~S.,  2020,
  \mn@doi [\apjl] {10.3847/2041-8213/ab8d2c}, 894, L24

\bibitem[\protect\citeauthoryear{{Foreman-Mackey}, {Hogg}, {Lang}  \&
  {Goodman}}{{Foreman-Mackey} et~al.}{2013}]{Foreman-Mackey:2013aa}
{Foreman-Mackey} D.,  {Hogg} D.~W.,  {Lang} D.,   {Goodman} J.,  2013, \pasp,
  125, 306

\bibitem[\protect\citeauthoryear{{Gauthier}, {Chen}  \& {Tinker}}{{Gauthier}
  et~al.}{2009}]{Gauthier2009}
{Gauthier} J.-R.,  {Chen} H.-W.,   {Tinker} J.~L.,  2009, \mn@doi [\apj]
  {10.1088/0004-637X/702/1/50}, \href
  {https://ui.adsabs.harvard.edu/abs/2009ApJ...702...50G} {702, 50}

\bibitem[\protect\citeauthoryear{{Gauthier}, {Chen}  \& {Tinker}}{{Gauthier}
  et~al.}{2010}]{Gauthier2010}
{Gauthier} J.-R.,  {Chen} H.-W.,   {Tinker} J.~L.,  2010, \mn@doi [\apj]
  {10.1088/0004-637X/716/2/1263}, \href
  {https://ui.adsabs.harvard.edu/abs/2010ApJ...716.1263G} {716, 1263}

\bibitem[\protect\citeauthoryear{{Green} et~al.,}{{Green}
  et~al.}{2012}]{Green2012}
{Green} J.~C.,  et~al., 2012, \mn@doi [\apj] {10.1088/0004-637X/744/1/60},
  \href {https://ui.adsabs.harvard.edu/abs/2012ApJ...744...60G} {744, 60}

\bibitem[\protect\citeauthoryear{{Gronke}, {Oh}, {Ji}  \& {Norman}}{{Gronke}
  et~al.}{2022}]{Gronke:2022aa}
{Gronke} M.,  {Oh} S.~P.,  {Ji} S.,   {Norman} C.,  2022, \mn@doi [\mnras]
  {10.1093/mnras/stab3351}, 511, 859

\bibitem[\protect\citeauthoryear{{Haardt} \& {Madau}}{{Haardt} \&
  {Madau}}{2001}]{Haardt:2001aa}
{Haardt} F.,  {Madau} P.,  2001, in {Neumann} D.~M.,  {Tran} J.~T.~V.,  eds,
  Clusters of Galaxies and the High Redshift Universe Observed in X-rays. p.~64

\bibitem[\protect\citeauthoryear{{Haislmaier}, {Tripp}, {Katz}, {Prochaska},
  {Burchett}, {O'Meara}  \& {Werk}}{{Haislmaier}
  et~al.}{2021}]{Haislmaier:2021}
{Haislmaier} K.~J.,  {Tripp} T.~M.,  {Katz} N.,  {Prochaska} J.~X.,  {Burchett}
  J.~N.,  {O'Meara} J.~M.,   {Werk} J.~K.,  2021, \mn@doi [\mnras]
  {10.1093/mnras/staa3544}, \href
  {https://ui.adsabs.harvard.edu/abs/2021MNRAS.502.4993H} {502, 4993}

\bibitem[\protect\citeauthoryear{{Ho}, {Martin}  \& {Schaye}}{{Ho}
  et~al.}{2020}]{Ho:2020aa}
{Ho} S.~H.,  {Martin} C.~L.,   {Schaye} J.,  2020, \mn@doi [\apj]
  {10.3847/1538-4357/abbe88}, \href
  {https://ui.adsabs.harvard.edu/abs/2020ApJ...904...76H} {904, 76}

\bibitem[\protect\citeauthoryear{{Huang}, {Chen}, {Johnson}  \&
  {Weiner}}{{Huang} et~al.}{2016}]{Huang:2016}
{Huang} Y.-H.,  {Chen} H.-W.,  {Johnson} S.~D.,   {Weiner} B.~J.,  2016,
  \mn@doi [\mnras] {10.1093/mnras/stv2327}, \href
  {https://ui.adsabs.harvard.edu/abs/2016MNRAS.455.1713H} {455, 1713}

\bibitem[\protect\citeauthoryear{{Huang}, {Chen}, {Shectman}, {Johnson},
  {Zahedy}, {Helsby}, {Gauthier}  \& {Thompson}}{{Huang}
  et~al.}{2021}]{Huang:2021aa}
{Huang} Y.-H.,  {Chen} H.-W.,  {Shectman} S.~A.,  {Johnson} S.~D.,  {Zahedy}
  F.~S.,  {Helsby} J.~E.,  {Gauthier} J.-R.,   {Thompson} I.~B.,  2021, \mn@doi
  [\mnras] {10.1093/mnras/stab360}, 502, 4743

\bibitem[\protect\citeauthoryear{{Hunter}}{{Hunter}}{2007}]{Matplotlib}
{Hunter} J.~D.,  2007, \mn@doi [Computing in Science and Engineering]
  {10.1109/MCSE.2007.55}, \href
  {https://ui.adsabs.harvard.edu/abs/2007CSE.....9...90H} {9, 90}

\bibitem[\protect\citeauthoryear{{Hussain}, {Khaire}, {Srianand}, {Muzahid}  \&
  {Pathak}}{{Hussain} et~al.}{2017}]{Hussain:2017aa}
{Hussain} T.,  {Khaire} V.,  {Srianand} R.,  {Muzahid} S.,   {Pathak} A.,
  2017, \mnras, 466, 3133

\bibitem[\protect\citeauthoryear{{Ji}, {Oh}  \& {Masterson}}{{Ji}
  et~al.}{2019}]{Ji:2019aa}
{Ji} S.,  {Oh} S.~P.,   {Masterson} P.,  2019, \mn@doi [\mnras]
  {10.1093/mnras/stz1248}, 487, 737

\bibitem[\protect\citeauthoryear{{Johnson}, {Chen}  \& {Mulchaey}}{{Johnson}
  et~al.}{2013}]{Johnson:2013aa}
{Johnson} S.~D.,  {Chen} H.-W.,   {Mulchaey} J.~S.,  2013, \mnras, 434, 1765

\bibitem[\protect\citeauthoryear{{Johnson}, {Chen}  \& {Mulchaey}}{{Johnson}
  et~al.}{2015}]{Johnson:2015aa}
{Johnson} S.~D.,  {Chen} H.-W.,   {Mulchaey} J.~S.,  2015, \mnras, 449, 3263

\bibitem[\protect\citeauthoryear{{Johnson}, {Chen}, {Mulchaey}, {Schaye}  \&
  {Straka}}{{Johnson} et~al.}{2017}]{Johnson:2017aa}
{Johnson} S.~D.,  {Chen} H.-W.,  {Mulchaey} J.~S.,  {Schaye} J.,   {Straka}
  L.~A.,  2017, \apjl, 850, L10

\bibitem[\protect\citeauthoryear{{Karwin}, {Murgia}, {Moskalenko},
  {Fillingham}, {Burns}  \& {Fieg}}{{Karwin} et~al.}{2021}]{Karwin:2021aa}
{Karwin} C.~M.,  {Murgia} S.,  {Moskalenko} I.~V.,  {Fillingham} S.~P.,
  {Burns} A.-K.,   {Fieg} M.,  2021, \mn@doi [\prd]
  {10.1103/PhysRevD.103.023027}, 103, 023027

\bibitem[\protect\citeauthoryear{{Keeney} et~al.,}{{Keeney}
  et~al.}{2017}]{Keeney:2017aa}
{Keeney} B.~A.,  et~al., 2017, \apjs, 230, 6

\bibitem[\protect\citeauthoryear{{Kere{\v s}}, {Katz}, {Weinberg}  \&
  {Dav{\'e}}}{{Kere{\v s}} et~al.}{2005}]{Keres:2005aa}
{Kere{\v s}} D.,  {Katz} N.,  {Weinberg} D.~H.,   {Dav{\'e}} R.,  2005, \mn@doi
  [\mnras] {10.1111/j.1365-2966.2005.09451.x}, 363, 2

\bibitem[\protect\citeauthoryear{{Khaire} \& {Srianand}}{{Khaire} \&
  {Srianand}}{2019}]{Khaire:2019aa}
{Khaire} V.,  {Srianand} R.,  2019, \mnras, 484, 4174

\bibitem[\protect\citeauthoryear{{Kim}, {Carswell}  \& {Ranquist}}{{Kim}
  et~al.}{2016}]{Kim2016}
{Kim} T.~S.,  {Carswell} R.~F.,   {Ranquist} D.,  2016, \mn@doi [\mnras]
  {10.1093/mnras/stv2847}, \href
  {https://ui.adsabs.harvard.edu/abs/2016MNRAS.456.3509K} {456, 3509}

\bibitem[\protect\citeauthoryear{{Lehner}, {Wotta}, {Howk}, {O'Meara},
  {Oppenheimer}  \& {Cooksey}}{{Lehner} et~al.}{2019}]{Lehner:2019aa}
{Lehner} N.,  {Wotta} C.~B.,  {Howk} J.~C.,  {O'Meara} J.~M.,  {Oppenheimer}
  B.~D.,   {Cooksey} K.~L.,  2019, \apj, 887, 5

\bibitem[\protect\citeauthoryear{{Li}, {Bryan}, {Ruszkowski}, {Voit}, {O'Shea}
  \& {Donahue}}{{Li} et~al.}{2015}]{Li:2015aa}
{Li} Y.,  {Bryan} G.~L.,  {Ruszkowski} M.,  {Voit} G.~M.,  {O'Shea} B.~W.,
  {Donahue} M.,  2015, \apj, 811, 73

\bibitem[\protect\citeauthoryear{{Li}, {Bregman}, {Wang}, {Crain}  \&
  {Anderson}}{{Li} et~al.}{2018}]{Li:2018aa}
{Li} J.-T.,  {Bregman} J.~N.,  {Wang} Q.~D.,  {Crain} R.~A.,   {Anderson}
  M.~E.,  2018, \mn@doi [\apjl] {10.3847/2041-8213/aab2af}, 855, L24

\bibitem[\protect\citeauthoryear{{Li}, {Li}, {Bryan}, {Ostriker}  \&
  {Quataert}}{{Li} et~al.}{2020}]{Li:2020aa}
{Li} M.,  {Li} Y.,  {Bryan} G.~L.,  {Ostriker} E.~C.,   {Quataert} E.,  2020,
  \mn@doi [\apj] {10.3847/1538-4357/ab9c22}, \href
  {https://ui.adsabs.harvard.edu/abs/2020ApJ...898...23L} {898, 23}

\bibitem[\protect\citeauthoryear{{Liang} \& {Chen}}{{Liang} \&
  {Chen}}{2014}]{Liang:2014aa}
{Liang} C.~J.,  {Chen} H.-W.,  2014, \mnras, 445, 2061

\bibitem[\protect\citeauthoryear{{Liang} \& {Remming}}{{Liang} \&
  {Remming}}{2020}]{Liang:2020aa}
{Liang} C.~J.,  {Remming} I.,  2020, \mn@doi [\mnras] {10.1093/mnras/stz3403},
  491, 5056

\bibitem[\protect\citeauthoryear{{Madau} \& {Dickinson}}{{Madau} \&
  {Dickinson}}{2014}]{Madau:2014aa}
{Madau} P.,  {Dickinson} M.,  2014, \araa, 52, 415

\bibitem[\protect\citeauthoryear{{Maller} \& {Bullock}}{{Maller} \&
  {Bullock}}{2004}]{Maller:2004}
{Maller} A.~H.,  {Bullock} J.~S.,  2004, \mn@doi [\mnras]
  {10.1111/j.1365-2966.2004.08349.x}, \href
  {https://ui.adsabs.harvard.edu/abs/2004MNRAS.355..694M} {355, 694}

\bibitem[\protect\citeauthoryear{{McCourt}, {Oh}, {O'Leary}  \&
  {Madigan}}{{McCourt} et~al.}{2018}]{McCourt:2018aa}
{McCourt} M.,  {Oh} S.~P.,  {O'Leary} R.,   {Madigan} A.-M.,  2018, \mnras,
  473, 5407

\bibitem[\protect\citeauthoryear{{Meiring}, {Tripp}, {Werk}, {Howk}, {Jenkins},
  {Prochaska}, {Lehner}  \& {Sembach}}{{Meiring} et~al.}{2013}]{Meiring:2013aa}
{Meiring} J.~D.,  {Tripp} T.~M.,  {Werk} J.~K.,  {Howk} J.~C.,  {Jenkins}
  E.~B.,  {Prochaska} J.~X.,  {Lehner} N.,   {Sembach} K.~R.,  2013, \apj, 767,
  49

\bibitem[\protect\citeauthoryear{{Naab} \& {Ostriker}}{{Naab} \&
  {Ostriker}}{2017}]{Naab2017}
{Naab} T.,  {Ostriker} J.~P.,  2017, \mn@doi [\araa]
  {10.1146/annurev-astro-081913-040019}, \href
  {https://ui.adsabs.harvard.edu/abs/2017ARA&A..55...59N} {55, 59}

\bibitem[\protect\citeauthoryear{{Nelson}, {Vogelsberger}, {Genel}, {Sijacki},
  {Kere{\v s}}, {Springel}  \& {Hernquist}}{{Nelson}
  et~al.}{2013}]{Nelson:2013aa}
{Nelson} D.,  {Vogelsberger} M.,  {Genel} S.,  {Sijacki} D.,  {Kere{\v s}} D.,
  {Springel} V.,   {Hernquist} L.,  2013, \mn@doi [\mnras]
  {10.1093/mnras/sts595}, 429, 3353

\bibitem[\protect\citeauthoryear{{Nelson} et~al.,}{{Nelson}
  et~al.}{2020}]{Nelson:2020aa}
{Nelson} D.,  et~al., 2020, \mn@doi [\mnras] {10.1093/mnras/staa2419}, \href
  {https://ui.adsabs.harvard.edu/abs/2020MNRAS.498.2391N} {498, 2391}

\bibitem[\protect\citeauthoryear{Oliphant}{Oliphant}{2015}]{Numpy}
Oliphant T.~E.,  2015, Guide to NumPy, 2nd edn.
CreateSpace Independent Publishing Platform, USA

\bibitem[\protect\citeauthoryear{{Oppenheimer} \& {Schaye}}{{Oppenheimer} \&
  {Schaye}}{2013}]{Oppenheimer:2013aa}
{Oppenheimer} B.~D.,  {Schaye} J.,  2013, \mnras, 434, 1043

\bibitem[\protect\citeauthoryear{{Osip}, {Floyd}  \& {Covarrubias}}{{Osip}
  et~al.}{2008}]{Osip:2008aa}
{Osip} D.~J.,  {Floyd} D.,   {Covarrubias} R.,  2008, in {McLean} I.~S.,
  {Casali} M.~M.,  eds,  Society of Photo-Optical Instrumentation Engineers
  (SPIE) Conference Series Vol. 7014, Ground-based and Airborne Instrumentation
  for Astronomy II. p. 70140A, \mn@doi{10.1117/12.790011}

\bibitem[\protect\citeauthoryear{{Price-Whelan} et~al.,}{{Price-Whelan}
  et~al.}{2018}]{astropy:2018}
{Price-Whelan} A.~M.,  et~al., 2018, \mn@doi [\aj] {10.3847/1538-3881/aabc4f},
  \href {https://ui.adsabs.harvard.edu/#abs/2018AJ....156..123T} {156, 123}

\bibitem[\protect\citeauthoryear{{Prochaska} et~al.,}{{Prochaska}
  et~al.}{2017}]{Prochaska:2017aa}
{Prochaska} J.~X.,  et~al., 2017, \apj, 837, 169

\bibitem[\protect\citeauthoryear{{Qu} \& {Bregman}}{{Qu} \&
  {Bregman}}{2018}]{Qu:2018aa}
{Qu} Z.,  {Bregman} J.~N.,  2018, \mn@doi [\apj] {10.3847/1538-4357/aaafd4},
  856, 5

\bibitem[\protect\citeauthoryear{{Rahmati}, {Schaye}, {Crain}, {Oppenheimer},
  {Schaller}  \& {Theuns}}{{Rahmati} et~al.}{2016}]{Rahmati2016}
{Rahmati} A.,  {Schaye} J.,  {Crain} R.~A.,  {Oppenheimer} B.~D.,  {Schaller}
  M.,   {Theuns} T.,  2016, \mn@doi [\mnras] {10.1093/mnras/stw453}, \href
  {https://ui.adsabs.harvard.edu/abs/2016MNRAS.459..310R} {459, 310}

\bibitem[\protect\citeauthoryear{{Rauch}, {Sargent}, {Womble}  \&
  {Barlow}}{{Rauch} et~al.}{1996}]{Rauch1996}
{Rauch} M.,  {Sargent} W.~L.~W.,  {Womble} D.~S.,   {Barlow} T.~A.,  1996,
  \mn@doi [\apjl] {10.1086/310187}, \href
  {https://ui.adsabs.harvard.edu/abs/1996ApJ...467L...5R} {467, L5}

\bibitem[\protect\citeauthoryear{{Rauch}, {Sargent}, {Barlow}  \&
  {Carswell}}{{Rauch} et~al.}{2001}]{Rauch2001bb}
{Rauch} M.,  {Sargent} W. L.~W.,  {Barlow} T.~A.,   {Carswell} R.~F.,  2001,
  \mn@doi [\apj] {10.1086/323523}, \href
  {https://ui.adsabs.harvard.edu/abs/2001ApJ...562...76R} {562, 76}

\bibitem[\protect\citeauthoryear{{Rudie}, {Steidel}, {Pettini}, {Trainor},
  {Strom}, {Hummels}, {Reddy}  \& {Shapley}}{{Rudie}
  et~al.}{2019}]{Rudie:2019aa}
{Rudie} G.~C.,  {Steidel} C.~C.,  {Pettini} M.,  {Trainor} R.~F.,  {Strom}
  A.~L.,  {Hummels} C.~B.,  {Reddy} N.~A.,   {Shapley} A.~E.,  2019, \mn@doi
  [\apj] {10.3847/1538-4357/ab4255}, 885, 61

\bibitem[\protect\citeauthoryear{{Sameer} et~al.,}{{Sameer}
  et~al.}{2021}]{Sameer:2021aa}
{Sameer} et~al., 2021, \mn@doi [\mnras] {10.1093/mnras/staa3754}, 501, 2112

\bibitem[\protect\citeauthoryear{{Savage}, {Lehner}, {Wakker}, {Sembach}  \&
  {Tripp}}{{Savage} et~al.}{2005}]{Savage:2005aa}
{Savage} B.~D.,  {Lehner} N.,  {Wakker} B.~P.,  {Sembach} K.~R.,   {Tripp}
  T.~M.,  2005, \mn@doi [\apj] {10.1086/429985}, 626, 776

\bibitem[\protect\citeauthoryear{{Savage}, {Kim}, {Wakker}, {Keeney}, {Shull},
  {Stocke}  \& {Green}}{{Savage} et~al.}{2014}]{Savage:2014aa}
{Savage} B.~D.,  {Kim} T.-S.,  {Wakker} B.~P.,  {Keeney} B.,  {Shull} J.~M.,
  {Stocke} J.~T.,   {Green} J.~C.,  2014, \mn@doi [\apjs]
  {10.1088/0067-0049/212/1/8}, 212, 8

\bibitem[\protect\citeauthoryear{{Shull}, {Danforth}  \& {Tilton}}{{Shull}
  et~al.}{2014}]{Shull:2014aa}
{Shull} J.~M.,  {Danforth} C.~W.,   {Tilton} E.~M.,  2014, \apj, 796, 49

\bibitem[\protect\citeauthoryear{{Simcoe}, {Sargent}, {Rauch}  \&
  {Becker}}{{Simcoe} et~al.}{2006}]{Simcoe:2006aa}
{Simcoe} R.~A.,  {Sargent} W. L.~W.,  {Rauch} M.,   {Becker} G.,  2006, \apj,
  637, 648

\bibitem[\protect\citeauthoryear{{Singh}, {Majumdar}, {Nath}  \&
  {Silk}}{{Singh} et~al.}{2018}]{Singh:2018}
{Singh} P.,  {Majumdar} S.,  {Nath} B.~B.,   {Silk} J.,  2018, \mn@doi [\mnras]
  {10.1093/mnras/sty1276}, \href
  {https://ui.adsabs.harvard.edu/abs/2018MNRAS.478.2909S} {478, 2909}

\bibitem[\protect\citeauthoryear{{Sparre}, {Pfrommer}  \&
  {Vogelsberger}}{{Sparre} et~al.}{2019}]{Sparre:2019aa}
{Sparre} M.,  {Pfrommer} C.,   {Vogelsberger} M.,  2019, \mn@doi [\mnras]
  {10.1093/mnras/sty3063}, 482, 5401

\bibitem[\protect\citeauthoryear{{Stern} et~al.,}{{Stern}
  et~al.}{2021}]{Stern:2021aa}
{Stern} J.,  et~al., 2021, \mn@doi [\apj] {10.3847/1538-4357/abd776}, \href
  {https://ui.adsabs.harvard.edu/abs/2021ApJ...911...88S} {911, 88}

\bibitem[\protect\citeauthoryear{{Stocke}, {Keeney}, {Danforth}, {Shull},
  {Froning}, {Green}, {Penton}  \& {Savage}}{{Stocke}
  et~al.}{2013}]{Stocke:2013}
{Stocke} J.~T.,  {Keeney} B.~A.,  {Danforth} C.~W.,  {Shull} J.~M.,  {Froning}
  C.~S.,  {Green} J.~C.,  {Penton} S.~V.,   {Savage} B.~D.,  2013, \mn@doi
  [\apj] {10.1088/0004-637X/763/2/148}, \href
  {https://ui.adsabs.harvard.edu/abs/2013ApJ...763..148S} {763, 148}

\bibitem[\protect\citeauthoryear{{Tumlinson} et~al.,}{{Tumlinson}
  et~al.}{2011}]{Tumlinson:2011aa}
{Tumlinson} J.,  et~al., 2011, Science, 334, 948

\bibitem[\protect\citeauthoryear{{Tumlinson} et~al.,}{{Tumlinson}
  et~al.}{2013}]{Tumlinson:2013aa}
{Tumlinson} J.,  et~al., 2013, \apj, 777, 59

\bibitem[\protect\citeauthoryear{{Tumlinson}, {Peeples}  \& {Werk}}{{Tumlinson}
  et~al.}{2017}]{Tumlinson:2017aa}
{Tumlinson} J.,  {Peeples} M.~S.,   {Werk} J.~K.,  2017, \mn@doi [\araa]
  {10.1146/annurev-astro-091916-055240}, 55, 389

\bibitem[\protect\citeauthoryear{{Turner}, {Schaye}, {Crain}, {Rudie},
  {Steidel}, {Strom}  \& {Theuns}}{{Turner} et~al.}{2017}]{Turner:2017aa}
{Turner} M.~L.,  {Schaye} J.,  {Crain} R.~A.,  {Rudie} G.,  {Steidel} C.~C.,
  {Strom} A.,   {Theuns} T.,  2017, \mn@doi [\mnras] {10.1093/mnras/stx1616},
  \href {https://ui.adsabs.harvard.edu/abs/2017MNRAS.471..690T} {471, 690}

\bibitem[\protect\citeauthoryear{{Upton Sanderbeck}, {McQuinn}, {D'Aloisio}  \&
  {Werk}}{{Upton Sanderbeck} et~al.}{2018}]{Upton-Sanderbeck:2018aa}
{Upton Sanderbeck} P.~R.,  {McQuinn} M.,  {D'Aloisio} A.,   {Werk} J.~K.,
  2018, \apj, 869, 159

\bibitem[\protect\citeauthoryear{{Virtanen} et~al.,}{{Virtanen}
  et~al.}{2020}]{Scipy}
{Virtanen} P.,  et~al., 2020, \mn@doi [Nature Methods]
  {10.1038/s41592-019-0686-2}, \href
  {https://ui.adsabs.harvard.edu/abs/2020NatMe..17..261V} {17, 261}

\bibitem[\protect\citeauthoryear{{Voit}, {Donahue}, {Zahedy}, {Chen}, {Werk},
  {Bryan}  \& {O'Shea}}{{Voit} et~al.}{2019}]{Voit:2019aa}
{Voit} G.~M.,  {Donahue} M.,  {Zahedy} F.,  {Chen} H.-W.,  {Werk} J.,  {Bryan}
  G.~L.,   {O'Shea} B.~W.,  2019, \mn@doi [\apjl] {10.3847/2041-8213/ab2766},
  879, L1

\bibitem[\protect\citeauthoryear{{Wakker}, {Hernandez}, {French}, {Kim},
  {Oppenheimer}  \& {Savage}}{{Wakker} et~al.}{2015}]{Wakker2015aa}
{Wakker} B.~P.,  {Hernandez} A.~K.,  {French} D.~M.,  {Kim} T.-S.,
  {Oppenheimer} B.~D.,   {Savage} B.~D.,  2015, \mn@doi [\apj]
  {10.1088/0004-637X/814/1/40}, \href
  {https://ui.adsabs.harvard.edu/abs/2015ApJ...814...40W} {814, 40}

\bibitem[\protect\citeauthoryear{{Werk} et~al.,}{{Werk}
  et~al.}{2014}]{Werk:2014aa}
{Werk} J.~K.,  et~al., 2014, \apj, 792, 8

\bibitem[\protect\citeauthoryear{{Werk} et~al.,}{{Werk}
  et~al.}{2016}]{Werk:2016aa}
{Werk} J.~K.,  et~al., 2016, \apj, 833, 54

\bibitem[\protect\citeauthoryear{{Werner}, {McNamara}, {Churazov}  \&
  {Scannapieco}}{{Werner} et~al.}{2019}]{Werner:2019aa}
{Werner} N.,  {McNamara} B.~R.,  {Churazov} E.,   {Scannapieco} E.,  2019,
  \mn@doi [\ssr] {10.1007/s11214-018-0571-9}, \href
  {https://ui.adsabs.harvard.edu/abs/2019SSRv..215....5W} {215, 5}

\bibitem[\protect\citeauthoryear{{Zahedy}, {Chen}, {Rauch}, {Wilson}  \&
  {Zabludoff}}{{Zahedy} et~al.}{2016}]{Zahedy:2016aa}
{Zahedy} F.~S.,  {Chen} H.-W.,  {Rauch} M.,  {Wilson} M.~L.,   {Zabludoff} A.,
  2016, \mn@doi [\mnras] {10.1093/mnras/stw484}, 458, 2423

\bibitem[\protect\citeauthoryear{{Zahedy}, {Chen}, {Johnson}, {Pierce},
  {Rauch}, {Huang}, {Weiner}  \& {Gauthier}}{{Zahedy}
  et~al.}{2019}]{Zahedy:2019aa}
{Zahedy} F.~S.,  {Chen} H.-W.,  {Johnson} S.~D.,  {Pierce} R.~M.,  {Rauch} M.,
  {Huang} Y.-H.,  {Weiner} B.~J.,   {Gauthier} J.-R.,  2019, \mn@doi [\mnras]
  {10.1093/mnras/sty3482}, 484, 2257

\bibitem[\protect\citeauthoryear{{Zahedy}, {Chen}, {Boettcher}, {Rauch},
  {French}  \& {Zabludoff}}{{Zahedy} et~al.}{2020}]{Zahedy:2020}
{Zahedy} F.~S.,  {Chen} H.-W.,  {Boettcher} E.,  {Rauch} M.,  {French} K.~D.,
  {Zabludoff} A.~I.,  2020, \mn@doi [\apjl] {10.3847/2041-8213/abc48d}, \href
  {https://ui.adsabs.harvard.edu/abs/2020ApJ...904L..10Z} {904, L10}

\bibitem[\protect\citeauthoryear{{Zahedy} et~al.,}{{Zahedy}
  et~al.}{2021}]{Zahedy:2021aa}
{Zahedy} F.~S.,  et~al., 2021, \mn@doi [\mnras] {10.1093/mnras/stab1661}, 506,
  877

\bibitem[\protect\citeauthoryear{{van de Voort} \& {Schaye}}{{van de Voort} \&
  {Schaye}}{2012}]{vandeVoort:2012aa}
{van de Voort} F.,  {Schaye} J.,  2012, \mn@doi [\mnras]
  {10.1111/j.1365-2966.2012.20949.x}, \href
  {https://ui.adsabs.harvard.edu/abs/2012MNRAS.423.2991V} {423, 2991}

\bibitem[\protect\citeauthoryear{{van de Voort}, {Schaye}, {Booth}, {Haas}  \&
  {Dalla Vecchia}}{{van de Voort} et~al.}{2011}]{vandeVoort:2011aa}
{van de Voort} F.,  {Schaye} J.,  {Booth} C.~M.,  {Haas} M.~R.,   {Dalla
  Vecchia} C.,  2011, \mn@doi [\mnras] {10.1111/j.1365-2966.2011.18565.x},
  \href {https://ui.adsabs.harvard.edu/abs/2011MNRAS.414.2458V} {414, 2458}

\makeatother
\end{thebibliography}


\bsp	
\label{lastpage}
\end{document}